\documentclass[12pt]{article}
\usepackage{float}
\usepackage{authblk}
\usepackage{amssymb,amsmath}
\usepackage{graphicx}
\usepackage{caption}
\usepackage{subcaption}
\usepackage{cite}
\usepackage{xcolor}
\usepackage{setspace}
\usepackage{appendix}
\usepackage{makecell}
\usepackage{array}
\usepackage{tabularx}
\newcolumntype{M}[1]{>{\centering\arraybackslash}m{#1}}
\usepackage[colorlinks=true,urlcolor=blue,citecolor=blue,linkcolor=red,linktocpage=true,pdfproducer=medialab]{hyperref}
\usepackage[a4paper,width=17cm,height=25cm]{geometry}
\usepackage{morefloats}
\usepackage{siunitx}
\makeatletter
\renewcommand{\@dotsep}{10000}
\makeatother
\allowdisplaybreaks[4]

\renewcommand{\arraystretch}{1.2}
\begin{document}
\title{Reinvestigating the semileptonic $B\to D^{(\ast)}\tau\bar{\nu}_{\tau}$ decays in the model independent scenarios and leptoquark models}
\author{
\parbox{\textwidth}{\centering Zhuo-Ran Huang\thanks{\texttt huangzhuoran@126.com }\vspace{-0.7em}\\
    \textit{Department of Physics, College of Physics, Mechanical and Electrical Engineering, Jishou University, Jishou 416000, China}
    \vspace{0.7em} \\
    Faisal Munir Bhutta\thanks{\texttt faisal.munir@sns.nust.edu.pk }, Nimra Farooq\thanks{\texttt nimrafarooq\_1995@hotmail.com }, and
    M. Ali Paracha\thanks{\texttt aliparacha@sns.nust.edu.pk}\\
    \textit{School of Natural Sciences (SNS), National University of Sciences and Technology (NUST), Sector H-12, Islamabad, Pakistan}
    \vspace{0.5em} \\
    Ying Li\thanks{\texttt liying@ytu.edu.cn} \\
    \textit{Department of Physics,Yantai University,Yantai 264005, China}
}
}
\maketitle
\begin{abstract}
In this work, we revisit the possible new physics (NP) solutions by analyzing the observables associated with $B\to D^{(\ast)}\tau\bar{\nu}_{\tau}$ decays. To explore the structure of new physics, the form factors of $B\to D^{(\ast)}$ decays play a crucial role. In this study, we utilize the form factor results obtained from a simultaneous fit to Belle data and lattice QCD calculations. Using these form factors, we conduct a global fit of the new physics Wilson coefficients, incorporating the most recent experimental data. Additionally, we use the form factors and Wilson coefficients to make predictions for physical observables, including the lepton flavor universality ratio $R_{D^{(\ast)}}$, polarized observables, and the normalized angular coefficients $\langle I\rangle$'s related to the four-fold decays $B\to D^{\ast}(\to D\pi, D\gamma)\tau\bar{\nu}_{\tau}$. These predictions are calculated in both model-independent scenarios and for three different leptoquark models.
\end{abstract}
\newpage
\section{Introduction}
The observables associated with $B\to D^{(\ast)}\tau\bar{\nu}_{\tau}$ semileptonic decay have attracted a lot of attention, particularly the lepton flavor universality (LFU) ratio, $R_{D^{(\ast)}}$, since it has been reported that the global average measurements by different experiments such as Babar~\cite{BaBar:2012obs,BaBar:2013mob}, Belle\cite{Belle:2015qfa, Belle:2016dyj, Belle:2017ilt, Belle:2019rba}, LHCb Run 1 and Run 2 ~\cite{LHCb:2015gmp, LHCb:2017smo, LHCb:2017rln}, and very recently by Belle II collaboration show a departure from the Standard Model (SM) prediction by about $3.3\sigma$ (for review of the anomalies, see for instance ~\cite{Capdevila:2023yhq, Bifani:2018zmi, Li:2018lxi}). Many efforts have been made to address these anomalies both in model independent and model dependent ways. As the predicted SM values of the ratio $R(D^{(\ast)})$ are smaller than the measured ones, the adopted model independent~\cite{Huang:2018nnq,Huang:2021fuc} and model dependent approaches~\cite{Cheung:2020sbq} tend to enhance the $B\to D^{(\ast)}\tau\bar{\nu}_{\tau}$ rate. Among the models used to address these anomalies are two-Higgs-doublet models (2HDM) \cite{Crivellin:2012ye, Celis:2012dk, Crivellin:2015hha, Wei:2017ago, Chen:2017eby}, $W^{\prime}$ models ~\cite{Asadi:2018wea, Greljo:2018ogz, Gomez:2019xfw}, and leptoquark (LQ) models~\cite{Sakaki:2013bfa, Li:2016vvp, Bansal:2018nwp}. Furthermore, the measured values are also used to constrain the values of new physics Wilson coefficients~\cite{Iguro:2024hyk}.

In this work, we explore the signatures of NP following the model independent approach and LQ models. To explore the said NP approaches, we consider both lepton flavor dependent (LFD) and LFU observables associated with $B\to D^{(\ast)}\tau\bar{\nu}_{\tau}$ decays. Moreover, we also explore the NP through the angular coefficients which appear in the four-fold distribution of $B\to D^{\ast}(\to D\pi, D\gamma)\tau\bar{\nu}_{\tau}$ decays\cite{Colangelo:2024mxe,Colangelo:2019axi,Colangelo:2021dnv,Kapoor:2024ufg}. The framework which we use to analyze the structure of NP is the low energy effective field theory which contains the four fermion operators assuming the neutrinos to be left handed. For the case of model independent scenarios the various Dirac structures of four fermion operators contribute in the effective Hamiltonian. These structures can be originated in different NP models, for instance, the right-handed contribution appears in left-right symmetric models\cite{Kou:2013gna,Pati:1974yy,Mohapatra:1974hk,Mohapatra:1974gc,Mohapatra:1980yp,Blanke:2011ry,Barenboim:1996nd,Kiers:2002cz}. However the scalar contributions appears in a model with extra charged scalar particles\cite{Kalinowski:1990ba,Hou:1992sy,Crivellin:2013wna}, and tensor operators arise after Fierz rearrangement on scalar operators\cite{Fedele:2023ewe}. To analyze the structure of NP through model independent scenarios and LQ models, angular distributions are expected to be a useful tool.

Belle has utilized the angular distribution data from the $B\to D^{\ast}\ell\bar{\nu}_{\ell}$ decay to perform a simultaneous fit of both CKM matrix $|V_{cb}|$ and the relevant transition form factors\cite{Belle:2017rcc,Belle:2018ezy}. The four-fold angular distribution can be expressed in terms of four kinematical variables, the dilepton invariant mass squared, $q^2$, and the three angles. The expression of four-fold distributions for $B\to D^{\ast}(\to D\pi,D\gamma)\tau\bar{\nu}_{\tau}$ will be presented in Section \ref{obs}. In the analysis of the said decay, the CKM parameter $|V_{cb}|$ is multiplied with a scaling parameter $\mathcal{F}(\omega)\eta_{\text{EW}}$ at zero recoil of the form factors, and at the zero recoil $\mathcal{F}(1)=h_{A_{1}}$\cite{Kapoor:2024ufg}. It is important to mention here that with the inclusion of NP effects, the number of parameters gets increased such that the experimental data by itself cannot fit the form factors, CKM parameter $V_{cb}$, and NP parameters simultaneously.

In our analysis, we use the “Lattice+ experiment” form factors obtained recently in \cite{Ray:2023xjn}, to first update the constraints on the NP Wilson coefficients and later to analyze the implications of the obtained NP patterns on the observables associated with $B\to D^{(\ast)}\tau\bar{\nu}_{\tau}$ decays. For $B\to D\ell\bar{\nu}$ decay, the Belle collaboration\cite{Belle:2015pkj}, provided measurements for differential decay width for the kinematical variable $w$ in 10 bins for both neutral and charged $B$ mesons with the muon or electron as a final state lepton. Later on the results on angular distribution appearing in four-fold $B\to D^{\ast}\ell\bar{\nu}$ decay were also presented\cite{Belle:2018ezy,Belle:2023bwv}.

The organization of the paper is as follows:  In Sec. \ref{TH}, we present the effective Hamiltonian within the SM and beyond it for $B\to D^{(\ast)}\ell\bar{\nu}_{\ell}$ decays. In Sec.~\ref{FFP}, the matrix elements in terms of transition form factors are given, where the form factors are discussed in the BGL parametrization. The semi-analytical expressions for the observables associated with $B\to D^{(\ast)}\ell\bar{\nu}_{\ell}$ decays are presented in Sec.~\ref{obs}. In Sec.~\ref{NA}, we present our phenomenological analysis of the observables in these decays both in the model independent and Leptoquark (LQ) scenarios. Finally, we conclude our study in Sec.~\ref{Conc}.

\section{Effective Hamiltonian for \texorpdfstring {$B\to D^{(\ast)}\tau\bar{\nu}_{\tau}$}{} Decay in SM and Beyond}\label{TH}
To investigate the structure of NP through $B\to D^{(\ast)}\tau\bar{\nu}_{\tau}$, a framework of low energy effective field theory has been used. In this framework the heavy degrees of freedom are integrated out and the effective Hamiltonian emerges in the form of Wilson coefficients and four fermion operators.

At quark level, the decay modes $B\to D^{(\ast)}\tau\bar{\nu}_{\tau}$ are governed by $b\to c\tau\bar{\nu}_{\tau}$ transition, the effective Hamiltonian in SM and beyond can be written as:
\begin{equation}
   \mathcal{H}_{\text{eff}} =\frac{4G_{F}}{\sqrt{2}} V_{cb}\left[ (1 + C_{V_1})O_{V_1} + C_{V_2}O_{V_2} + C_{S_1}O_{S_1} + C_{S_2}O_{S_2} + C_TO_T \right] + \text{h.c.} \,,
      \label{eq:lag}
\end{equation}
where $G_{F}$ is the Fermi coupling constant, $C_{X}(X=V_{1},V_{2},S_{1},S_{2},T)$ represents the Wilson coefficients and $O_{X}$ are four fermion operators with different Chiral and Lorentz structure. The explicit form the operators $O_{X}$ with left and right handed chiralities can be expressed as,
\begin{eqnarray}
 &O_{S_1} =  (\bar{c}_L b_R)(\bar{\tau}_R \nu_{L}) \,, \,\,\,
    O_{S_2} =  (\bar{c}_R b_L)(\bar{\tau}_R \nu_{L}) \,, \nonumber \\
  &  O_{V_1} = (\bar{c}_L \gamma^\mu b_L)(\bar{\tau}_L \gamma_\mu \nu_{L}) \,,  \,\,\,
     O_{V_2} = (\bar{c}_R \gamma^\mu b_R)(\bar{\tau}_L \gamma_\mu \nu_{L}) \,, \nonumber \\
  & O_T =  (\bar{c}_R \sigma^{\mu\nu} b_L)(\bar{\tau}_R \sigma_{\mu\nu} \nu_{L}) \,.
   \label{eq:operators}
\end{eqnarray}
It is important to note here that only the operator $O_{V_{1}}$ is present in the SM and the Wilson coefficients of $O_{X}$ can get modified at the short-distance scale. The effective Hamiltonian given in Eq.(\ref{eq:lag}) can be used to calculate the physical observables such as lepton flavor universality ratio $R_{D^{(\ast)}}$, lepton polarization asymmetry $P_{\tau}(D^{(\ast)})$, longitudinal helicity fraction $F^{D^{\ast}}_{L}$ of $D^{\ast}$ meson, forward backward asymmetry $\mathcal{A}^{\text{FB}}_{D^{(\ast)}}$, and normalized angular coefficients $\langle I^{n}_{\lambda}\rangle$ associated with $B\to D^{\ast}(\to D\pi,D\gamma)\tau\bar{\nu}_{\tau}$ decays. These observables can be expressed in terms of Wilson coefficients $C_{X}$ and hadronic matrix elements of decay modes $B\to D^{(\ast)}$. The hadronic matrix element of the concerned decay can be evaluated in the  Helicity framework as\cite{Tanaka:2012nw,Sakaki:2014sea}
\begin{equation}
 \begin{split}
  H^{\lambda_{M}}_{V_{1,2,\lambda}}(q^{2})=&\varepsilon^{\ast}_{\mu}(\lambda)\langle M(\lambda_{M})|\bar c\gamma^{\mu}(1\mp\gamma_{5})b|B\rangle \, ,\\
   H^{\lambda_{M}}_{S_{1,2,\lambda}}(q^{2})=&\langle M(\lambda_{M})|\bar c(1\pm\gamma_{5})b|B\rangle \, ,\\
    H^{\lambda_{M}}_{T,\lambda\lambda^{\prime}}(q^{2})=-& H^{\lambda_{M}}_{T,\lambda^{\prime}\lambda}(q^{2})= \varepsilon^{\ast}_{\mu}(\lambda)\varepsilon^{\ast}_{\nu}(\lambda^{\prime})\langle M(\lambda_{M})|\bar c\sigma^{\mu\nu}(1-\gamma_{5})b|B\rangle \, ,
 \end{split}
\label{eq:Helicity}
\end{equation}
where $\lambda_{M}$ and $\lambda$ represent the final state vector $D^{\ast}$ meson and exchange particle helicities. The amplitudes given in Eq.(\ref{eq:Helicity}) can be expressed in terms of transition form factors and can be used to analyze the above mentioned physical observables. In next section we will briefly discuss the form factors of $B\to D^{(\ast)}$ used to investigate the structure of NP.

\section {\texorpdfstring {$B\to D$}{} and \texorpdfstring {$B\to D^{\ast}$}{} Transition Form Factors}\label{FFP}
For $B\to D$ and $B\to D^{\ast}$ decays, we use  Boyd, Grinstein and Lebed (BGL) parametrization\cite{Boyd:1994tt} for hadronic form factors, and we use them as our numerical inputs to investigate the structure of NP. For these two decays the vector and axial-vector operators are through the matrix elements:
\begin{align}
\label{eq:Fp0_parametrization}
 \langle D(k)|\bar{c}\gamma_\mu b|B(p)\rangle & = \left[(p+k)_\mu-\frac{m_{B}^2-m_{D}^2}{q^2}q_\mu\right] F_1(q^2)+q_\mu\frac{m_{B}^2-m_{D}^2}{q^2}F_0(q^2) \,,\\
  \langle D^{\ast}(k) | \bar{c} \gamma^\mu b | B(p) \rangle
  & = -\frac{2i V(q^2)}{m_{B} + m_{D^{\ast}} }\, \varepsilon^{\mu\nu\rho\sigma}\, \epsilon_\nu^*\, {p}_\rho\, {k}_\sigma\,,\label{eq:FFVq} \\
  \langle D^{\ast}(k) | \bar c \gamma^\mu\gamma^5 b | B(p) \rangle
  & = 2m_{D^{\ast}} A_0(q^2) \frac{ \epsilon^* \cdot q}{q^2} q^\mu + (m_{B} + m_{D^{\ast}}) A_1(q^2) \left[ \epsilon^{*\mu} - \frac{\epsilon^* \cdot q} {q^2} q^\mu \right] \notag \\
  &~~~ - A_2(q^2) \frac{\epsilon^* \cdot q}{m_{B} + m_{D^{\ast}} } \left[ p^{\mu} + k^{\mu} - \frac{ m_{B}^2 - m_{D^{\ast}}^2}{q^2} q^\mu \right] \,,\label{eq:FFA1q}
\end{align}
  %
For the same decays the matrix elements through tensor operators can be defined as:
\begin{align}\label{eq:FTff1a}
\langle D(k)|\bar c\sigma_{\mu\nu} b|B(p)\rangle & = -i ( p_\mu k_\nu - k_\mu p_\nu ) \frac{ 2F_T(q^2)}{m_{B}+m_{D}} \,,\\
 \langle D^{\ast}(k) |\bar c \sigma^{\mu\nu} q_\nu b| B(p) \rangle \label{eq:FFT1}
  & = -2T_1(q^2)\, \varepsilon^{\mu\nu\rho\sigma}  \epsilon_\nu^*\, {p}_{\rho}\, {k}_{\sigma}\,, \\[0.5em]
  \langle D^{\ast}(k) |\bar c \sigma^{\mu\nu}\gamma_5 q_\nu b| B(p) \rangle
  & = - T_2(q^2) \Big[(m_{B}^2-m_{D^{\ast}}^2) \epsilon^{*\mu} - (\epsilon^* \cdot q) (p + k)^\mu\Big] \notag \\
  &~~~ - T_3(q^2) (\epsilon^* \cdot q) \left[q^\mu-\frac{q^2}{m_{B}^2-m_{D^{\ast}}^2} (p + k)^\mu \right] \,.\label{eq:FTff1b}
\end{align}
  %
The form factors given in Eqs.(\ref{eq:Fp0_parametrization}) to (\ref{eq:FTff1b}) can be expressed in terms of BGL parametrization.

For $B\to D$ decay mode, the form factors can be expanded in terms of BGL parametrization as\cite{Boyd:1994tt,Ray:2023xjn}
\begin{equation}
\mathcal{F}_i (z) = \frac{1}{P_i (z) \phi_i (z)} \sum_{j=0}^{N} a_{j}^i z^j,
\label{eq:FF-BGL}
\end{equation}
where the parameter $z$ can be related with factor $w$ as
\begin{equation}\label{eq:z}
z = \frac{\sqrt{w+1}-\sqrt{2}}{\sqrt{w+1}+\sqrt{2}}.
\end{equation}
The factor $w$ can be related with kinematical variable $q^{2}$ as
\begin{equation}
q^2 = m_B^2 + m_D^2 - 2 m_B m_D w
\end{equation}
The function $P_{i}(z)$ given in Eq. (\ref{eq:FF-BGL}) is known as Blasche factor, and its analytical expression can be written as
\begin{equation}
P_i(z) = \prod_p \frac{z-z_p}{1 - z z_p}. \label{eq:Blaschke-fact}
\end{equation}
The function $P_{i}(z)$ can be used to eliminate the poles at $z=z_{p}$\cite{Ray:2023xjn}. The function $\phi_{i}(z)$ given in Eq. (\ref{eq:FF-BGL}) can be analytic function of $q^{2}$ and can be written for form factors $F_{1}(q^{2})$ and $F_{0}(q^{2})$ as
\begin{eqnarray}
	\phi_{F_1} &=& \frac{8r^2}{m_B} \sqrt{\frac{8 n_I}{3\pi \tilde{\chi}_{1^-}^T (0)}} \frac{(1+z)^2(1-z)^{1/2}}{\left[(1+r)(1-z) + 2\sqrt{r}(1+z)\right]^5}, \nonumber \\
   	\phi_{F_0} &=& r(1-r^2) \sqrt{\frac{8 n_I}{\pi \tilde{\chi}_{1^-}^L (0)}} \frac{(1-z^2)(1-z)^{1/2}}{\left[(1+r)(1-z) + 2\sqrt{r}(1+z)\right]^4},
\end{eqnarray}
where $r = m_D/m_B$.

For $B\to D^{\ast}$ transition, we first convert the form factors given in Eq. (\ref{eq:FFVq}), Eq. (\ref{eq:FFA1q}), Eq. (\ref{eq:FFT1}), and Eq. (\ref{eq:FTff1b}) into HQET parametrization\cite{Sakaki:2013bfa} and use the HQET parametrization in the form of BGL parametrization of form factors as,
\begin{eqnarray}
&&h_{V}=gm_{B}\sqrt{r_{\ast}}\label{eq:hV},\\
&&h_{A_{1}}=\frac{f}{m_{B}\sqrt{r_{\ast}}(w+1)}\label{eq:hA1},\\
&&h_{A_{2}}=\frac{(w-r_{\ast})F_{1}-m_{B}[(1+r^{2}_{\ast}-2r_{\ast}w)f+m_{B}r_{\ast}(w^{2}-1)F_{2}]}{m^{2}_{B}\sqrt{r_{\ast}}(w^{2}-1)(1+r^{2}_{\ast}-2r_{\ast}w)}\label{eq:hA2},\\
&&h_{A_{3}}=\frac{m_{B}[w(1+r^{2}_{\ast}-2r_{\ast}w)f+m_{B}r^{2}_{\ast}(w^{2}-1)F_{2}]+(r_{\ast}w-1)F_{1}}{m^{2}_{B}\sqrt{r_{\ast}}(w^{2}-1)(1+r^{2}_{\ast}-2r_{\ast}w)}
\end{eqnarray}
where $r_{\ast}=m_{D^{\ast}}/m_{B}$. The function $\phi_{i}(z)$ for the $B\to D^{\ast}$ transition form factors can be written as\cite{Boyd:1997kz}
\begin{eqnarray}
&&\phi_f = \frac{4r_{\ast}}{m_B^2} \sqrt{\frac{n_I}{6\pi \chi_{1^+}^T (0)}} \frac{(1+z)(1-z)^{3/2}}{\left[(1+r_{\ast})(1-z) + 2\sqrt{r_{\ast}}(1+z)\right]^4}, \nonumber \\
&&\phi_g =16r_{\ast}^2 \sqrt{\frac{n_I}{3\pi \tilde{\chi}_{1^-}^T (0)}} \frac{(1+z)^2(1-z)^{-1/2}}{\left[(1+r_{\ast})(1-z) + 2\sqrt{r_{\ast}}(1+z)\right]^4}, \nonumber\\
&&\phi_{{F}_1}=\frac{4r_{\ast}}{m_B^3} \sqrt{\frac{n_I}{6\pi \chi_{1^+}^T (0)}} \frac{(1+z)(1-z)^{5/2}}{\left[(1+r_{\ast})(1-z) + 2\sqrt{r_{\ast}}(1+z)\right]^5}, \nonumber \\
&&\phi_{{F}_2}= 8\sqrt{2}r_{\ast}^2 \sqrt{\frac{n_I}{\pi \tilde{\chi}_{1^+}^L (0)}} \frac{(1+z)^2 (1-z)^{-1/2}}{\left[(1+r_{\ast})(1-z) + 2\sqrt{r_{\ast}}(1+z)\right]^4}.
\end{eqnarray}
These form factors are then used as input parameters to investigate the structure of NP through the physical observables discussed in Section \ref{obs}.
\section{Physical Observables}\label{obs}
As mentioned in Sect.~\ref{TH}, the physical observables which we use to test NP in this work are $R_{D^{(\ast)}}$, $P_{\tau}(D^{(\ast)})$, $F^{D^{\ast}}_{L}$ of the $D^{\ast}$ meson, $\mathcal{A}^{\text{FB}}_{D^{(\ast)}}$, and normalized angular coefficients $\langle I^{n}_{\lambda}\rangle$ for $B\to D^{\ast}(\to D\pi,D\gamma)\tau\bar{\nu}_{\tau}$ decays. 
The expressions of these observables are first computed in the helicity frame work (c.f. Eq. \ref{eq:Helicity}), which are further reduced to the semi-analytical expressions, in terms of NP Wilson coefficients, by using the input parameters and the form factors, as given below

\begin{align}
R_{D}&=R_{D}^{SM}\Bigl\{\big|1+C_{V_{1}}+C_{V_{2}}\big|^{2}
+1.09\big|C_{S_{1}}+C_{S_{2}}\big|^2+1.54\rm Re\big[\left(1+C_{V_{1}}+C_{V_{2}}\right)\notag \\
&\left(C_{S_{1}}+C_{S_{2}}\right)^{*}\big]+0.41\big|C_{T}\big|^2+0.77\rm Re\big[\left(1+C_{V_{1}}+C_{V_{2}}\right)C_{T}^{*}\big]\Bigl\}, \label{eq4}\\
R_{D^*}&=R_{D^*}^{SM}\Bigl\{\big| 1+C_{V_{1}}\big|^2+ \big| C_{V_{2}}\big|^2+0.04 \big|C_{S_{1}}-C_{S_{2}}\big|^2+13.6\big|C_{T}\big|^2\notag\\  &-1.82\rm Re\big[\left(1+C_{V_{1}}\right)C_{V_{2}}^{{*}}\big]+6.21\rm Re[\left(C_{V_{2}}\right)C_{T}^{*}\big]-5.0\rm Re\big[\left(1+C_{V_{1}}\right)C_{T}^{*}\big]\notag\\
&+0.12\rm Re \big[ \left( 1+C_{V_{1}}-C_{V_{2}} \right) \left( C_{S_{1}}-C_{S_{2}} \right)^{*} \big] \Bigl\},\label{eq5}\\
P_{\tau}(D)&=\Bigl(\frac{R_{D}}{R_{D}^{SM}}\Bigl)^{-1}\Bigl\{0.32\big| 1+C_{V_{1}}+C_{V_{2}}\big|^2+1.07\big|C_{S_{1}}+C_{S_{2}}\big|^2+0.02\big|C_{T}\big|^2\notag\\
&-0.25\rm Re\big[\left(1+C_{V_{1}}+C_{V_{2}}\right)C_{T}^{*}\big]
+4.76 \rm Re\big[\left(1+C_{V_{1}}+C_{V_{2}}\right)\left(C_{S_{1}}+C_{S_{2}}\right)^{*}\big]\Bigl\},\label{PtauD}\\
\mathcal{A}^{\text{FB}}_{D}&=\Bigl(\frac{R_{D}}{R_{D}^{SM}}\Bigl)^{-1}\Bigl\{0.36\big| 1+C_{V_{1}}+C_{V_{2}}\big|^2+0.61\rm Re\big[\left(1+C_{V_{1}}+C_{V_{2}}\right)C_{T}^{*}\big] \notag\\
& +0.44\rm Re\big[\left(1+C_{V_{1}}+C_{V_{2}}\right)\left(C_{S_{1}}+C_{S_{2}}\right)^{*}\big]+0.81\rm Re\big[\left(C_{S_{1}}+C_{S_{2}}\right)C_T^{*}\big]\Bigl\},\label{AFBD}\\
P_{\tau}(D^{\ast})&=\Bigl(\frac{R_{D}}{R_{D}^{SM}}\Bigl)^{-1}\Bigl\{-0.50\left[\big| 1+C_{V_{1}}\big|^2+\big|C_{V_{2}}\big|^{2}\right]+0.04\big|C_{S_{1}}-C_{S_{2}}\big|^2+0.85\big|C_{T}\big|^2\notag\\
&+1.65\rm Re\big[\left(1+C_{V_{1}}\right)\left(C_{T}\right)^{*}\big] +0.11 \rm Re\big[\left(1+C_{V_{1}}-C_{V_{2}}\right)\left(C_{S_{1}}-C_{S_{2}}\right)^{*}\big]\notag\\
&-2.04 \rm Re\big[C_{V_{2}}C_{T}^{\ast}\big]+0.88 \rm Re\big[(1+C_{V_{1}})C_{V_{2}}^{*}\bigl]\Bigl\},\label{PtauDst}\\
\mathcal{A}^{\text{FB}}_{D^{\ast}}&=\Bigl(\frac{R_{D}}{R_{D}^{SM}}\Bigl)^{-1}\Bigl\{-0.051\big| 1+C_{V_{1}}\big|^2+0.17\big|C_{V_{2}}\big|^{2}+3.00\big|C_{T}\big|^2
\notag\\
&-0.87\rm Re\big[\left(1+C_{V_{1}}\right)C_{T}^{*}\big]+0.09 \rm Re\big[\left(1+C_{V_{1}}-C_{V_{2}}\right)\left(C_{S_{1}}-C_{S_{2}}\right)^{*}\big]\notag\\
&+1.99\rm Re\big[C_{V_{2}}C_{T}^{\ast}\big]-0.12\left( \rm Re\big[(1+C_{V_{1}})C_{V_{2}}^{*}\bigl]+\rm Re\big[(1+C^{\ast}_{V_{1}})C_{V_{2}}\bigl]\right)\Bigl\},\label{AFBDst}\\
F_{L}^{D^{\ast}}&=\Bigl(\frac{R_{D}}{R_{D}^{SM}}\Bigl)^{-1}\Bigl\{0.45\big| 1+C_{V_{1}}-C_{V_{2}}\big|^2+0.04\big| C_{S_{1}}-C_{S_{2}}\big|^2+2.83\big|C_{T}\big|^2\notag\\
&+0.117\rm Re \big[ \left( 1+C_{V_{1}}-C_{V_{2}} \right) \left( C_{S_{1}}-C_{S_{2}} \right)^{*} \big]-0.185 \rm Re[\left(1+C_{V_{1}}-C_{V_{2}}\right)C^{\ast}_{T}]\Bigl\}.\label{FDst}
\end{align}
For $B\to D^{\ast}(\to D\pi)\tau\bar{\nu}_{\tau}$ decay, the four fold distributions can be written as,
\begin{eqnarray}
 \frac{d^4\Gamma\left(B\to D^{\ast}\,(\to D\pi)\tau\bar{\nu}_{\tau}\right)}{dq^2 \ d\cos{\theta_{\tau}} \ d\cos {\theta}_{D^{\ast}} \ d\phi} &=& \frac{9}{32 \pi} \mathcal{B}(D^{\ast}\to D\pi)\notag\bigg[I^{\pi}_{1s}\sin^2\theta_{D^{\ast}}+I^{\pi}_{1c}\cos^2\theta_{D^{\ast}}\notag\\
&+&\Big(I^{\pi}_{2s}\sin^2\theta_{D^{\ast}}+I^{\pi}_{2c}\cos^2\theta_{D^{\ast}}\Big)\cos{2\theta_{\tau}}
\notag\\
&+&\Big(I^{\pi}_{6s}\sin^2\theta_{D^{\ast}}+I^{\pi}_{6c}\cos^2\theta_{D^{\ast}}\Big)\cos{\theta_{\tau}}\notag\\
&+&\Big(I^{\pi}_{3}\cos{2\phi}
+I^{\pi}_{9}\sin{2\phi}\Big)\sin^2\theta_{D^{\ast}}\sin^2\theta_{\tau}\notag
\\
&+&\Big(I^{\pi}_{4}\cos{\phi}+I^{\pi}_{8}\sin{\phi}\Big)\sin2\theta_{D^{\ast}}\sin2\theta_{\tau}\notag
\\
&+&\Big(I^{\pi}_{5}\cos{\phi}+I^{\pi}_{7}\sin{\phi}\Big)\sin2\theta_{D^{\ast}}\sin\theta_{\tau}\bigg], \label{fullad}
\end{eqnarray}
where the expressions of normalized angular coefficients, defined as $\langle I^{\pi}_{\lambda}\rangle=\frac{\mathcal{B}(D^{\ast}\to D\pi)I^{\pi}_{\lambda}}{d\Gamma\left(B\to D^{\ast}\,(\to D\pi)\tau\bar{\nu}_{\tau}\right)/dq^2}$, can be written in terms of NP Wilson coefficients as,
\begin{align}
\langle I^{\pi}_{1s}\rangle&=\langle I_{1s}^{\pi SM}\rangle\left(\frac{d\Gamma^{\text{NP}}}{dq^{2}}\right)^{-1}\Bigl\{\big|1+C_{V_{1}}\big|^{2}+\big|C_{V_{2}}\big|^{2}
+20.0\big|C_{T}\big|^2+7.99\rm Re\big[C_{V_{2}}C^{\ast}_{T}\big]
\notag\\
& -5.65\rm Re\big[\left(1+C^{\ast}_{V_{1}}\right)C_{T}\big]-1.65 \rm Re\big[\left(1+C^{\ast}_{V_{1}}\right)C_{V_{2}}\big]\Bigl\},\label{eq.I1s}
\\
\langle I^{\pi}_{1c}\rangle&=\langle I_{1c}^{\pi SM}\rangle\left(\frac{d\Gamma^{\text{NP}}}{dq^{2}}\right)^{-1}\Bigl\{\big|1+C_{V_{1}}\big|^{2}+\big|C_{V_{2}}\big|^{2}
+0.108\big|C_{P}\big|^{2}+11.50\big|C_{T}\big|^2\notag\\
&+0.31 \rm Re\big[\left(1+C^{\ast}_{V_{1}}\right)C_{P}\big]-0.065\rm Re\big[C_{P}C^{\ast}_{V_{2}}\big]+5.45\rm Re\big[C_{V_{2}}C_{T}^{\ast}\big]\notag\\
&-5.45\rm Re\big[\left(1+C^{\ast}_{V_{1}}\right)C_{T}\big]- 2.0\rm Re\big[\left(1+C^{\ast}_{V_{1}}\right)C_{V_{2}}\big]\Bigl\},\label{eq.I1c}
\\
\langle I^{\pi}_{2s}\rangle&=\langle I_{2s}^{\pi SM}\rangle\left(\frac{d\Gamma^{\text{NP}}}{dq^{2}}\right)^{-1}\Bigl\{\big|1+C_{V_{1}}\big|^{2}+\big|C_{V_{2}}\big|^{2}
-24.0\big|C_{T}\big|^2
- 1.71 \rm Re\big[\left(1+C^{\ast}_{V_{1}}\right)C_{V_{2}}\big]\Bigl\},\label{eq.I2s}
\\
\langle I^{\pi}_{2c}\rangle&=\langle I_{2c}^{\pi SM}\rangle\left(\frac{d\Gamma^{\text{NP}}}{dq^{2}}\right)^{-1} \Bigl\{\big|1+C_{V_{1}}\big|^{2}+\big|C_{V_{2}}\big|^{2}
-14.85\big|C_{T}\big|^2
- 2.0\rm Re\big[\left(1+C^{\ast}_{V_{1}}\right)C_{V_{2}}\big]\Bigl\},\label{eq.I2c}
\\
\langle I^{\pi}_{3}\rangle&=\langle I_{3}^{\pi SM}\rangle\left(\frac{d\Gamma^{\text{NP}}}{dq^{2}}\right)^{-1}  \Bigl\{\big|1+C_{V_{1}}\big|^{2}+\big|C_{V_{2}}\big|^{2}
-17.86\big|C_{T}\big|^2
- 2.33\rm Re\big[\left(1+C^{\ast}_{V_{1}}\right)C_{V_{2}}\big]\Bigl\},\label{eq.I3}
\\
\langle I^{\pi}_{4}\rangle&=\langle I_{4}^{\pi SM}\rangle\left(\frac{d\Gamma^{\text{NP}}}{dq^{2}}\right)^{-1} \Bigl\{\big|1+C_{V_{1}}\big|^{2}+\big|C_{V_{2}}\big|^{2}
-17.94\big|C_{T}\big|^2
- 2.00\rm Re\big[\left(1+C^{\ast}_{V_{1}}\right)C_{V_{2}}\big]\Bigl\},\label{eq.I4}
\\
\langle I^{\pi}_{5}\rangle&=\langle I_{5}^{\pi SM}\rangle\left(\frac{d\Gamma^{\text{NP}}}{dq^{2}}\right)^{-1}\Bigl\{\big|1+C_{V_{1}}\big|^{2}+\big|C_{V_{2}}\big|^{2}
-8.73\big|C_{T}\big|^2-3.01\rm Re\big[C_{P}C^{\ast}_{T}\big]\notag\\
&+0.41\rm Re\big[\left(1+C^{\ast}_{V_{1}}\right)C_{P}\big]-2.08 \rm Re\big[\left(1+C^{\ast}_{V_{1}}\right)C_{T}\big]-0.41 \rm Re\big[C_{P}C^{\ast}_{V_{2}}\big]\notag\\
&-0.65\rm Re\big[C_{V_{2}}C^{\ast}_{T}\big]-1.00 \rm Re\big[\left(1+C^{\ast}_{V_{1}}\right)C_{V_{2}}\big]\Bigl\},\label{eq.I5}
\\
\langle I^{\pi}_{6s}\rangle&=\langle I_{6s}^{\pi SM}\rangle\left(\frac{d\Gamma^{\text{NP}}}{dq^{2}}\right)^{-1} \Bigl\{\big|1+C_{V_{1}}\big|^{2}-\big|C_{V_{2}}\big|^{2}-21.9\big|C_{T}\big|^2
+2.05\rm Re\big[\left(1+C^{\ast}_{V_{1}}\right)C_{T}\big]\notag\\
&-9.1 \rm Re\big[C_{V_{2}}C^{\ast}_{T}\big]\Bigl\},\label{eq.I6s}
\\
\langle I^{\pi}_{6c}\rangle&=\langle I_{6c}^{\pi SM}\rangle\left(\frac{d\Gamma^{\text{NP}}}{dq^{2}}\right)^{-1} \Bigl\{\big|1+C_{V_{1}}\big|^{2}+\big|C_{V_{2}}\big|^{2}
-4.24\rm Re\big[C_{P}C^{\ast}_{T}\big]
+0.77\rm Re\big[\left(1+C^{\ast}_{V_{1}}\right)C_{P}\big]\notag\\
&-5.23 \rm Re\big[\left(1+C_{V_{1}}\right)C^{\ast}_{T}\big]-0.77\rm Re\big[C_{P}C^{\ast}_{V_{2}}\big]+5.23 \rm Re\big[C_{V_{2}}C^{\ast}_{T}\big]\notag\\
&-2.0\rm Re\big[\left(1+C^{\ast}_{V_{1}}\right)C_{V_{2}}\big]\Bigl\}.\label{eq.I6c}
\end{align}
Similarly, the four fold distribution for $B\to D^{\ast}(\to D\gamma)\tau\bar{\nu}_{\tau}$ can be expressed as follows,
\begin{eqnarray}
 \frac{d^4\Gamma\left(B\to D^{\ast}\,(\to D\gamma)\tau\bar{\nu}_{\tau}\right)}{dq^2 \ d\cos{\theta_{\tau}} \ d\cos {\theta}_{D^{\ast}} \ d\phi} &=& \frac{9}{32 \pi} \mathcal{B}(D^{\ast}\to D\gamma)
\bigg[I^{\gamma}_{1s}\sin^2\theta_{D^{\ast}}+I^{\gamma}_{1c}\cos^2\theta_{D^{\ast}}\notag\\
&&+\Big(I^{\gamma}_{2s}\sin^2\theta_{D^{\ast}}+I^{\gamma}_{2c}\cos^2\theta_{D^{\ast}}\Big)\cos{2\theta_{\tau}}
\notag\\
&&+\Big(I^{\gamma}_{6s}\sin^2\theta_{D^{\ast}}+I^{\gamma}_{6c}\cos^2\theta_{D^{\ast}}\Big)\cos{\theta_{\tau}}\notag\\
&&+\Big(I^{\gamma}_{3}\cos{2\phi}
+I^{\gamma}_{9}\sin{2\phi}\Big)\sin^2\theta_{D^{\ast}}\sin^2\theta_{\tau}\notag
\\
&&+\Big(I^{\gamma}_{4}\cos{\phi}+I^{\gamma}_{8}\sin{\phi}\Big)\sin2\theta_{D^{\ast}}\sin2\theta_{\tau}\notag
\\
&&+\Big(I^{\gamma}_{5}\cos{\phi}+I^{\gamma}_{7}\sin{\phi}\Big)\sin2\theta_{D^{\ast}}\sin\theta_{\tau}\bigg], \label{fulladgma}
\end{eqnarray}
where the normalized angular coefficients, defined as $\langle I^{\gamma}_{n\lambda}\rangle=\frac{\mathcal{B}(D^{\ast}\to D\gamma)I^{\gamma}_{n\lambda}}{d\Gamma\left(B\to D^{\ast}\,(\to D\gamma)\tau\bar{\nu}_{\tau}\right)/dq^2}$, can be written in terms of NP Wilson coefficients as,
\begin{align}
\langle I^{\gamma}_{1s}\rangle &=\langle I_{1s}^{\gamma SM}\rangle \left(\frac{d\Gamma^{\text{NP}}}{dq^{2}}\right)^{-1}\Bigl\{\big|1+C_{V_{1}}\big|^{2}+\big|C_{V_{2}}\big|^{2}+0.144\big|C_{P}\big|^{2}
+11.48\big|C_{T}\big|^2\notag\\
&+0.36 \rm Re\big[\left(1+C^{\ast}_{V_{1}}\right)C_{P}\big]-0.37\rm Re\big[C_{P}C^{\ast}_{V_{2}}\big]
-5.44\rm Re\big[\left(1+C^{\ast}_{V_{1}}\right)C_{T}\big]\notag\\
&+5.44\rm Re\big[C_{V_{2}}C^{\ast}_{T}\big]
- 2.00 \rm Re\big[\left(1+C^{\ast}_{V_{1}}\right)C_{V_{2}}\big]\Bigl\},\label{eq.I1sgma}\\
\langle I^{\gamma}_{1c}\rangle &=\langle I_{1c}^{\gamma SM}\rangle\left(\frac{d\Gamma^{\text{NP}}}{dq^{2}}\right)^{-1} \Bigl\{\big|1+C_{V_{1}}\big|^{2}+\big|C_{V_{2}}\big|^{2}
+15.80\big|C_{T}\big|^2-5.64\rm Re\big[\left(1+C^{\ast}_{V_{1}}\right)C_{T}\big]
\notag\\
&+7.94\rm Re\big[C_{V_{2}}C_{T}^{\ast}\big]-1.66\rm Re\big[\left(1+C^{\ast}_{V_{1}}\right)C_{V_{2}}\big]\Bigl\},\label{eq.I1cgma}\\
\langle I^{\gamma}_{2s}\rangle&=\langle I_{2s}^{\gamma SM}\rangle\left(\frac{d\Gamma^{\text{NP}}}{dq^{2}}\right)^{-1}\Bigl\{\big|1+C_{V_{1}}\big|^{2}+\big|C_{V_{2}}\big|^{2}
-14.86\big|C_{T}\big|^2
- 2.00\rm Re\big[\left(1+C^{\ast}_{V_{1}}\right)C_{V_{2}}\big]\Bigl\},\label{eq.I2sgma}\\
\langle I^{\gamma}_{2c}\rangle&=\langle I_{2c}^{\gamma SM}\rangle\left(\frac{d\Gamma^{\text{NP}}}{dq^{2}}\right)^{-1}\Bigl\{\big|1+C_{V_{1}}\big|^{2}+\big|C_{V_{2}}\big|^{2}
-24.03\big|C_{T}\big|^2
- 1.71\rm Re\big[\left(1+C^{\ast}_{V_{1}}\right)C_{V_{2}}\big]\Bigl\},\label{eq.I2cgma}\\
\langle I^{\gamma}_{3}\rangle&=\langle I_{3}^{\gamma SM}\rangle \left(\frac{d\Gamma^{\text{NP}}}{dq^{2}}\right)^{-1}\Bigl\{\big|1+C_{V_{1}}\big|^{2}+\big|C_{V_{2}}\big|^{2}
-17.82\big|C_{T}\big|^2
- 2.33\rm Re\big[\left(1+C^{\ast}_{V_{1}}\right)C_{V_{2}}\big]\Bigl\},\label{eq.I3gma}\\
\langle I^{\gamma}_{4}\rangle&=\langle I_{4}^{\gamma SM}\rangle\left(\frac{d\Gamma^{\text{NP}}}{dq^{2}}\right)^{-1} \Bigl\{\big|1+C_{V_{1}}\big|^{2}+\big|C_{V_{2}}\big|^{2}
-17.90\big|C_{T}\big|^2
- 2.00\rm Re\big[\left(1+C^{\ast}_{V_{1}}\right)C_{V_{2}}\big]\Bigl\},\label{eq.I4gma}\\
\langle I^{\gamma}_{5}\rangle &=\langle I_{5}^{\gamma SM}\rangle \left(\frac{d\Gamma^{\text{NP}}}{dq^{2}}\right)^{-1}\Bigl\{\big|1+C_{V_{1}}\big|^{2}+\big|C_{V_{2}}\big|^{2}
-8.67\big|C_{T}\big|^2-3.02\rm Re\big[C_{P}C^{\ast}_{T}\big]\notag\\
&+0.41\rm Re\big[\left(1+C^{\ast}_{V_{1}}\right)C_{P}\big]- 2.08 \rm Re\big[\left(1+C^{\ast}_{V_{1}}\right)C_{T}\big]-0.40 \rm Re\big[C_{P}C^{\ast}_{V_{2}}\big]\notag\\
&-0.66\rm Re\big[C_{V_{2}}C^{\ast}_{T}\big]-1.002\rm Re\big[\left(1+C^{\ast}_{V_{1}}\right)C_{V_{2}}\big]\Bigl\},\label{eq.I5gma}\\
\langle I^{\gamma}_{6s}\rangle&=\langle I_{6s}^{\gamma SM}\rangle\left(\frac{d\Gamma^{\text{NP}}}{dq^{2}}\right)^{-1} \Bigl\{\big|1+C_{V_{1}}\big|^{2}+\big|C_{V_{2}}\big|^{2}
-4.26\rm Re\big[C_{P}C^{\ast}_{T}\big]
+0.77\rm Re\big[\left(1+C^{\ast}_{V_{1}}\right)C_{P}\big]\notag\\
& -5.24 \rm Re\big[\left(1+C_{V_{1}}\right)C^{\ast}_{T}\big]+0.76\rm Re\big[C_{P}C^{\ast}_{V_{2}}\big]+5.24 \rm Re\big[C_{V_{2}}C^{\ast}_{T}\big]\notag\\
&-2.00\rm Re\big[\left(1+C^{\ast}_{V_{1}}\right)C_{V_{2}}\big]\Bigl\},\label{eq.I6sgma}\\
\langle I^{\gamma}_{6c}\rangle &=\langle I_{6c}^{\gamma SM}\rangle\left(\frac{d\Gamma^{\text{NP}}}{dq^{2}}\right)^{-1}\Bigl\{\big|1+C_{V_{1}}\big|^{2}-\big|C_{V_{2}}\big|^{2}
-21.8\big|C_{T}\big|^2+2.04\rm Re\big[\left(1+C^{\ast}_{V_{1}}\right)C_{T}\big]\notag\\
&
-9.09\rm Re\big[C_{V_{2}}C_{T}^{\ast}\big]\Bigl\}.\label{eq.I6cgma}
\end{align}

\section{Numerical Analysis}\label{NA}
In this section, we investigate the NP solutions through the observables associated with $B\to D^{(\ast)}\tau\bar{\nu}_{\tau}$ decays. To analyze NP signatures, the main input parameters are hadronic form factors, which are discussed in Section \ref{FFP}. Further, to obtain the NP Wilson coefficients, a global fit is performed on the Wilson coefficients of specific operators without assuming a particular type of new physics (NP) at the high-energy scale, in a model-independent approach. We define the $\chi^{2}$ as follows,
\begin{equation}
\chi^2 (C_X) = \sum_{m,n=1}^{\text{data}} \left( O^{th}(C_X) - O^{exp} \right)_m
\left( V^{exp} + V^{th} \right)^{-1}_{mn}
\left( O^{th}(C_X) - O^{exp} \right)_n.
\end{equation}\label{chi}

\begin{table*}[!htbp]\small%
\centering
\caption{Experimental data used in the fits.}
\begin{tabular}{c c c c c }
\hline
\hline
    & $R_D$ & $R_{D^*}$& Correlation & $P_\tau(D^*)$\\
\hline
BaBar \cite{Lees:2012xj,Lees:2013uzd}       &$0.440(58)(42)$    &$0.332(24)(18)$   &$-0.27$  &$-$\\
Belle \cite{Huschle:2015rga}       &$0.375(64)(26)$    &$0.293(38)(15)$   &$-0.49$ &$-$ \\
Belle \cite{Sato:2016svk}      &$-$    &$0.302(30)(11)$  &$-$&$-$ \\
Belle \cite{Hirose:2016wfn}       &$-$    &$0.270(35)(_{-0.025}^{+0.028})$   &$0.33$ &$-0.38(51)(_{-0.16}^{+0.21})$ \\
LHCb  \cite{Aaij:2017uff,Aaij:2017deq,LHCb:2023uiv}       &$-$    &$0.257(12)(18)$   &$-$ &$-$\\
Belle \cite{Abdesselam:2019dgh}   &$0.307(37)(16)$    &$0.283(18)(14)$   &$-0.54$ &$-$\\
LHCb  \cite{Aaij:2015yra,LHCb:2023zxo}                            & $0.281(18)(24)$ & 0.441(60)(66)& -0.43 & $-$\\
Belle II \cite{Belle-II:2024ami} &     $-$ & $0.262(_{-0.039}^{+0.041})(_{-0.032}^{+0.035})$ & $-$ & $-$\\
LHCb \cite{Chen:2024hln} &  $0.249(43)(47)$    &  $0.402(81)(85)$ & $-0.39$ & $-$\\
\hline
\hline
\end{tabular}
\centering
\begin{tabular}{c c c c}
\hline
\hline
     &$R_{J/\psi}$ & $F_L^{D^{*}}$ & $R_{\Lambda_c}$\\
\hline
LHCb \cite{Aaij:2017tyk}       &$0.71(17)(18)$ & $-$ & $-$\\
Belle \cite{Adamczyk:2019wyt,Abdesselam:2019wbt}& $-$& $0.60(8)(4)$ & $-$\\
LHCb \cite{LHCb:2022piu} & $-$ & $-$ & $0.242(26)(40)(59)$ \\
LHCb \cite{LHCb:2023ssl} & $-$ & $0.43(6)(3)$ & $-$\\
CMS  \cite{Riti:2024lrs}             &  $0.17(_{-0.17}^{+0.18})(_{-0.22}^{+0.21})(_{-0.18}^{+0.19})$   &  $-$            &   $-$  \\
\hline
\hline
\end{tabular}
\label{tab:exdata}
\end{table*}
To obtain the $\chi^{2}$ fit values of Wilson coefficients $C_{V_{1}},C_{V_{2}},C_{S_{1}},C_{S_{2}}$ and $C_{T}$ given in Eq. (\ref{eq:lag}), we use the experimental values and theoretical predictions of the observables $R_{D^{(\ast)}}$, $P_{\tau}(D^{\ast})$, $R_{J/\psi}$, $F^{D^{\ast}}_{L}$, and $R_{\Lambda_{c}}$. For experimental values, we take into account the results of $R_{D^{(*)}}$ reported by BaBar, Belle and LHCb, $R_{J/\psi}$ by LHCb, and the longitudinal polarization fractions $P_\tau(D^*)$ and $F_L^{D^*}$ by Belle. All data and references are presented in Table~\ref{tab:exdata}. For theoretical predictions, we incorporate the covariance matrix $V^{th}$, which takes into account form factors uncertainties and their correlations. For these observables in the SM, we obtain $\chi^{2}_{SM}=37.5$.
Moreover, for NP the $\chi^{2}_{NP}/dof\approx \mathcal{O}(1)$ suggests an excellent fit to the data. However, the interpretation of this condition depends on the number of fitted Wilson coefficients. In this work, the goodness of fit for given NP scenarios is expressed by the "Pull" value. It depends on the number of fitted Wilson coefficients. For  Wilson coefficient fits, the pull w.r.t. SM is defined as\cite{Iguro:2024hyk},
\begin{eqnarray}
\text{Pull}=\sqrt{\chi^{2}_{SM}-\chi^{2}_{NP}}(\sigma).\label{pulldef}
\end{eqnarray}

\begin{table*}[!b]
\centering
\renewcommand{\arraystretch}{1.2}  
\setlength{\tabcolsep}{6pt}  
\scriptsize  
\caption{Best-fit values of the Wilson coefficients in different NP scenarios without $\mathcal B(B_c\to\tau\nu)<0.3$.}
\begin{tabular}{c c c c c c}
\hline\hline
\textbf{NP Scenario} & \textbf{Best-fit Value} & \textbf{$\chi^2/dof$} & \textbf{Corr.} & \textbf{p-value (\%)} & \textbf{Pull ($\sigma$)} \\
\hline
$V_1$  & $(1+{\rm Re}[C_{V_1}])^2 + ({\rm Im}[C_{V_1}])^2 = 1.116(31)$
       & $23.56/17$  & --  & $13.2$  & $3.73$ \\
\hline
$V_2$  & ${\rm Re}[C_{V_2}], {\rm Im}[C_{V_2}] = (-0.015(23), \pm 0.329(51))$
       & $23.13/17$  & $\pm 0.39$  & $14.5$  & $3.79$ \\
\hline
$S_1$  & ${\rm Re}[C_{S_1}], {\rm Im}[C_{S_1}] = (0.093(35), 0.000(433))$
       & $31.24/17$  & $0$  & $18.6$  & $2.50$ \\
\hline
$S_2$  & ${\rm Re}[C_{S_2}], {\rm Im}[C_{S_2}] = (-0.671(229), \pm 0.758(38))$
       & $23.25/17$  & $\pm 0.06$  & $14.1$  & $3.77$ \\
\hline
$T$  & ${\rm Re}[C_T], {\rm Im}[C_T] = (-0.009(52), \pm 0.086(113))$
       & $25.50/17$  & $\pm 0.99$  & $8.40$  & $3.46$ \\
\hline\hline
\end{tabular}
\label{tab:wcoef1a}
\end{table*}
Among the operators $O_{S_{1}},O_{S_{2}},O_{V_{1}},O_{V_{2}},O_{T}$, the operator $O_{V_{2}}$ originates minimally at dimension-8 in the Standard Model Effective Field Theory (SMEFT), while the remaining four operators can be generated at dimension 6. The Wilson coefficients $C_{V_{1},S_{1},S_{2},T}$ corresponding to the dimension 6 operators scale as $\frac{1}{\Lambda^{2}}$, and therefore generate NP at a relatively low energy scale. However, the Wilson coefficient $C_{V_{2}}$ corresponding to the dimension 8 operator scale as $\frac{1}{\Lambda^{4}}$, and therefore the NP generated due to $C_{V_{2}}$ is weaker.
\begin{table*}[!htbp]
\centering
\renewcommand{\arraystretch}{1.2}
\setlength{\tabcolsep}{6pt}  
\scriptsize  
\caption{Best-fit values of the Wilson coefficients in different NP scenarios, with $\mathcal B(B_c\to\tau\nu)<0.3$.}
\begin{tabular}{c c c c c c}
\hline\hline
\textbf{NP Scenario} & \textbf{Best-fit value} & \textbf{$\chi^2/dof$} & \textbf{Corr.} & \textbf{p-value ($\%$)} & \textbf{pull ($\sigma$)} \\
\hline
$V_1$  & $(1+{\rm Re}[C_{V_1}])^2 + ({\rm Im}[C_{V_1}])^2 = 1.116(31)$
       & $23.56/17$  & --  & $13.2$  & $3.73$ \\
\hline
$V_2$  & ${\rm Re}[C_{V_2}], {\rm Im}[C_{V_2}] = (-0.015(23), \pm 0.329(51))$
       & $23.13/17$  & $\pm 0.39$  & $14.5$  & $3.79$ \\
\hline
$S_1$  & ${\rm Re}[C_{S_1}], {\rm Im}[C_{S_1}] = (0.093(35), 0.000(433))$
       & $31.24/17$  & $0$  & $18.6$  & $2.50$ \\
\hline
$S_2$  & ${\rm Re}[C_{S_2}], {\rm Im}[C_{S_2}] = (-0.301(327), \pm 0.655(189))$
       & $25.95/17$  & $\mp 0.97$  & $75.3$  & $3.39$ \\
\hline
$T$  & ${\rm Re}[C_T], {\rm Im}[C_T] = (-0.009(52), \pm 0.086(113))$
       & $25.50/17$  & $\pm 0.99$  & $8.40$  & $3.46$ \\
\hline\hline
\end{tabular}
\label{tab:wcoef1}
\end{table*}

\begin{figure}[!t]
\begin{center}
\includegraphics[scale=0.22]{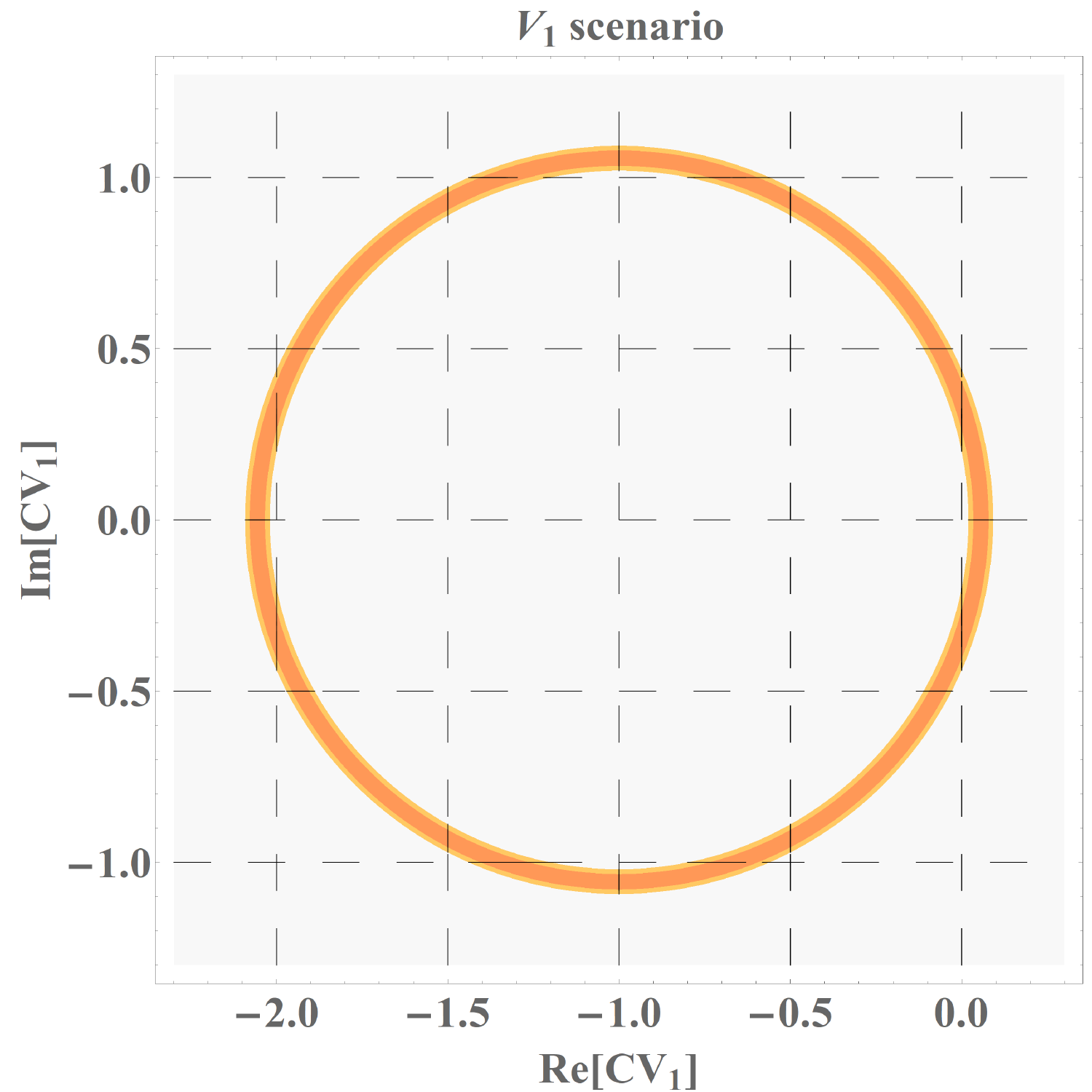}
\includegraphics[scale=0.22]{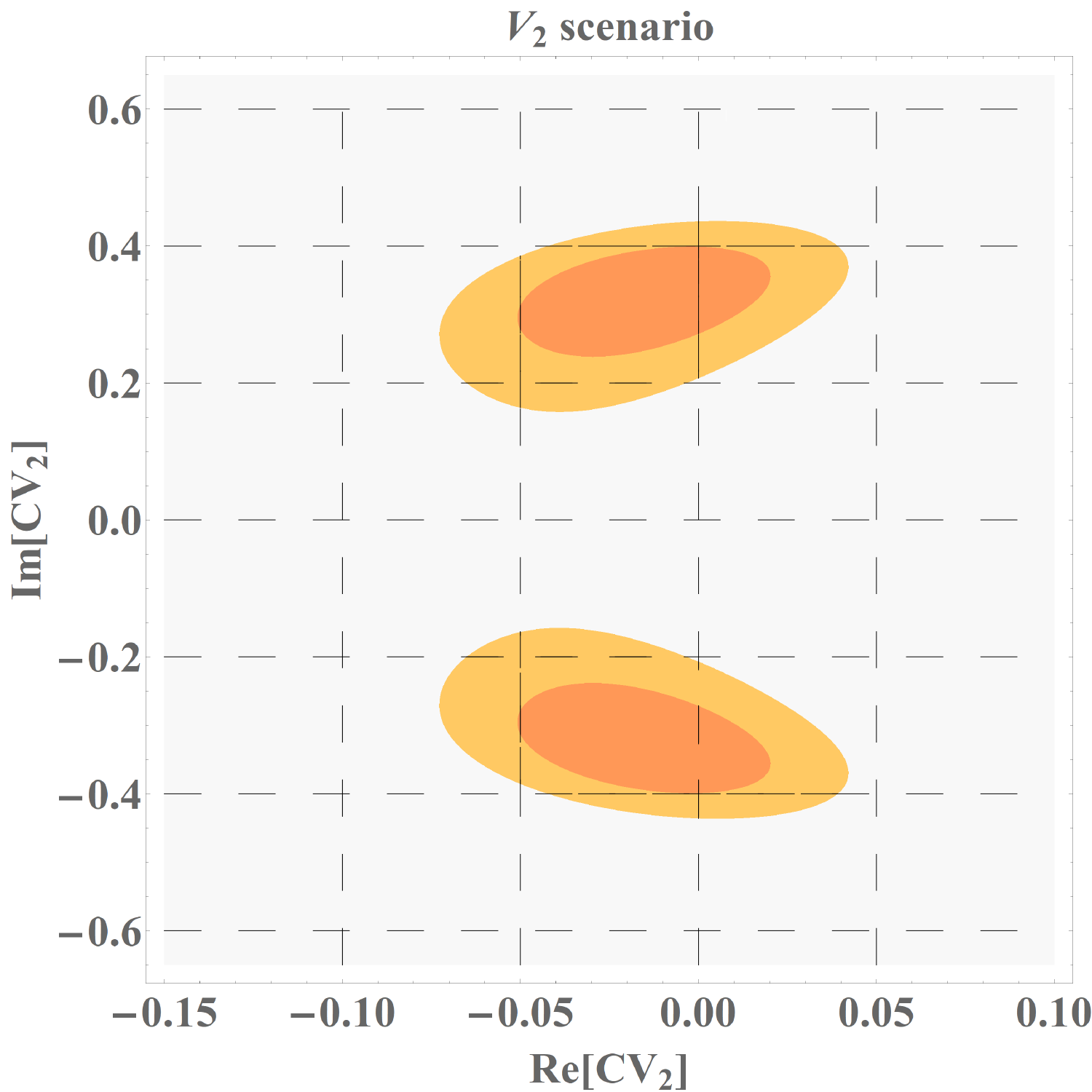}
\includegraphics[scale=0.22]{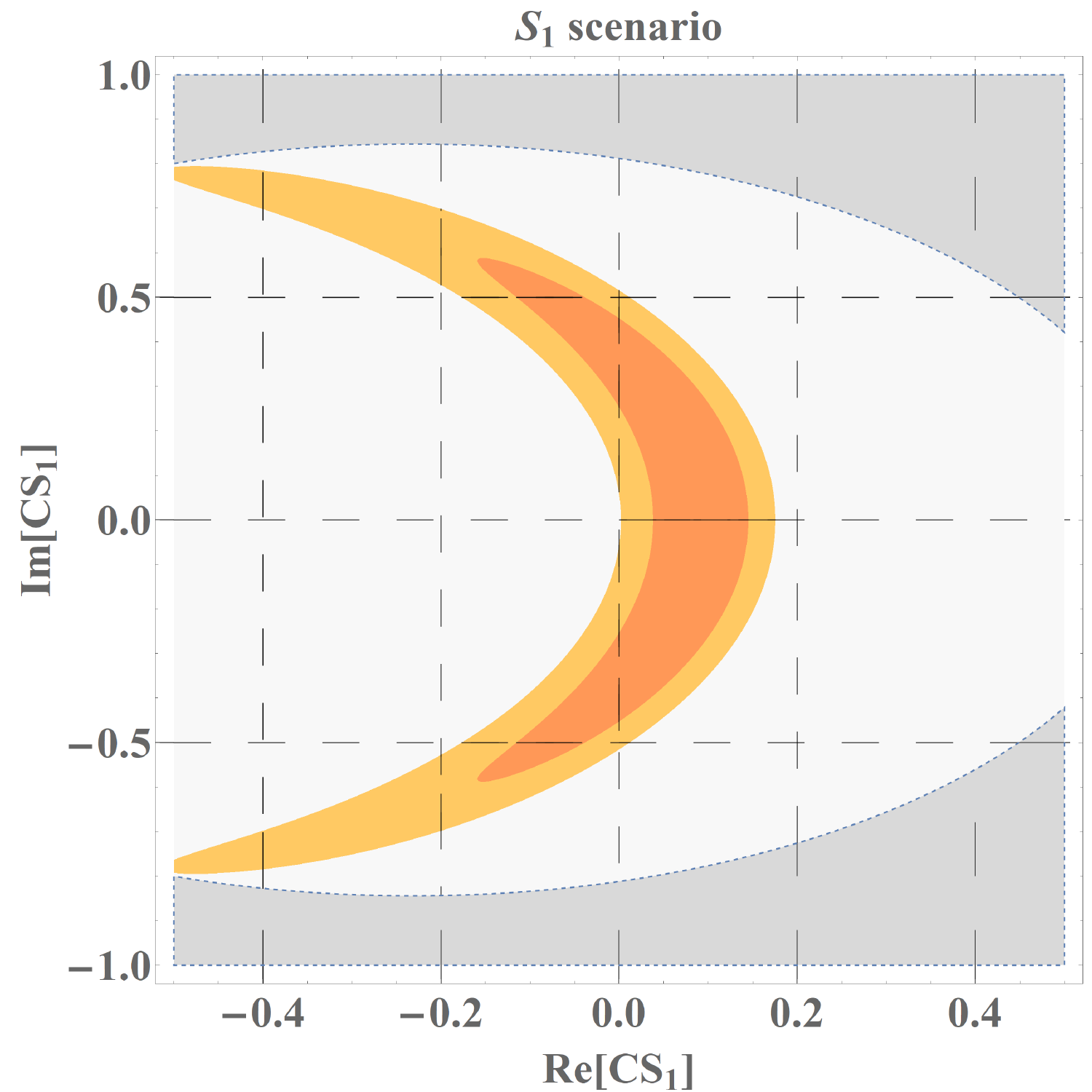}
\includegraphics[scale=0.22]{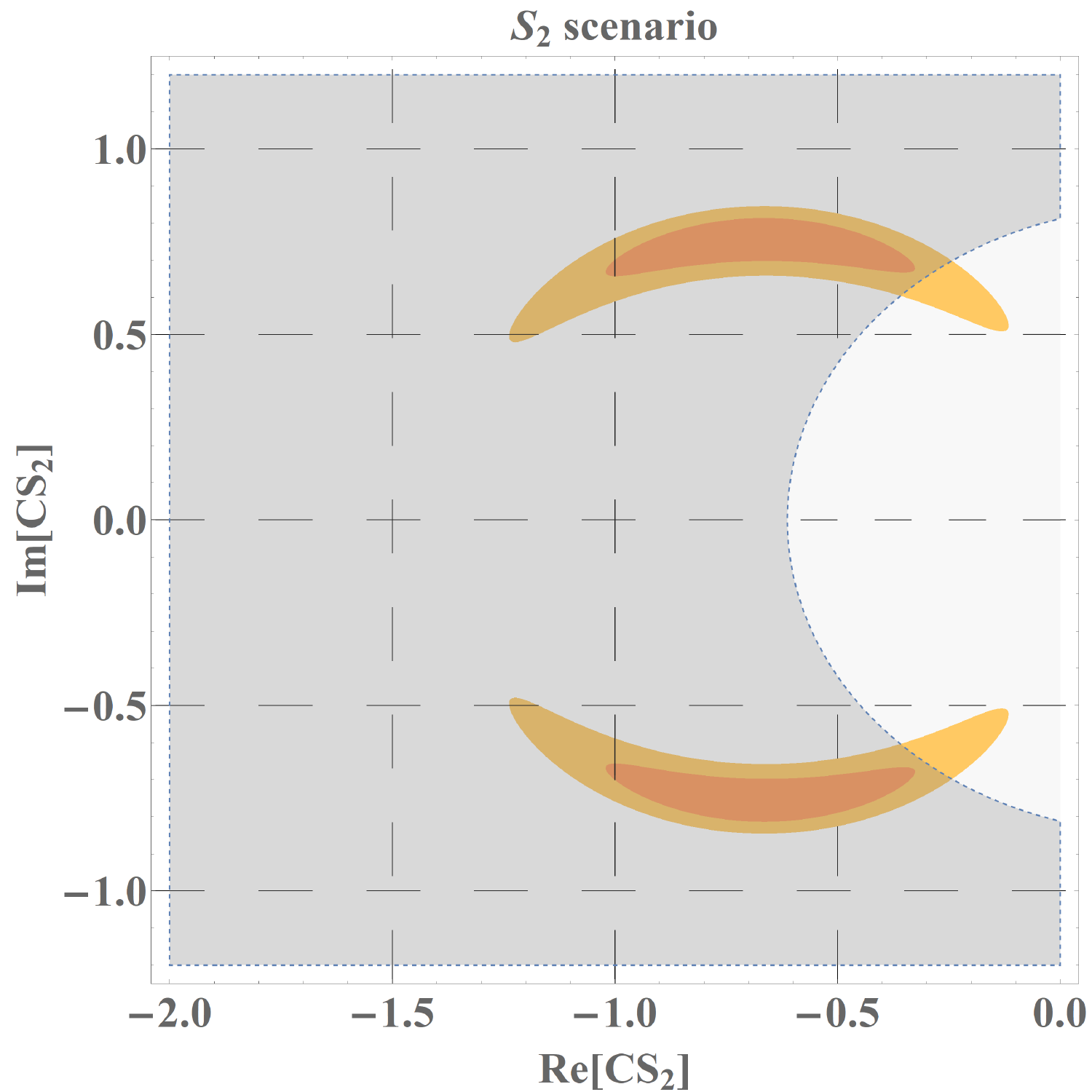}
\includegraphics[scale=0.22]{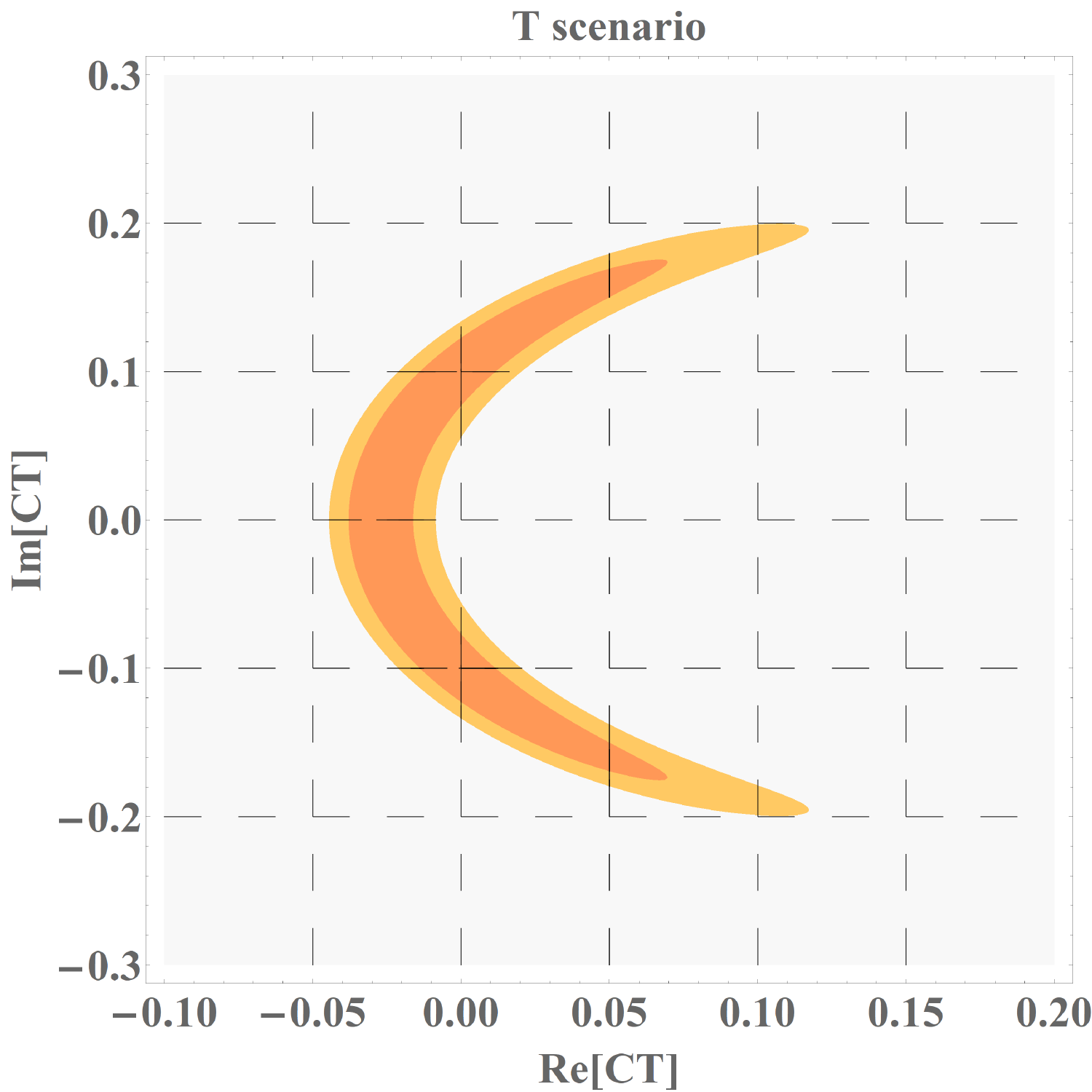}
\caption{
Constraints on the Wilson coefficients by the $b\to c\tau\nu$ data at  $1\sigma$ and $2\sigma$, and the limit on $\mathcal B(B_c \to \tau\nu)$. The forbidden regions by $\mathcal B(B_c \to \tau\nu)<30\%$ are in dark grey.}
\label{fig:constraint2}
\end{center}
\end{figure}

The resulting global $\chi^{2}$ best-fit values of the Wilson coefficients in model-independent scenarios without and with imposing $\mathcal{B}(B_{c}\to\tau\nu)< 30\%$, along with the p-values and the pulls, are presented in Table~\ref{tab:wcoef1a} and Table~\ref{tab:wcoef1}, respectively. For single operator scenarios, it has been observed that the scalar scenario $S_{1}$ is excluded at $2\sigma$ given that $\chi^{2}=27.6$, is the excluded limit at $95\%$ CL for dof = 17, while scenarios $S_{2}$, $V_{1}$, $V_{2}$ and $T$ are still allowed. The exclusion of the scalar $S_1$ is mainly due to its limitation in explaining the anomalies $R_D$ and $R_{D^*}$, which is significant at 3.48$\sigma$. This suggests that the remaining NP scenarios may still accommodate the observed anomalies. Besides the values of Wilson coefficients obtained through global fit, we also present the plots
of Wilson coefficients allowed by different measurements within $1\sigma$ and $2\sigma$ ranges in Figure~\ref{fig:constraint2}.


\subsection{Analysis of Physical Observables using Model independent Approach}\label{MOD1}

In this section, we present our predictions of the observables involved in $B\to D^{(\ast)}\tau\bar{\nu}_{\tau}$, such as LFU ratio $R_{D^{(\ast)}}$, longitudinal $\tau$ polarization $P_{\tau}(D^{(\ast)})$, longitudinal polarization of final state $D^{\ast}$ meson $F_{L}^{D^{\ast}}$, and $\mathcal{A}^{\text{FB}}_{D^{(\ast)}}$, all within model independent scenario. Furthermore, we also provide the predictions using normalized angular coefficients $\langle I^{n}_{\lambda}\rangle$  in the decays of $B\to D^{\ast}(\to D\pi,D\gamma)\tau\bar{\nu}_{\tau}$ in the model independent approach.

\begin{table*}[!b]\small%
\centering
\caption{Predictions for $R_D$, $R_{D^*}$, $P_\tau(D)$ and $P_\tau(D^*)$. The first and second uncertainties result from the form factors and the fitted Wilson coefficients.}
\begin{tabular}{ccccc}
\hline
\hline
Scenario & $R_D$& $R_{D^*}$   &$P_\tau(D)$ & $P_\tau(D^*)$ \\
\hline
SM             &$0.300(3)(0)$ & $0.251(1)(0)$ &$0.325(3)(0)$ &$-0.507(3)(0)$ \\
$V_1$       &$0.335(3)(10)$ & $0.280(1)(8)$ &$0.325(3)(0)$ &$-0.507(3)(0)$ \\
$V_2$       &$0.324(3)(20)$ & $0.285(1)(11)$ &$0.325(3)(0)$ &$-0.507(3)(0)$ \\
$S_2$      & $0.330(3)(73)$ &  $0.266(1)(15)$ &$0.386(3)(82)$
&$-0.422(5)(77)$   \\
$T$       &$0.299(3)(14)$ & $0.289(1)(71)$ &$0.328(3)(28)$ &$-0.449(3)(92)$ \\
\hline
\hline
\end{tabular}
\label{tab:obser1}
\end{table*}
\begin{table*}[!htbp]\small%
\centering
\caption{Predictions for $F_L^{D^*}$, $\mathcal{A}^{\text{FB}}_{D}$ and $\mathcal{A}^{\text{FB}}_{D^{\ast}}$. The first and second uncertainties result from the input parameters and the fitted Wilson coefficients.}
\begin{tabular}{cccc}
\hline
\hline
Scenario  &$F_L^{D^*}$ &$\mathcal{A}^{\text{FB}}_{D}$& $\mathcal{A}^{\text{FB}}_{D^{\ast}}$\\
\hline
SM          &$0.453(3)(0)$ & $0.360(0)(0)$ & $-0.052(4)(0)$\\
$V_1$       &$0.453(3)(0)$ & $0.360(0)(0)$ & $-0.052(4)(0)$\\
$V_2$        &$0.454(3)(2)$ & $0.360(0)(0)$ & $-0.013(3)(9)$\\
$S_2$     &$0.484(3)(28)$  & $0.205(0)(99)$   & $-0.022(4)(30)$ \\
$T$      &$0.428(3)(38)$ & $0.356(0)(16)$ & $-0.018(3)(35)$\\
\hline
\hline
\end{tabular}
\label{tab:obser12}
\end{table*}


Using the best-fit values of the Wilson coefficients presented in Tables \ref{tab:wcoef1a} and \ref{tab:wcoef1}, we obtain the predictions on the physical observables.The predicted numerical values of the above-mentioned physical observables are presented in Tables~\ref{tab:obser1}-\ref{table:anggma}. The errors given in these tables arise due to hadronic uncertainties and Wilson coefficients. Compared to SM, each of the NP scenarios except $S_1$ can explain the data of $R_{D^{(\ast)}}$. The polarization observables $P_{\tau}(D),P_{\tau}(D^{\ast})$, $F^{D^{\ast}}_{L}$, $\mathcal{A}^{\text{FB}}_{D^{(\ast)}}$, and the normalized angular coefficients $\langle I^{n}_{\lambda}\rangle$ arising from the four-fold distributions given in Eqs. (\ref{fullad}) and (\ref{fulladgma}), will be useful in discriminating different NP scenarios in the incoming and future collider experiments, such as Belle-II and LHCb. Among the observables mentioned above, $P_{\tau}(D^{\ast})$ and $F^{D^{\ast}}_{L}$ have already been measured in Belle~\cite{Hirose:2016wfn,Adamczyk:2019wyt,Abdesselam:2019wbt} as listed in Table~\ref{tab:exdata}, while $P_{\tau}(D)$ and $\mathcal{A}^{\text{FB}}_{D^{(\ast)}}$ have not been measured yet.

\begin{table}[b!]
\begin{center}
\caption{Predictions of averaged values of angular observables for the $B\to D^{\ast}(\to D\pi)\tau\bar{\nu}_{\tau}$ decay, for ${V_{1}}$, ${V_{2}}$, ${P}$ and ${T}$. The first and second errors presented arise from the uncertainties of the form factors and Wilson coefficients, respectively.}\label{table:angpi}
\begin{tabular}{|c||c|c|c|c|c|}
 \hline
 \textbf{~}&\textbf{SM}& \textbf{${V_{1}}$} &\textbf{${V_{2}}$}&\textbf{${P}$}&\textbf{${T}$}\\
 \hline
 $\langle I^{\pi}_{1s}\rangle$   & $0.364(6)(0)$& $0.364(7)(0)$&   $0.363(6)(11)$&   $0.373(6)(10)$&   $0.372(8)(13)$\\
 \hline
 $\langle I^{\pi}_{1c}\rangle$ & $0.594(12)(0)$   & $0.594(12)(0)$&   $0.595(12)(1)$&   $0.573(12)(20)$&   $0.579(16)(27)$\\
 \hline
$\langle I^{\pi}_{2s}\rangle$ & $0.058(0)(0)$   & $0.058(0)(0)$&   $0.058(0)(0)$&   $0.060(0)(1)$&   $0.035(0)(30)$\\
 \hline
$\langle I^{\pi}_{2c}\rangle$ & $-0.151(2)(0)$   & $-0.151(2)(0)$&   $-0.151(2)(0)$&   $-0.155(2)(4)$&   $-0.10(4)(60)$\\
 \hline
  $\langle I^{\pi}_{3}\rangle$   & $-0.100(2)(0)$   & $-0.100(2)(0)$&   $-0.101(2)(0)$&   $-0.103(1)(3)$&   $-0.065(1)(45)$\\
 \hline
 $\langle I^{\pi}_{4}\rangle$ & $-0.127(0)(0)$   & $-0.127(0)(0)$&   $-0.127(0)(0)$&   $-0.130(0)(3)$&   $-0.082(1)(58)$\\
 \hline
$\langle I^{\pi}_{5}\rangle$ & $0.284(6)(0)$   & $0.284(6)(0)$&   $0.280(6)(4)$&   $0.248(7)(18)$&   $0.212(8)(90)$\\
 \hline
$\langle I^{\pi}_{6s}\rangle$ & $-0.217(10)(0)$   & $-0.217(10)(0)$&   $-0.170(10)(12)$&   $-0.223(10)(6)$&   $-0.134(6)(95)$\\
 \hline
 $\langle I^{\pi}_{6c}\rangle$   & $0.40(11)(0)$   & $0.40(9)(0)$&   $0.40(9)(1)$&   $0.30(8)(65)$&   $0.344(9)(81)$\\
 \hline
\end{tabular}
\end{center}
\end{table}

\begin{table}[htbp!]
\begin{center}
\caption{Predictions of averaged values of angular observables for the $B\to D^{\ast}(\to D\gamma)\tau\bar{\nu}_{\tau}$ decay, for ${V_{1}}$, ${V_{2}}$, ${P}$ and ${T}$. The first and second errors presented arise from the uncertainties of the form factors and Wilson coefficients, respectively.}\label{table:anggma}
\begin{tabular}{|c||c|c|c|c|c|}
 \hline
 \textbf{~}&\textbf{SM}& \textbf{${V_{1}}$} &\textbf{${V_{2}}$}&\textbf{${P}$}&\textbf{${T}$}\\
 \hline
 $\langle I^{\gamma}_{1s}\rangle$ & $0.542(2)(0)$ & $0.542(2)(0)$   & $0.542(2)(5)$&   $0.536(3)(8)$&   $0.547(3)(8)$\\
 \hline
 $\langle I^{\gamma}_{1c}\rangle$& $0.165(5)(0)$& $0.165(5)(0)$   & $0.164(5)(0)$&   $0.173(5)(10)$&   $0.171(5)(11)$\\
 \hline
$\langle I^{\gamma}_{2s}\rangle$ &$-0.137(1)(0)$& $-0.137(1)(0)$   & $-0.137(1)(0)$&   $-0.144(1)(9)$&   $-0.110(3)(17)$\\
 \hline
$\langle I^{\gamma}_{2c}\rangle$ &$0.026(0)(0)$& $0.026(0)(0)$   & $0.026(0)(0)$&   $0.027(0)(1)$&   $0.020(0)(11)$\\
 \hline
  $\langle I^{\gamma}_{3}\rangle$  &$0.091(3)(0)$ & $0.091(3)(0)$   & $0.091(3)(0)$&   $0.096(3)(5)$&   $0.072(2)(38)$\\
 \hline
 $\langle I^{\gamma}_{4}\rangle$ &$0.115(2)(0)$& $0.115(2)(0)$   & $0.115(2)(0)$&   $0.121(2)(7)$&   $0.091(2)(43)$\\
 \hline
$\langle I^{\gamma}_{5}\rangle$ &$-0.257(7)(0)$ &$-0.257(7)(0)$   & $-0.253(7)(4)$&   $-0.228(7)(17)$&   $-0.221(9)(68)$\\
 \hline
$\langle I^{\gamma}_{6s}\rangle$ & $0.367(5)(0)$&$0.367(5)(0)$   & $0.367(5)(0)$&   $0.273(4)(62)$&   $0.342(7)(50)$\\
 \hline
 $\langle I^{\gamma}_{6c}\rangle$   &$-0.098(5)(0)$& $-0.098(5)(0)$  & $-0.08(4)(6)$&   $-0.103(5)(6)$&   $-0.074(5)(44)$\\
 \hline
\end{tabular}
\end{center}
\end{table}

\begin{figure}[!htbp]
\begin{center}
\includegraphics[scale=0.55]{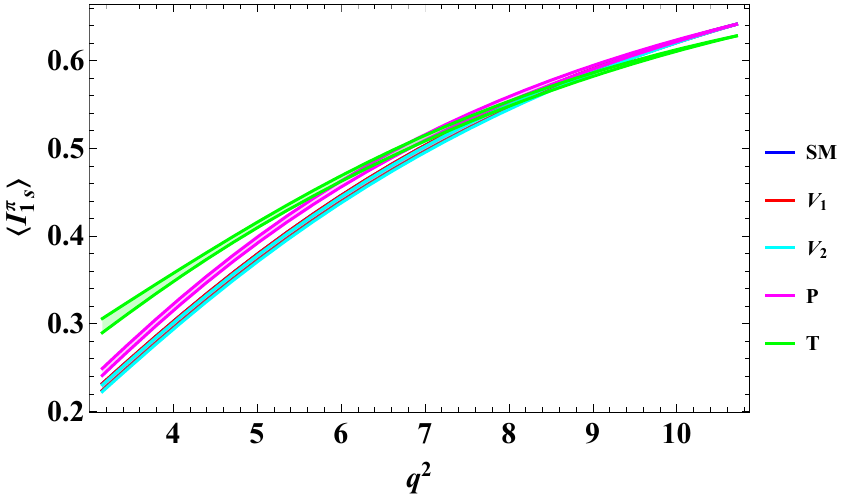}
\includegraphics[scale=0.55]{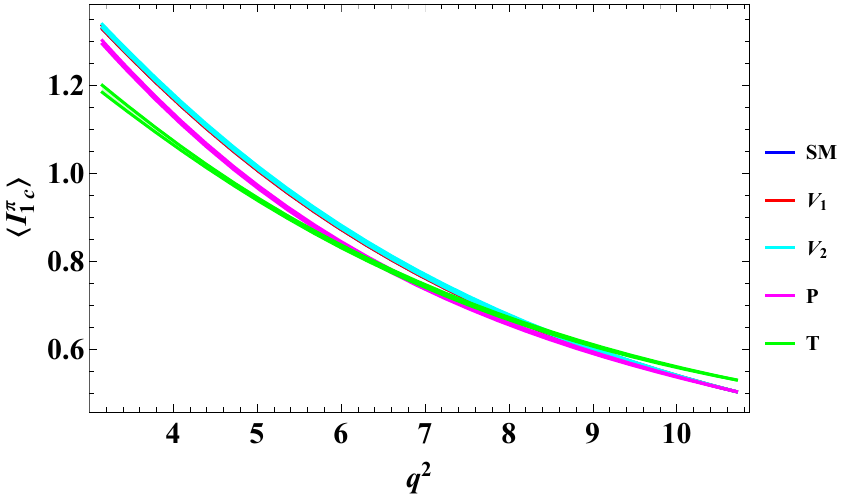}
\includegraphics[scale=0.55]{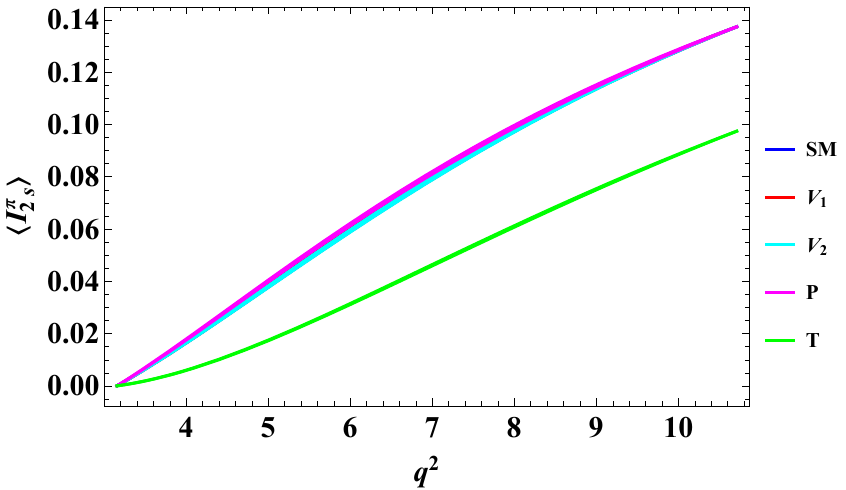}
\includegraphics[scale=0.55]{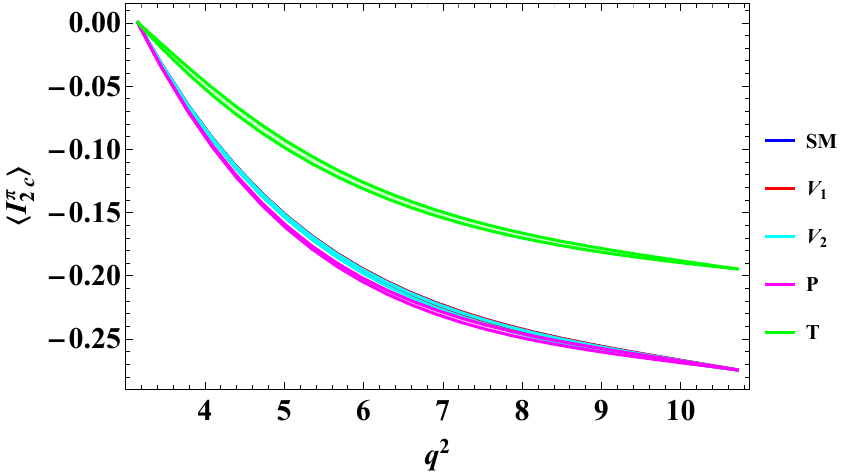}
\includegraphics[scale=0.55]{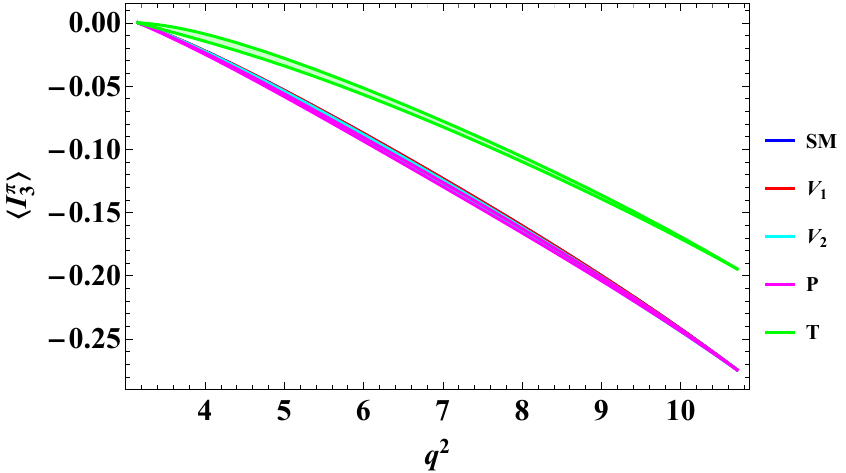}
\includegraphics[scale=0.55]{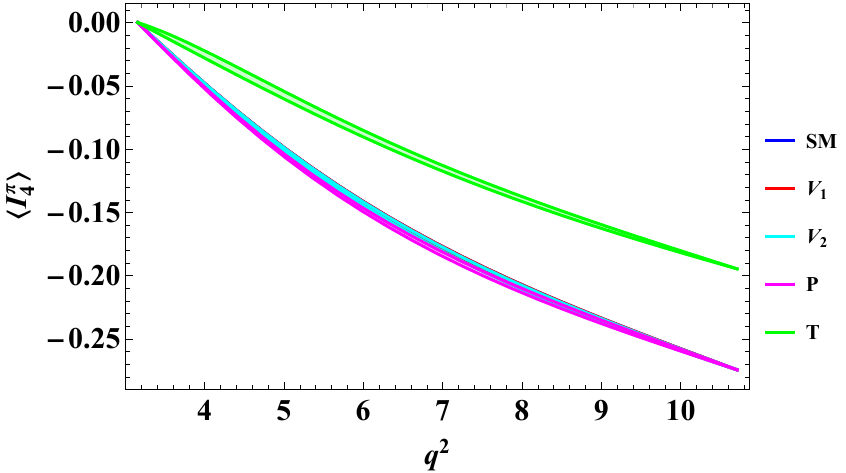}
\includegraphics[scale=0.55]{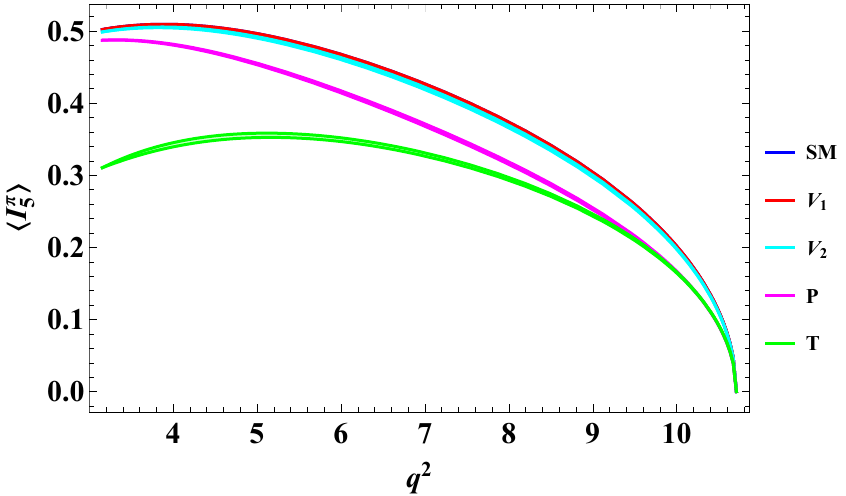}
\includegraphics[scale=0.55]{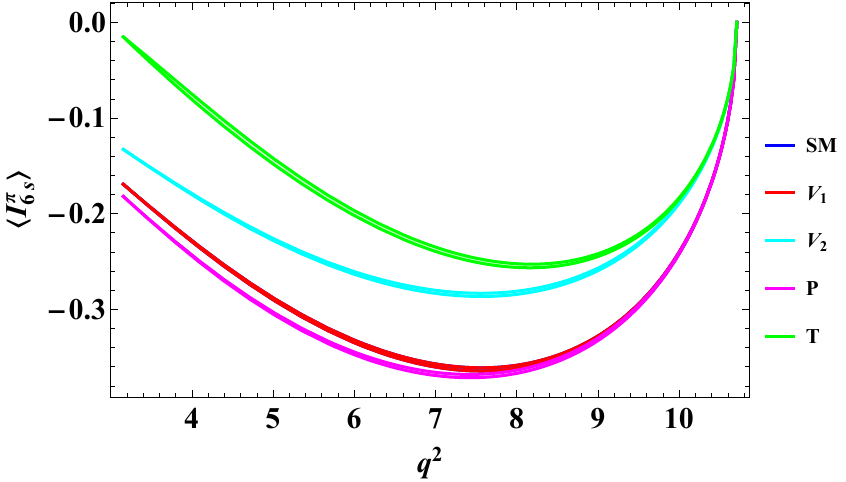}
\includegraphics[scale=0.55]{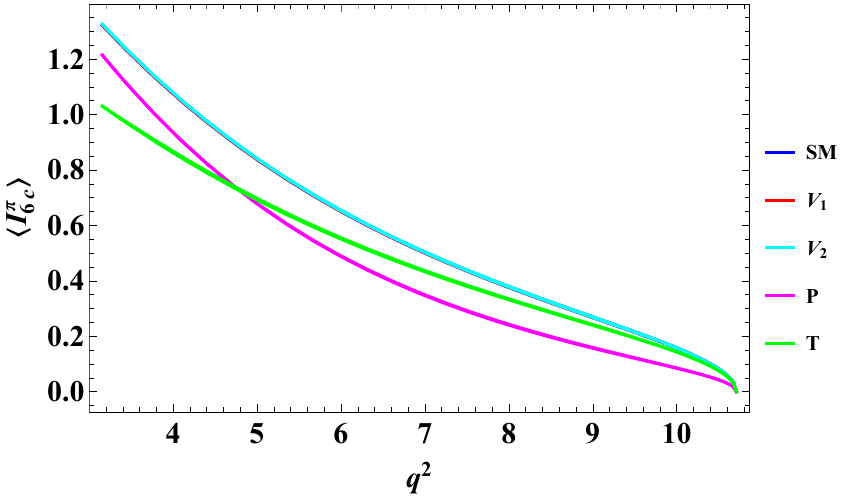}
\caption{Angular observables $\langle I^{\pi}_{1s}\rangle$, $\langle I^{\pi}_{1c}\rangle$, $\langle I^{\pi}_{2s}\rangle$, $\langle I^{\pi}_{2c}\rangle$, $\langle I^{\pi}_{3}\rangle$, $\langle I^{\pi}_{4}\rangle$, $\langle I^{\pi}_{5}\rangle$, $\langle I^{\pi}_{6s}\rangle$, and $\langle I^{\pi}_{6c}\rangle$ for the decay $B\to D^{\ast}(\to D\pi)\tau^{-}\bar{\nu}$, in SM and in the model independent scenarios.}
\label{fig:angcoeffDpiMI}
\end{center}
\end{figure}

\begin{figure}[!htbp]
\begin{center}
\includegraphics[scale=0.55]{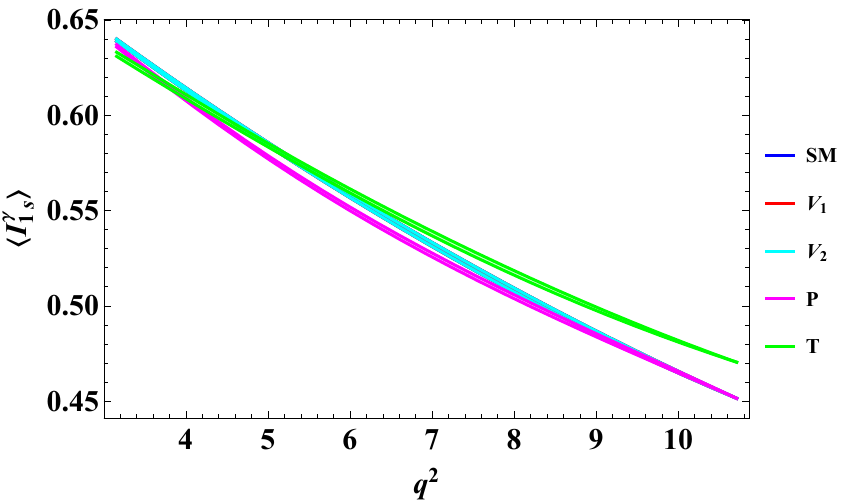}
\includegraphics[scale=0.55]{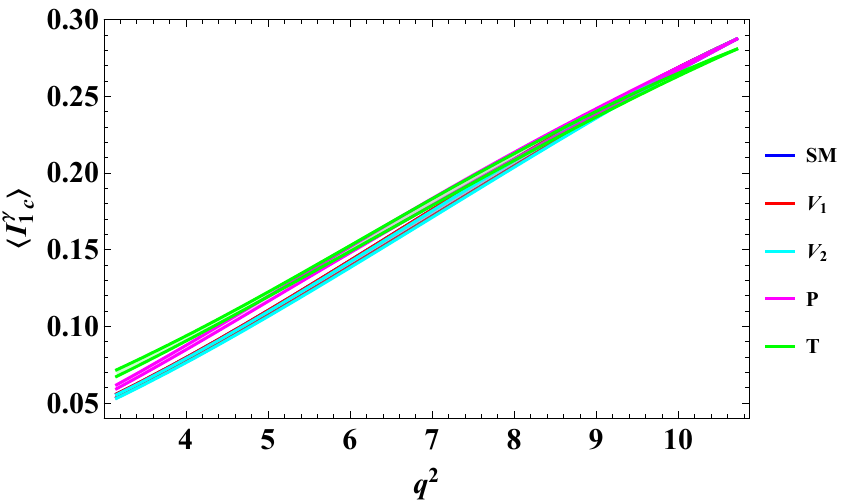}
\includegraphics[scale=0.55]{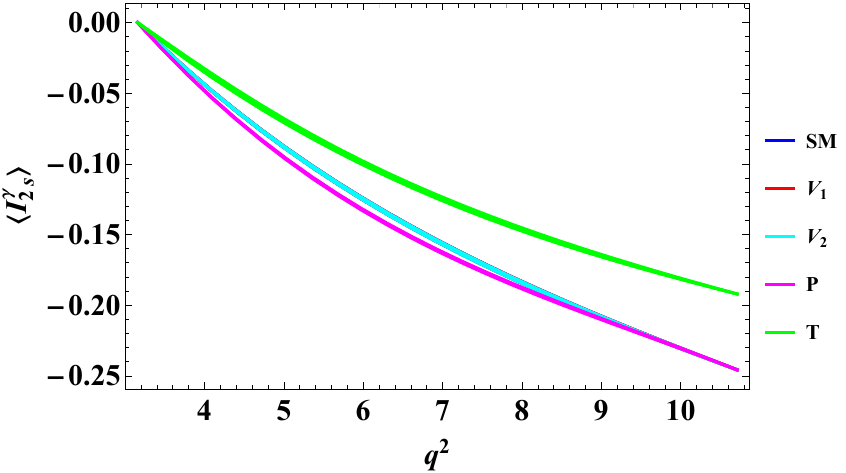}
\includegraphics[scale=0.55]{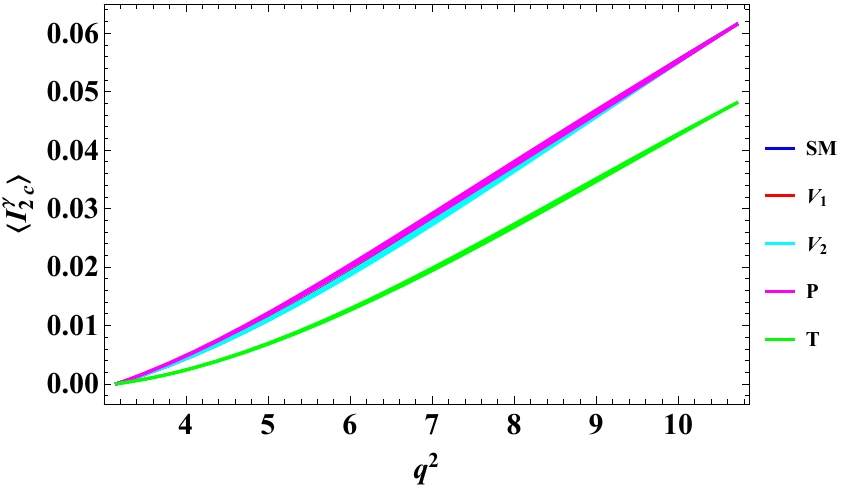}
\includegraphics[scale=0.55]{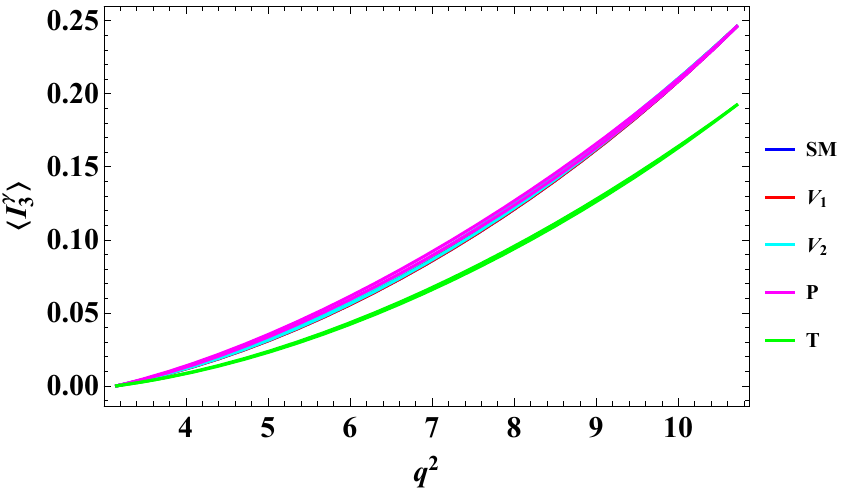}
\includegraphics[scale=0.55]{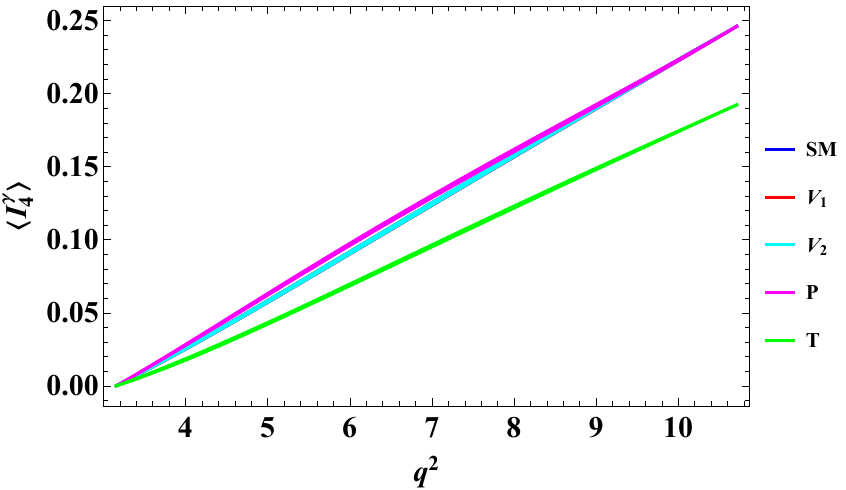}
\includegraphics[scale=0.55]{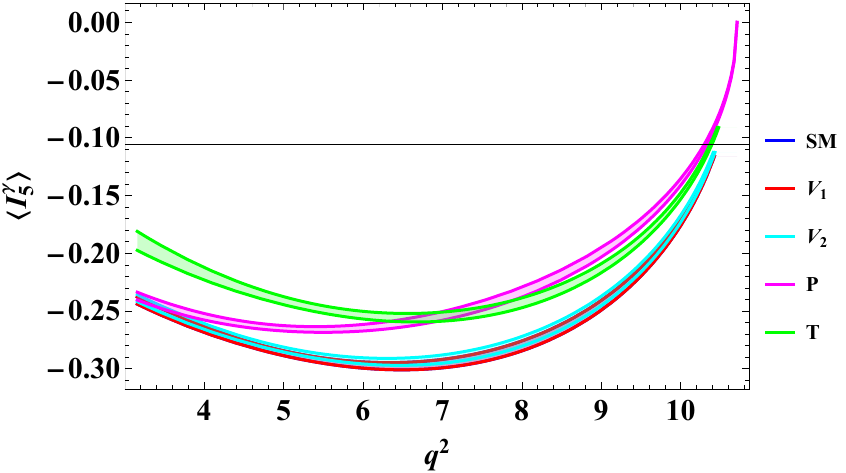}
\includegraphics[scale=0.55]{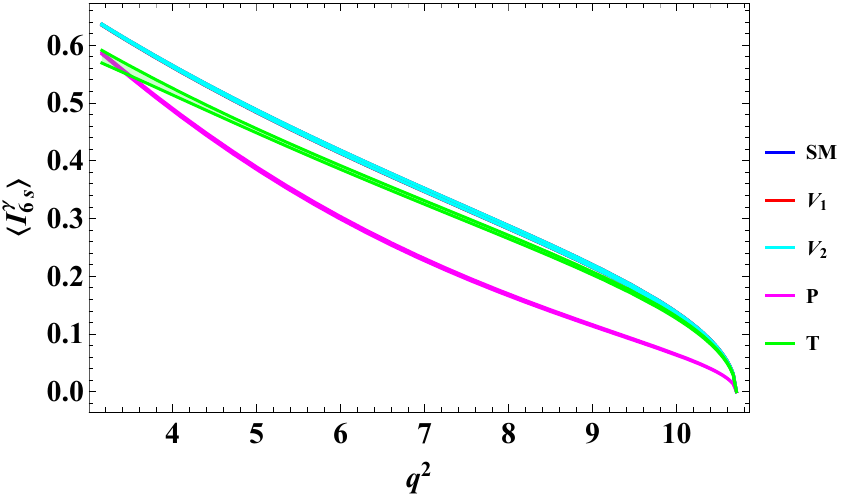}
\includegraphics[scale=0.55]{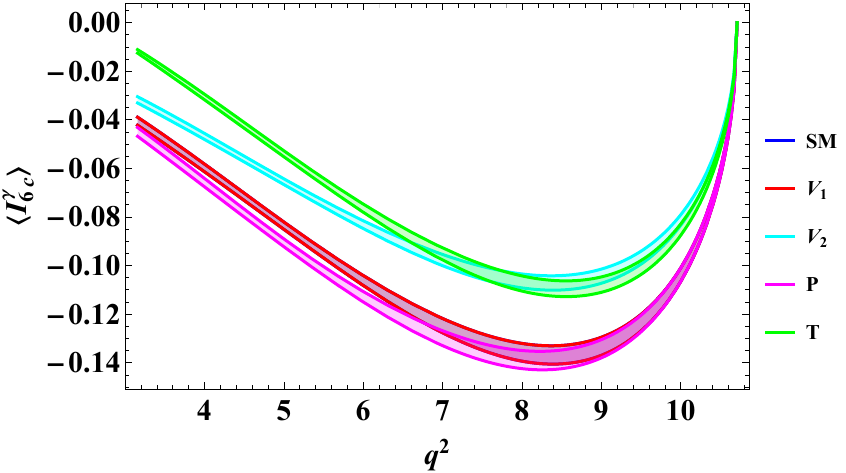}
\caption{Angular observables $\langle I^{\gamma}_{1s}\rangle$, $\langle I^{\gamma}_{1c}\rangle$, $\langle I^{\gamma}_{2s}\rangle$, $\langle I^{\gamma}_{2c}\rangle$, $\langle I^{\gamma}_{3}\rangle$, $\langle I^{\gamma}_{4}\rangle$, $\langle I^{\gamma}_{5}\rangle$, $\langle I^{\gamma}_{6s}\rangle$, and  $\langle I^{\gamma}_{6c}\rangle$ for the decay $B\to D^{\ast}(\to D\gamma)\tau^{-}\bar{\nu}$, in SM and in the model independent scenarios.}
\label{fig:angcoeffDgammaMI}
\end{center}
\end{figure}

In most NP scenarios, the predictions for $P_{\tau}(D)$ are closer to that of SM, but in the scenario $S_{2}$, we predict the larger value of $P_{\tau}(D)$ compared to that of SM. Similarly, in most scenarios, the predictions of $P_{\tau}(D^*)$ are also close to SM prediction, but for scenarios $S_{2}$ and $T$, the predicted value of $P_{\tau}(D^{\ast})$ is large compared to that of SM prediction, and hence the precise measurements on longitudinal  $\tau$ polarizations will be useful to discriminate different NP scenarios. For the longitudinal $D^*$ polarization, predictions of SM and most scenarios are close to $0.46$, but in $T$ scenario the prediction is smaller than that of SM value, while in $S_{2}$ scenario the value is bigger than that of SM value. In SM and NP scenarios $V_{1}$, $V_{2}$, $\mathcal{A}^{\text{FB}}_{D}$ is predicted to be close to $0.36$, except for scenarios $S_{2}$ and $T$, where the predicted values of $\mathcal{A}^{\text{FB}}_{D}$ are smaller than the SM value. For  $\mathcal{A}^{\text{FB}}_{D^{\ast}}$ only the scenario $V_{1}$ merges with SM, with the predicted value of $\mathcal{A}^{\text{FB}}_{D^{\ast}}\simeq-0.052$. However, the value of $\mathcal{A}^{\text{FB}}_{D^{\ast}}$ in other scenarios $V_{2}$, $S_{2}$, and $T$ shows departure from SM prediction. For both observables  $\mathcal{A}^{\text{FB}}_{D}$ and $\mathcal{A}^{\text{FB}}_{D^{\ast}}$ the numerical values are presented in Table~\ref{tab:obser12}, where the predicted scenarios can be discriminated in the future and current colliders.

Now, we turn to discuss the phenomenological analysis of the normalized angular coefficients of $B\to D^{\ast}(\to D\pi,D\gamma)\tau\bar{\nu}_{\tau}$ decays.
In Figures \ref{fig:angcoeffDpiMI} and \ref{fig:angcoeffDgammaMI}, we have plotted the normalized angular coefficients as a function of $q^2$ for SM and the model independent NP scenarios, whereas the averaged values of these observables are presented in Tables~\ref{table:angpi} and \ref{table:anggma}. In Figure~\ref{fig:angcoeffDpiMI}, we present the angular observables $\langle I_{1s}^{\pi}\rangle$, $\langle I_{1c}^{\pi}\rangle$, $\langle I_{2s}^{\pi}\rangle$, $\langle I_{2c}^{\pi}\rangle$, $\langle I^{\pi}_{3}\rangle$, $\langle I^{\pi}_{4}\rangle$, $\langle I^{\pi}_{5}\rangle$, $\langle I_{6s}^{\pi}\rangle$ and $\langle I_{6c}^{\pi}\rangle$ for the decay $B\to D^{\ast}(\to D\pi)\tau\bar{\nu}_{\tau}$ in SM and NP model independent scenarios that exhibit vectors $V_{1}$, $V_{2}$, pseudoscalar $P$ and tensor $T$ type couplings.
It is evident that the angular coefficients predictions of $\langle I_{1s}^{\pi}\rangle$, $\langle I_{1c}^{\pi}\rangle$, $\langle I_{2s}^{\pi}\rangle$, $\langle I_{2c}^{\pi}\rangle$, $\langle I^{\pi}_{3}\rangle$, $\langle I^{\pi}_{4}\rangle$, $\langle I^{\pi}_{5}\rangle$, and $\langle I_{6c}^{\pi}\rangle$ overlap with the predictions of the SM in the presence of vector-type NP couplings, $V_1$ and $V_2$. However, NP contributions from pseudoscalar ($P$) and tensor ($T$) couplings deviate significantly from the SM predictions. For the observables $\langle I_{1s}^{\pi} \rangle$ and $\langle I_{1c}^{\pi} \rangle$, the NP couplings $P$ and $T$ show significant overlap at higher $q^2$ values but exhibit deviations at lower $q^2$ values. For the angular coefficients $\langle I_{2s}^{\pi}\rangle$, $\langle I_{2c}^{\pi}\rangle$, $\langle I_{3}^{\pi}\rangle$, and
$\langle I_{4}^{\pi}\rangle$, it is clear that all scenarios are closer to SM values except for the coupling $T$ which is distinguishable from SM. Furthermore, for the angular coefficients $\langle I^{\pi}_{5}\rangle$ and $\langle I^{\pi}_{6c}\rangle$, $P$ and $T$ scenarios show significant deviations from the SM predictions for almost the entire range of $q^2$. Lastly, for $\langle I^{\pi}_{6s}\rangle$, $V_2$, $P$ and $T$ scenarios predictions vary from the SM values, whereas scenario, $V_1$ lies very close to the SM predictions.

\begin{table*}[!b]
\centering
\footnotesize  
\renewcommand{\arraystretch}{1.1}  
\setlength{\tabcolsep}{4pt}  
\caption{Predictions of observables $R_D$, $R_{D^*}$, $P_\tau(D)$, and $P_\tau(D^*)$, showing their most distinguishing values w.r.t the SM in the selected scenarios. The first and second uncertainties arise from the form factors and the fitted Wilson coefficients, respectively.}\label{tab:obserDis}
\begin{tabular}{lccccccc}
\hline\hline
\textbf{Scenario} & $R_D$ & $R_{D^*}$ & $P_\tau(D)$ & $P_\tau(D^*)$ & $F_L^{D^*}$ & $\mathcal{A}^{\text{FB}}_{D}$ & $\mathcal{A}^{\text{FB}}_{D^*}$ \\
\hline
SM        & $0.300(3)(0)$  & $0.251(1)(0)$  & $0.325(3)(0)$  & $-0.507(3)(0)$  & $0.453(3)(0)$  & $0.360(0)(0)$  & $-0.052(4)(0)$ \\
$V_1$     & $0.335(3)(10)$ & $--$  & $--$             & $--$             & $--$             & $--$             & $--$  \\
$V_2$     & $--$ & $--$ & $--$             & $--$             & $--$             & $--$             & $-0.013(3)(9)$ \\
$S_2$     & $--$ & $--$ & $0.386(3)(82)$ & $-0.422(5)(77)$ & $0.484(3)(28)$            & $0.205(0)(99)$            & $--$  \\
$T$       & $--$           & $0.289(1)(71)$ & $--$ & $--$ & $--$            & $--$           & $--$  \\
\hline\hline
\end{tabular}
\end{table*}


\begin{table}[ht!]
\begin{center}
\caption{Predictions of different angular observables for the $B\to D^{\ast}(\to D\pi,D\gamma)\tau\bar{\nu}_{\tau}$ decay, showing their most distinguishing values w.r.t the SM in the selected scenarios. The first and second errors presented arise from the uncertainties of the form factors and Wilson coefficients, respectively.}\label{table:angDis}
\begin{tabular}{|c||c|c|c|c|c|}
 \hline
 \textbf{~}&\textbf{SM}& \textbf{${V_{1}}$} &\textbf{${V_{2}}$}&\textbf{${P}$}&\textbf{${T}$}\\
 \hline
 $\langle I^{\pi}_{2s}\rangle$   & $0.058(0)(0)$&   $--$&   $--$& $0.060(0)(1)$&  $0.035(0)(30)$\\
 \hline
 $\langle I^{\pi}_{2c}\rangle$ & $-0.151(2)(0)$   & $--$&   $--$&   $-0.155(2)(4)$&   $-0.10(4)(60)$\\
 \hline
$\langle I^{\pi}_{3}\rangle$ & $-0.100(2)(0)$   & $--$&   $--$&   $--$&   $-0.065(1)(45)$\\
 \hline
$\langle I^{\pi}_{4}\rangle$ & $-0.127(0)(0)$   & $--$&   $--$&   $--$&   $-0.082(1)(58)$\\
 \hline
  $\langle I^{\pi}_{6s}\rangle$   & $-0.217(10)(0)$   & $--$&   $-0.170(10)(12)$&   $--$&   $-0.134(6)(95)$\\
 \hline
 $\langle I^{\pi}_{6c}\rangle$ & $0.40(11)(0)$   & $--$&   $--$&   $0.30(8)(65)$&   $0.344(9)(81)$\\
 \hline
$\langle I^{\gamma}_{2s}\rangle$ & $-0.137(1)(0)$   & $--$&   $--$&   $-0.144(1)(9)$&   $-0.110(3)(17)$\\
 \hline
$\langle I^{\gamma}_{3}\rangle$ & $0.091(3)(0)$   & $--$&   $--$&   $--$&   $0.072(2)(38)$\\
 \hline
 $\langle I^{\gamma}_{4}\rangle$   & $0.115(2)(0)$   & $--$&   $--$&   $--$&   $0.091(2)(43)$\\
 \hline
 $\langle I^{\gamma}_{6s}\rangle$   & $0.367(5)(0)$   & $--$&   $--$&   $0.273(4)(62)$&   $0.342(7)(50)$\\
 \hline
 $\langle I^{\gamma}_{6c}\rangle$   & $-0.098(5)(0)$   & $--$&   $-0.08(4)(6)$&   $-0.103(5)(6)$&   $-0.074(5)(44)$\\
 \hline
\end{tabular}
\end{center}
\end{table}

\par
Figure \ref{fig:angcoeffDgammaMI}, presents the angular observables $\langle I_{1s}^{\gamma}\rangle$, $\langle I_{1c}^{\gamma}\rangle$, $\langle I_{2s}^{\gamma}\rangle$, $\langle I_{2c}^{\gamma}\rangle$, $\langle I_{3}^{\gamma}\rangle$, $\langle I_{4}^{\gamma}\rangle$, $\langle I_{5}^{\gamma}\rangle$, $\langle I_{6s}^{\gamma}\rangle$ and $\langle I_{6c}^{\gamma}\rangle$ for the $B\to D^{\ast}(\to D\gamma)\tau\bar{\nu}_{\tau}$ decay in SM and model independent scenarios $V_{1}$, $V_{2}$, $P$ and $T$. For all angular observables, it is observed that the scenario $V_{1}$ and $V_{2}$ completely overlap with the SM except $\langle I_{6c}^{\gamma}\rangle$, where $V_{2}$ scenario is distinct from the SM and $V_{1}$ scenario predictions. The values within scenario $T$ are observed to be clearly distinct from SM and all other scenarios, for all angular observables, except for $\langle I^{\gamma}_{1s}\rangle$ and $\langle I^{\gamma}_{1c}\rangle$. However, for certain angular observables such as $\langle I^{\gamma}_{5}\rangle,\langle I^{\gamma}_{6s}\rangle$ and $\langle I^{\gamma}_{6c}\rangle$, scenario $T$ partially overlaps with $P$ and $V_{2}$ scenarios, respectively. Furthermore, scenario $P$ shows a clear departure from all other scenarios for the angular coefficient $\langle I^{\gamma}_{6s}\rangle$, whereas in $\langle I^{\gamma}_{6c}\rangle$ scenarios $V_{2}$ and $T$ overlap at high $q^{2}$, while scenarios $V_{1}$ and $P$ almost overlap for all values of $q^{2}$. Finally, we also give the summary tables of the above discussed observables with their most discriminant values w.r.t the SM, against the selected scenarios, in Tables \ref{tab:obserDis} and \ref{table:angDis}.


\subsection{Analysis of Physical Observables Using Leptoquark Models}\label{LQM1}
In this section, we analyze the physical observables for $B\to D^{\ast}(\to D\pi,D\gamma)\tau\bar{\nu}_{\tau}$ decay in SM and Leptoquark (LQ) models. There are 10 LQ models \cite{Tanabashi:2018oca,Dorsner:2016wpm,Buchmuller:1986zs}, that provide UV completion for SM fermions with left-handed neutrinos. Among them, three LQ models, namely $R_2$, $S_1$, and $U_1$, are presented in Table~\ref{tab:lq}, along with their quantum numbers, which can possibly explain the $b\to c\tau\nu$ anomalies as discussed in Refs.~\cite{Sakaki:2013bfa, Angelescu:2018tyl, Dorsner:2016wpm, Iguro:2018vqb}.

\begin{table}[!htbp]\small%
\centering
\caption{Quantum numbers and couplings of the leptoquarks which can explain the $b\rightarrow c\tau\nu$ anomalies. Here $F\equiv3B+L$.}
\begin{tabular}{ccccc}
\hline
\hline
&\begin{tabular}{c}SM quantum number\\$[\textrm{SU}(3)\times\textrm{SU}(2)\times\textrm{U}(1)]$\end{tabular}&$F$&Spin&Fermions coupled to\\
\hline
$R_2$&$(3,2,7/6)$&$0$&$0$&$\bar{c}_R\nu_L,\bar{b}_L\tau_R$\\
$S_1$&$(\bar{3},1,1/3)$&$-2$&$0$&$\bar{b}_L^c\nu_L,\bar{c}_L^c\tau_L,\bar{c}_R^c\tau_R$\\
$U_1$&$(3,1,2/3)$&$0$&$1$&$\bar{c}_L\gamma_{\mu}\nu_L,\bar{b}_L\gamma_{\mu}\tau_L,\bar{b}_R\gamma_{\mu}\tau_R$\\
\hline
\hline
\end{tabular}
\label{tab:lq}
\end{table}

\begin{figure}[!htb]
\begin{center}
\includegraphics[scale=0.22]{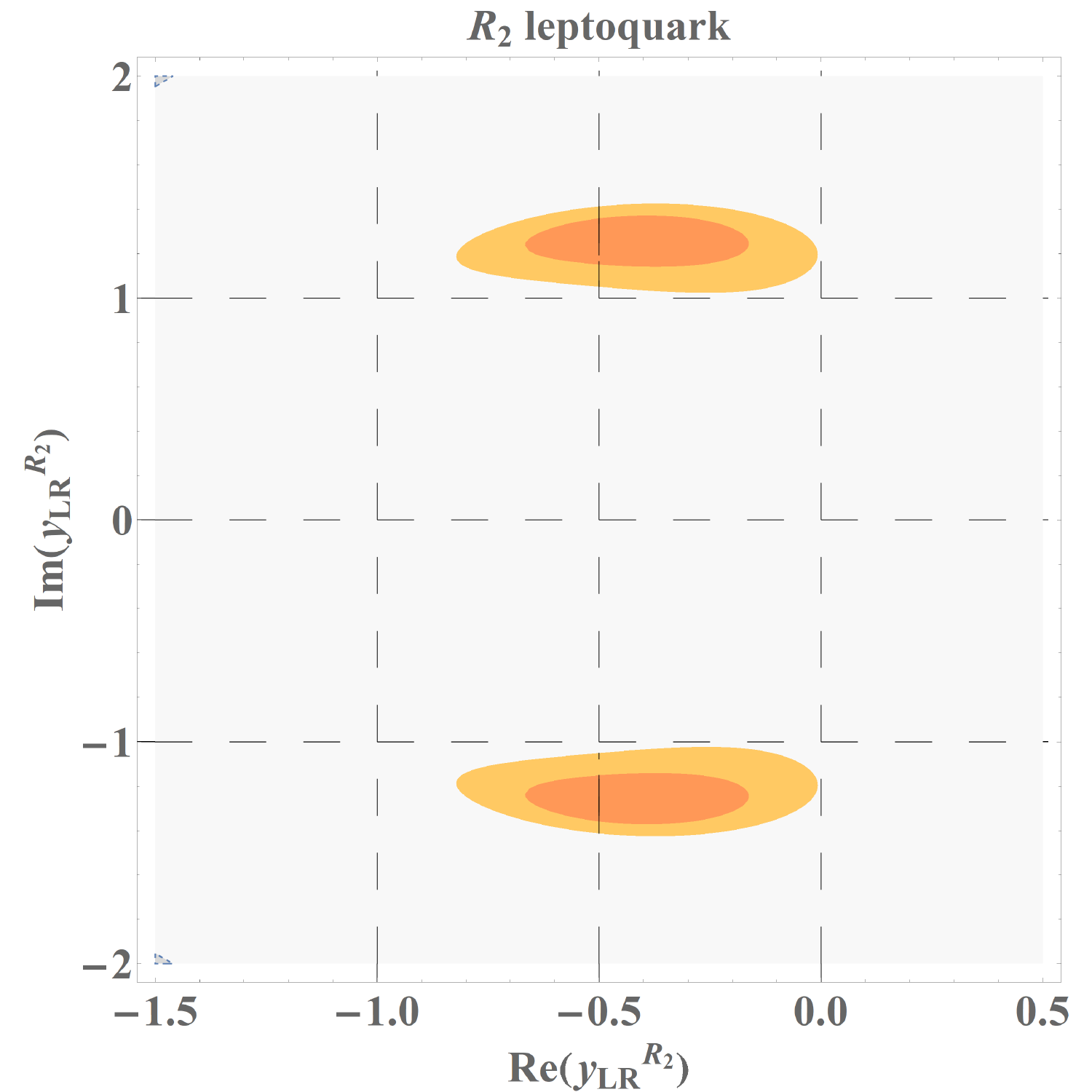}
\includegraphics[scale=0.22]{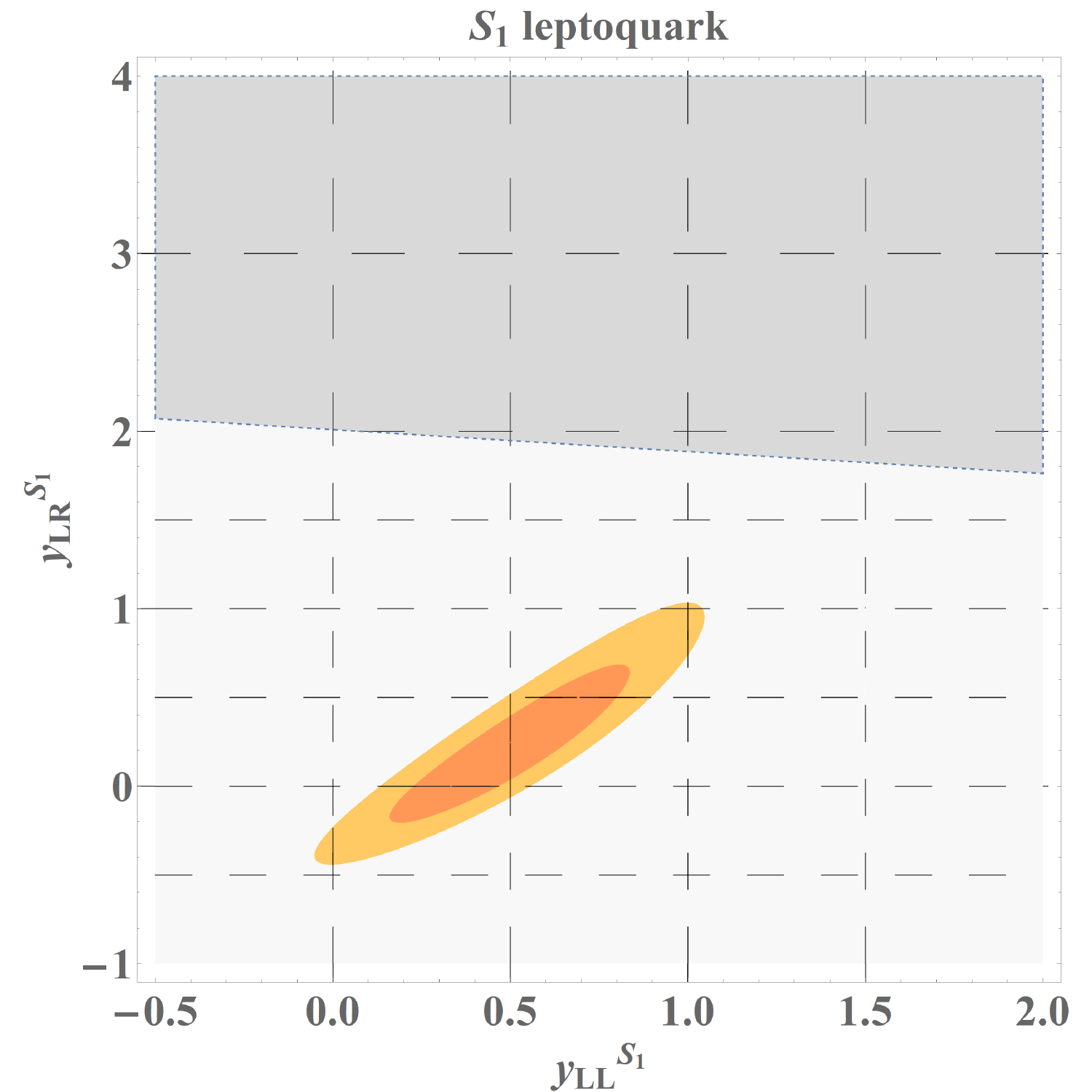}
\includegraphics[scale=0.22]{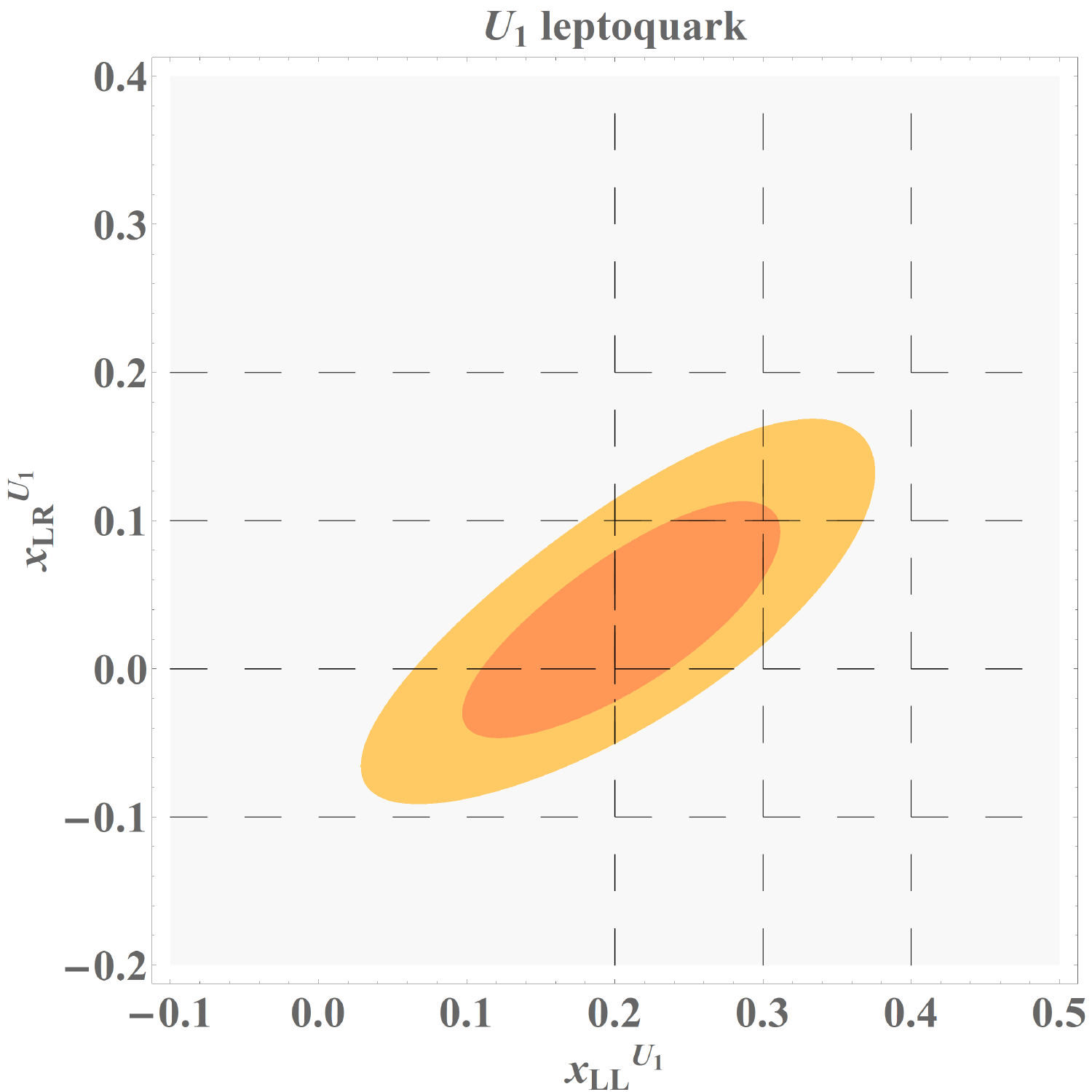}
\caption{Constraints on the leptoquark couplings by the $b\to c\tau\nu$ data at  $1\sigma$ and $2\sigma$, and the limit on $\mathcal B(B_c \to \tau\nu)$. The forbidden regions by $\mathcal B(B_c \to \tau\nu)<30\%$ are in dark grey.}
\label{fig:constraint2LQ}
\end{center}
\end{figure}

For a given LQ model, resulting global $\chi^{2}$ best-fit values of the NP couplings with and without imposing the constraint $\mathcal{B}(B_{c}\to\tau\nu_{\tau})<30\%$, along with the p-values and the pulls, are presented in Table~\ref{tab:wcoef3}. In addition to that, we have also plotted the allowed values of the Wilson coefficients, using different measurements, within the ranges $1\sigma$ and $2\sigma$, which are shown in Figure~{\ref{fig:constraint2LQ}}.
\begin{table*}[!htbp]
\centering
\renewcommand{\arraystretch}{1.2}  
\setlength{\tabcolsep}{6pt}  
\scriptsize  
\caption{Best-fit values of the NP couplings in the leptoquark models with and without $\mathcal B(B_c\to\tau\nu)<0.3$.}
\begin{tabular}{c c c c c c}
\hline\hline
\textbf{LQ Type} & \textbf{Best-fit values}  (with and without $\mathcal B(B_c\to\tau\nu)<0.3$) & \textbf{$\chi^2/dof$} & \textbf{Corr.} & \textbf{p-value (\%)} & \textbf{Pull ($\sigma$)} \\
\hline
$R_2$  & ${\rm Re}[y_L^{c\tau}(y_R^{b\tau})^*], {\rm Im}[y_L^{c\tau}(y_R^{b\tau})^*] = (-0.415(166), \pm 1.271(74))$
       & $22.45/17$  & $\pm 0.35$  & $16.8$ & $3.87$ \\
\hline
$S_1$  & ${y_L^{b\tau}(Vy_L^*)^{c\tau}, y_L^{b\tau}(y_R^{c\tau})^*} = (0.498(223), 0.209(291))$
       & $23.01/17$  & $0.91$  & $14.8$  & $3.80$ \\
$S_1$  & ${y_L^{b\tau}(Vy_L^*)^{c\tau}, y_L^{b\tau}(y_R^{c\tau})^*} = (-12.787(223), -0.209(291))$
       & $23.01/17$  & $0.91$  & $14.8$  & $3.80$ \\
\hline
$U_1$  & ${(Vx_L)^{c\tau}(x_L^{b\tau})^*, (Vx_L)^{c\tau}(x_R^{b\tau})^*} = (0.206(71), 0.030(52))$
       & $23.22/17$  & $0.77$  & $14.2$  & $3.77$ \\
$U_1$  & ${(Vx_L)^{c\tau}(x_L^{b\tau})^*, (Vx_L)^{c\tau}(x_R^{b\tau})^*} = (-6.350(71), -0.030(52))$
       & $23.22/17$  & $0.77$  & $14.2$  & $3.77$ \\
\hline\hline
\end{tabular}
\label{tab:wcoef3}
\end{table*}

\begin{figure}[!htbp]
\begin{center}
\includegraphics[scale=0.55]{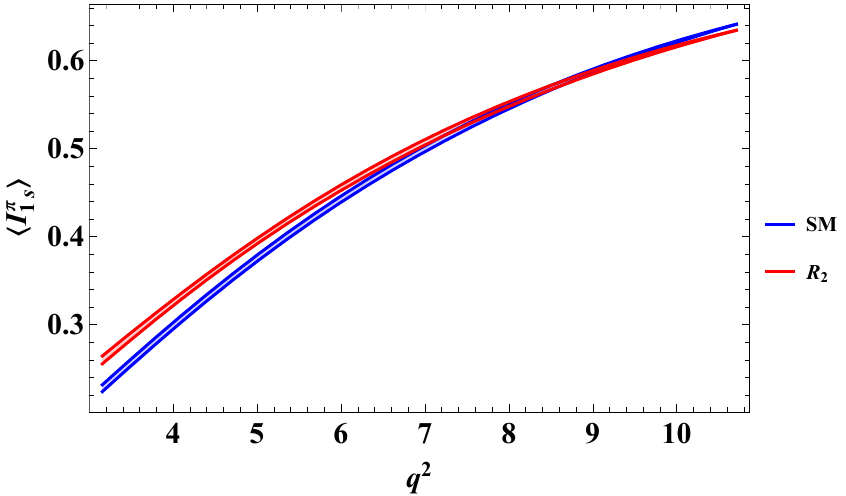}
\includegraphics[scale=0.55]{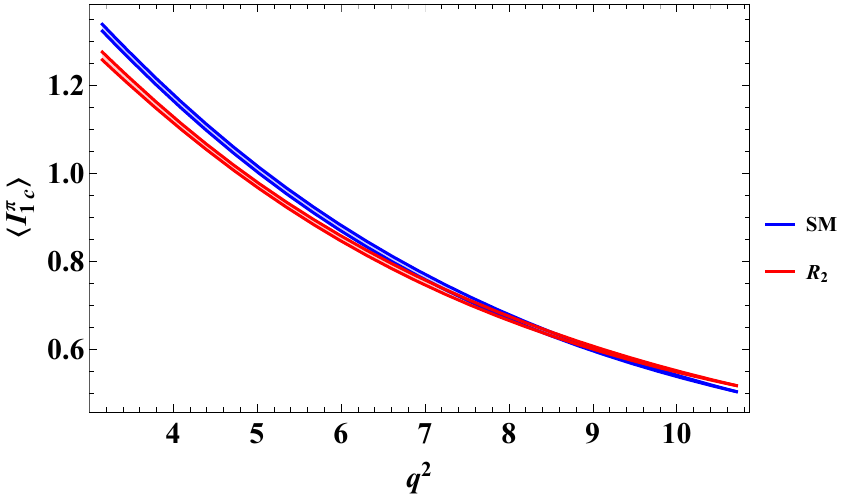}
\includegraphics[scale=0.55]{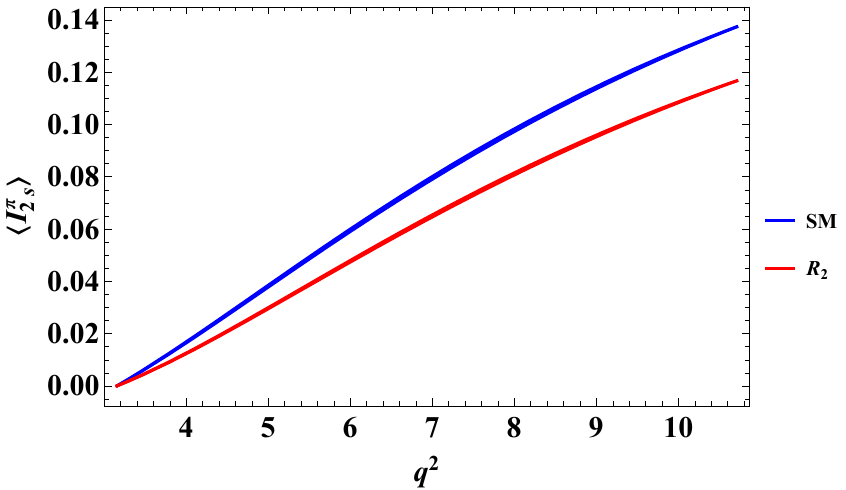}
\includegraphics[scale=0.55]{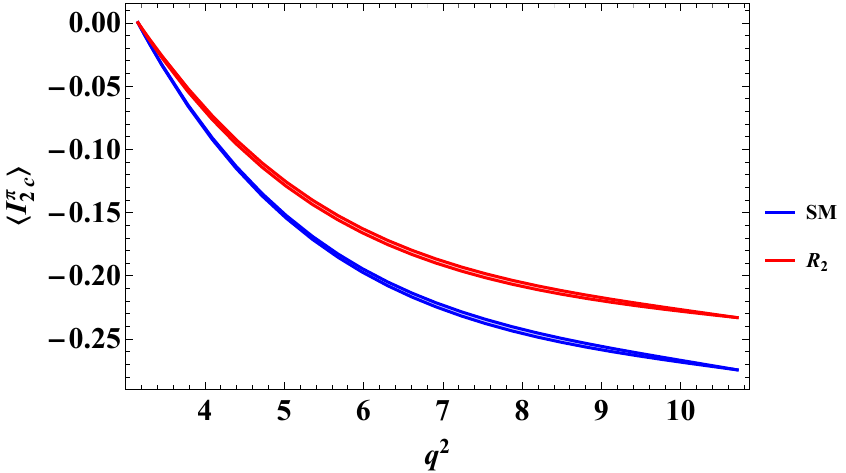}
\includegraphics[scale=0.55]{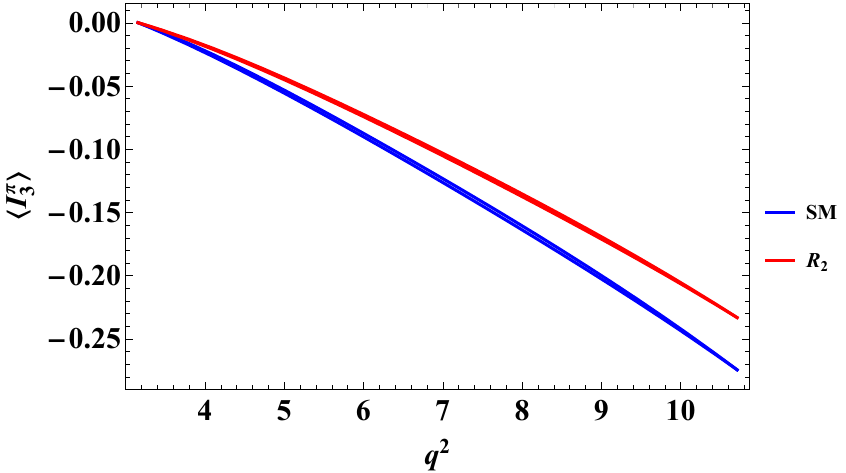}
\includegraphics[scale=0.55]{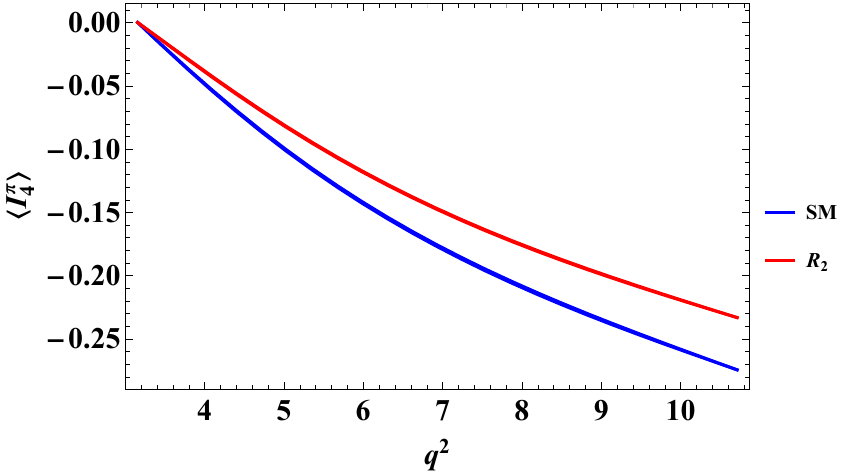}
\includegraphics[scale=0.55]{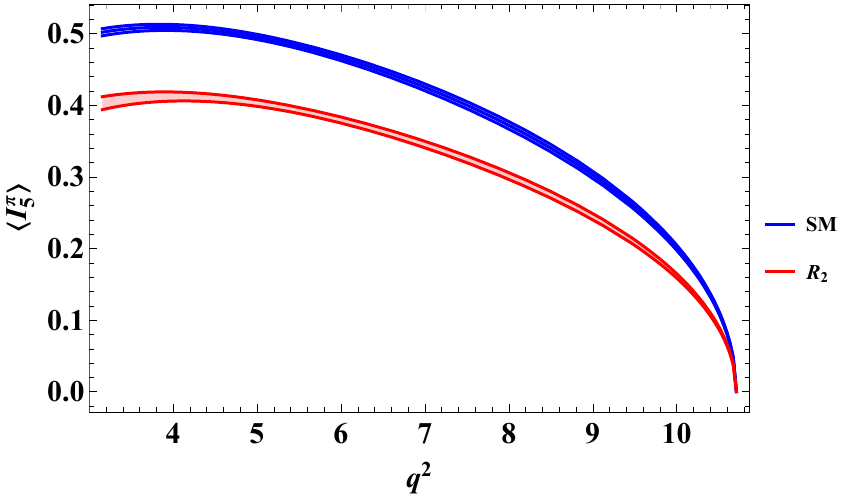}
\includegraphics[scale=0.55]{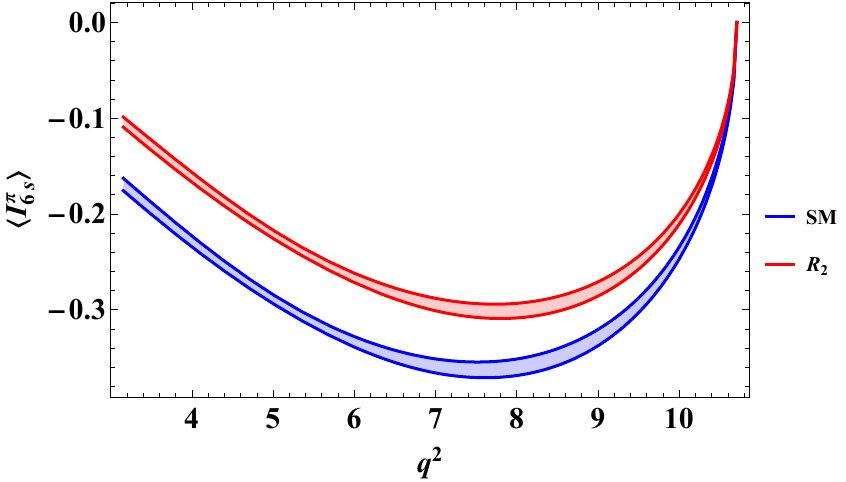}
\includegraphics[scale=0.55]{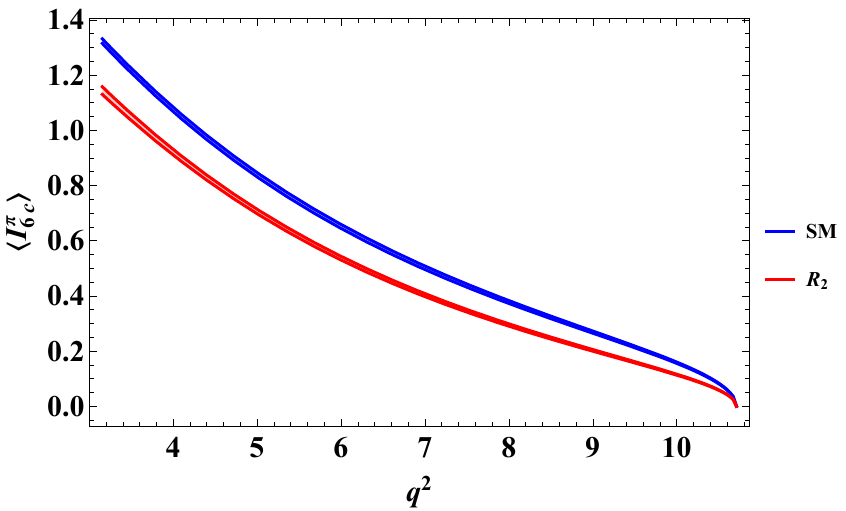}
\caption{Angular observables $\langle I^{\pi}_{1s}\rangle$, $\langle I^{\pi}_{1c}\rangle$, $\langle I^{\pi}_{2s}\rangle$, $\langle I^{\pi}_{2c}\rangle$, $\langle I^{\pi}_{3}\rangle$, $\langle I^{\pi}_{4}\rangle$, $\langle I^{\pi}_{5}\rangle$, $\langle I^{\pi}_{6s}\rangle$, and  $\langle I^{\pi}_{6c}\rangle$ for the decay $B\to D^{\ast}(\to D\pi)\tau^{-}\bar{\nu}$, in SM and in the $R_{2}$ Leptoquark model.}
\label{fig:angcoeffDpiSMR2LQ}
\end{center}
\end{figure}

\begin{figure}[!htbp]
\begin{center}
\includegraphics[scale=0.55]{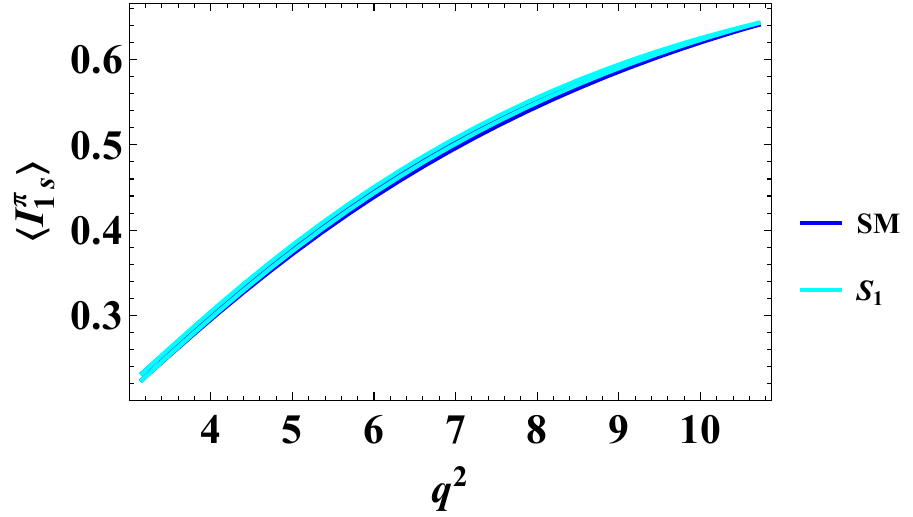}
\includegraphics[scale=0.55]{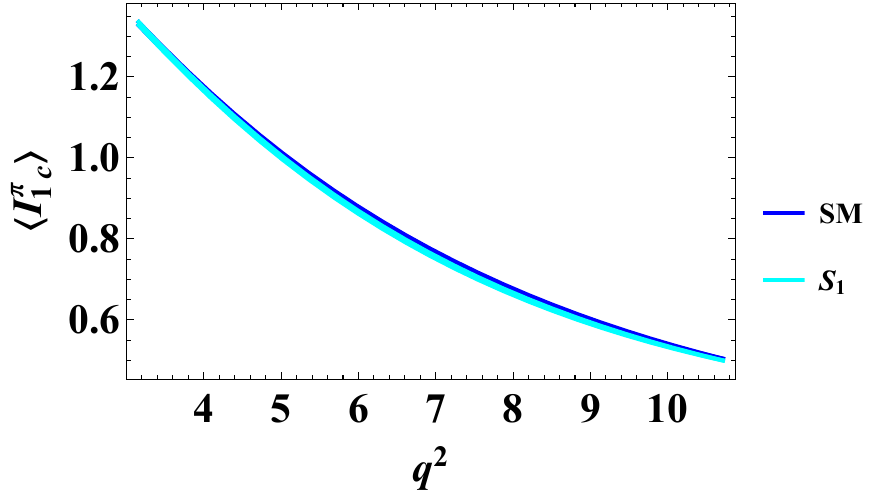}
\includegraphics[scale=0.55]{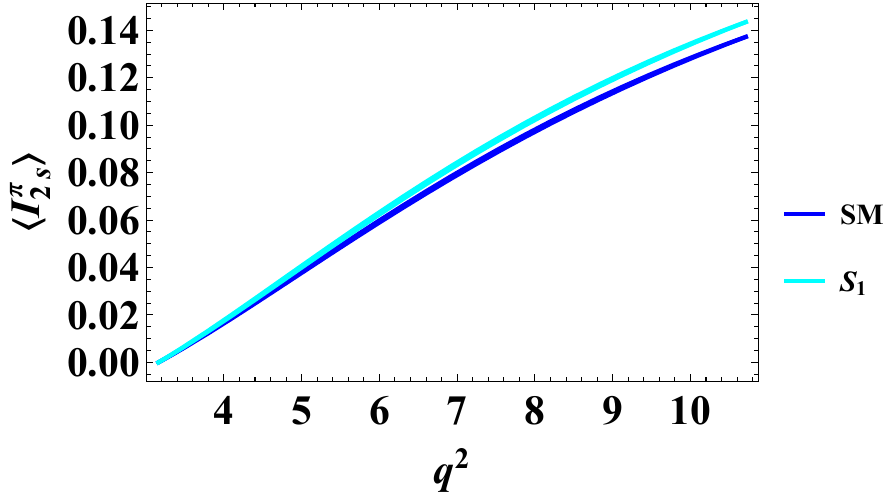}
\includegraphics[scale=0.55]{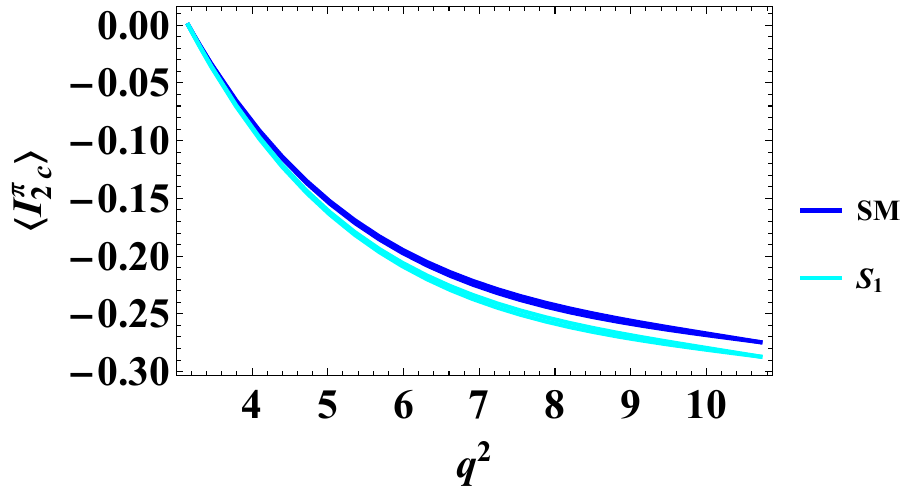}
\includegraphics[scale=0.55]{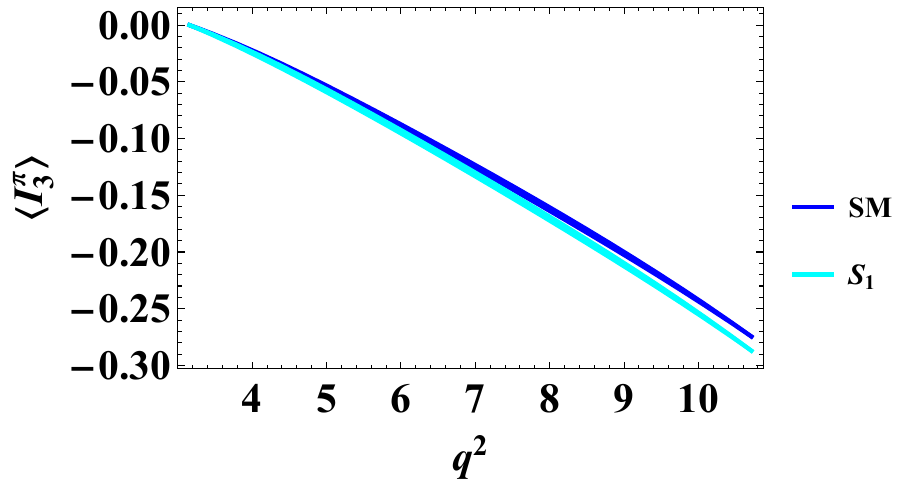}
\includegraphics[scale=0.55]{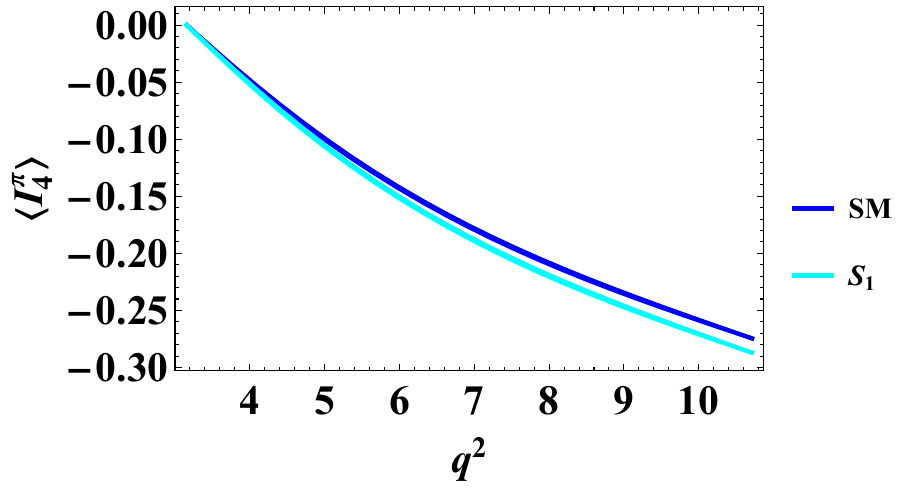}
\includegraphics[scale=0.55]{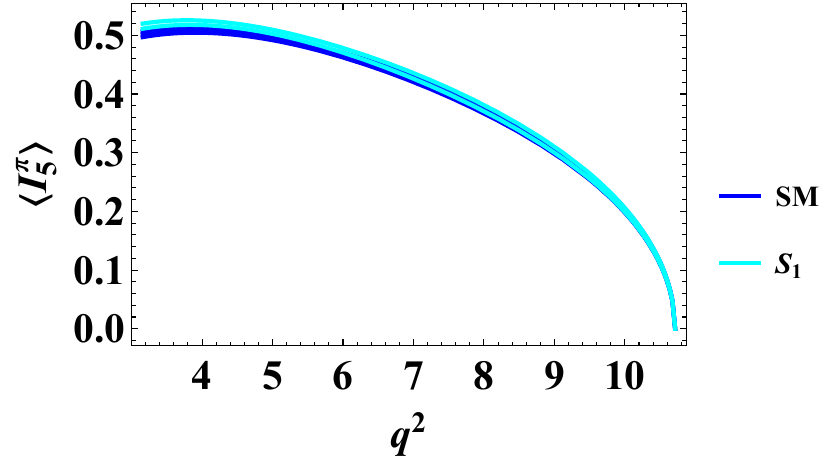}
\includegraphics[scale=0.55]{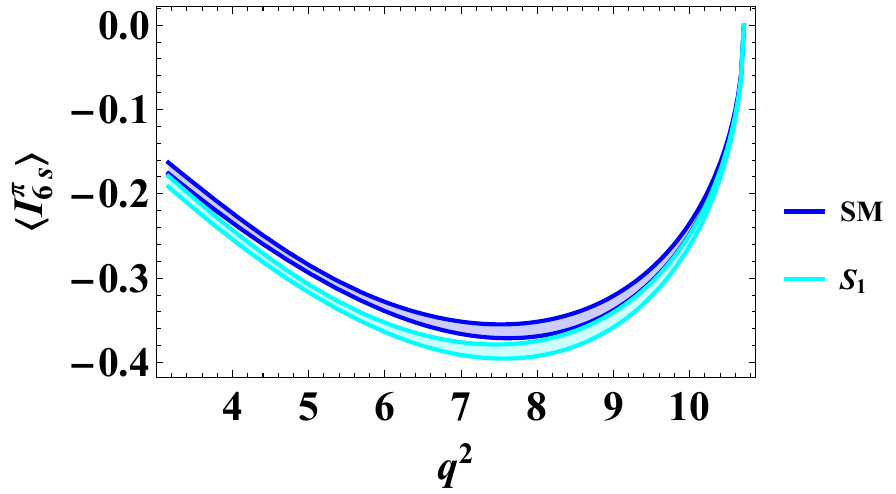}
\includegraphics[scale=0.55]{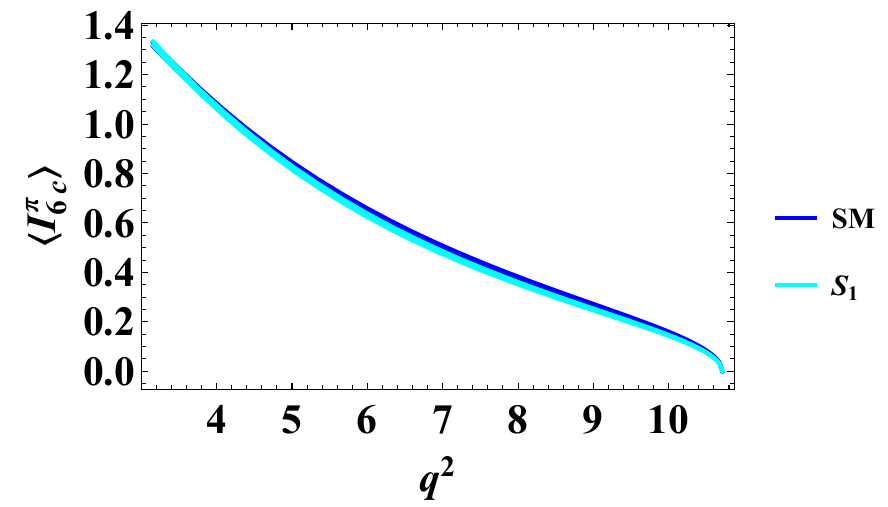}
\caption{Angular observables $\langle I^{\pi}_{1s}\rangle$, $\langle I^{\pi}_{1c}\rangle$, $\langle I^{\pi}_{2s}\rangle$, $\langle I^{\pi}_{2c}\rangle$, $\langle I^{\pi}_{3}\rangle$, $\langle I^{\pi}_{4}\rangle$, $\langle I^{\pi}_{5}\rangle$, $\langle I^{\pi}_{6s}\rangle$, and  $\langle I^{\pi}_{6c}\rangle$ for the decay $B\to D^{\ast}(\to D\pi)\tau^{-}\bar{\nu}$, in SM and in the $S_{1}$ Leptoquark model.}
\label{fig:angcoeffDpiSMS1LQ}
\end{center}
\end{figure}

\begin{figure}[!htbp]
\begin{center}
\includegraphics[scale=0.55]{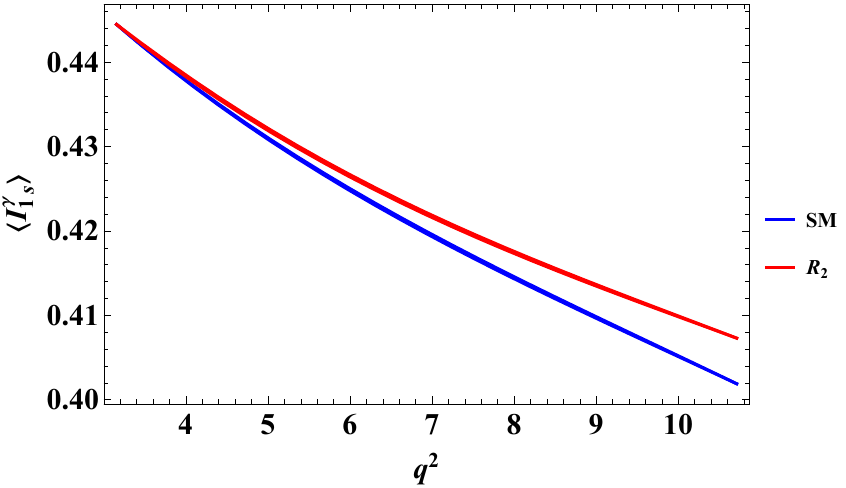}
\includegraphics[scale=0.55]{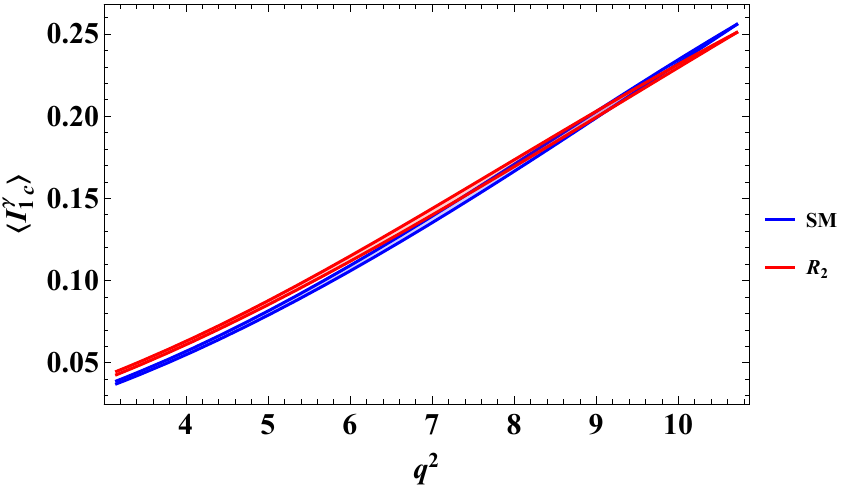}
\includegraphics[scale=0.55]{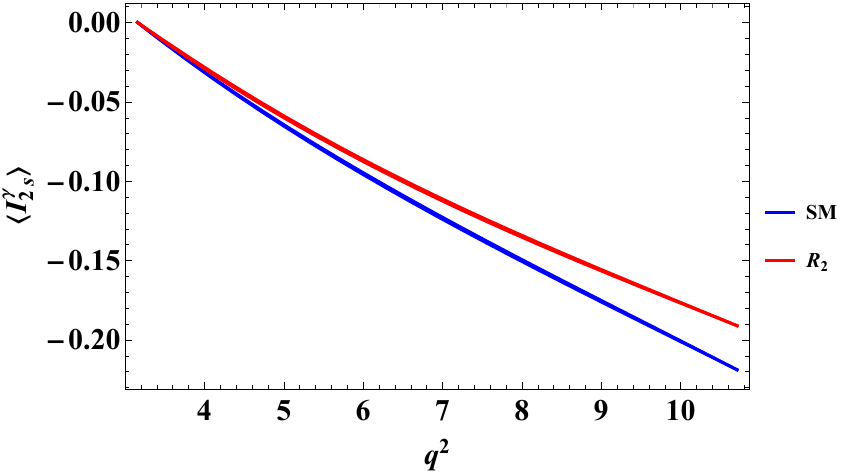}
\includegraphics[scale=0.55]{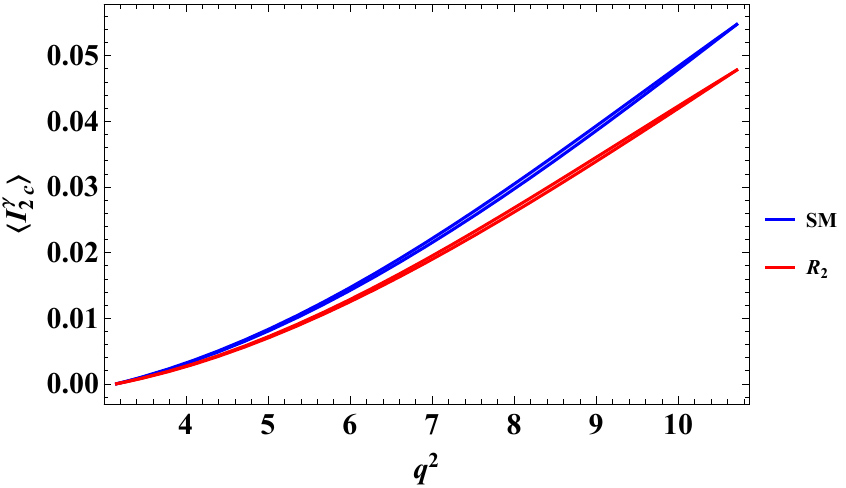}
\includegraphics[scale=0.55]{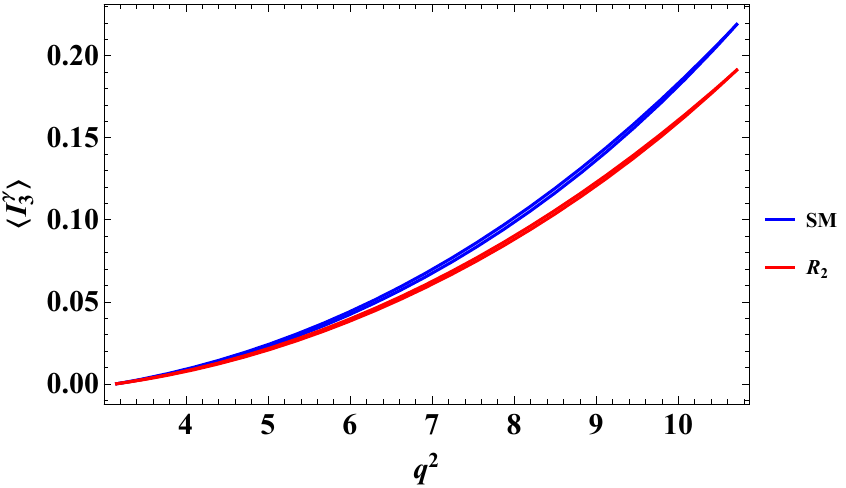}
\includegraphics[scale=0.55]{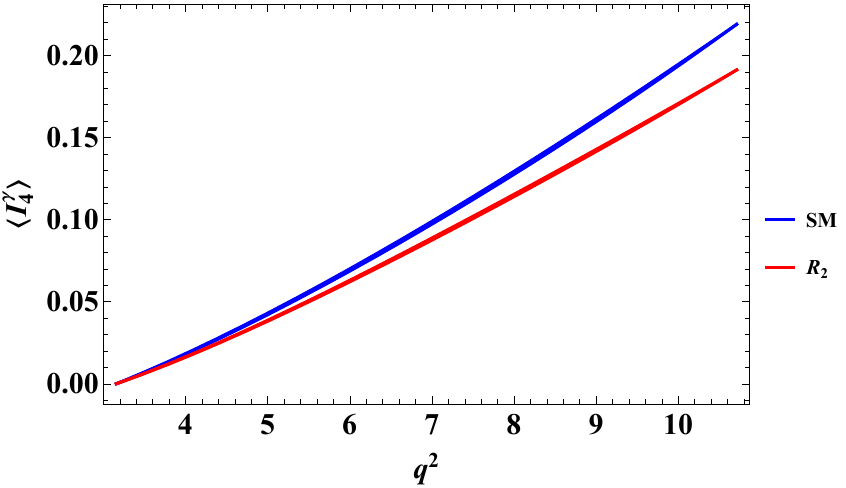}
\includegraphics[scale=0.55]{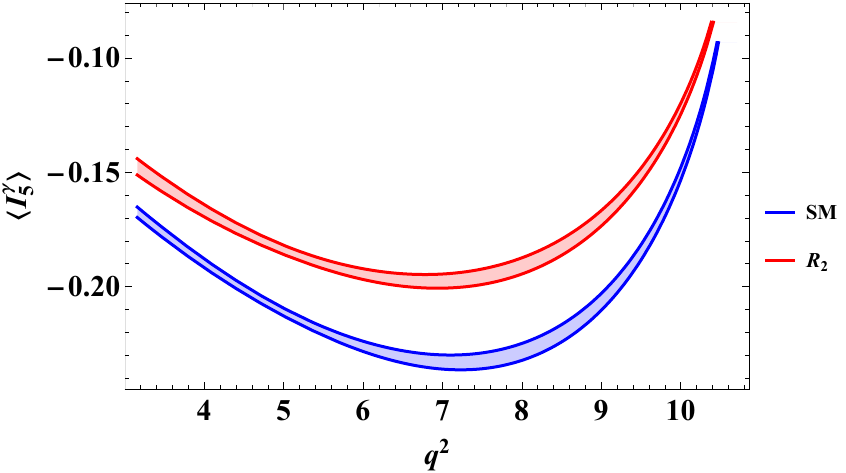}
\includegraphics[scale=0.55]{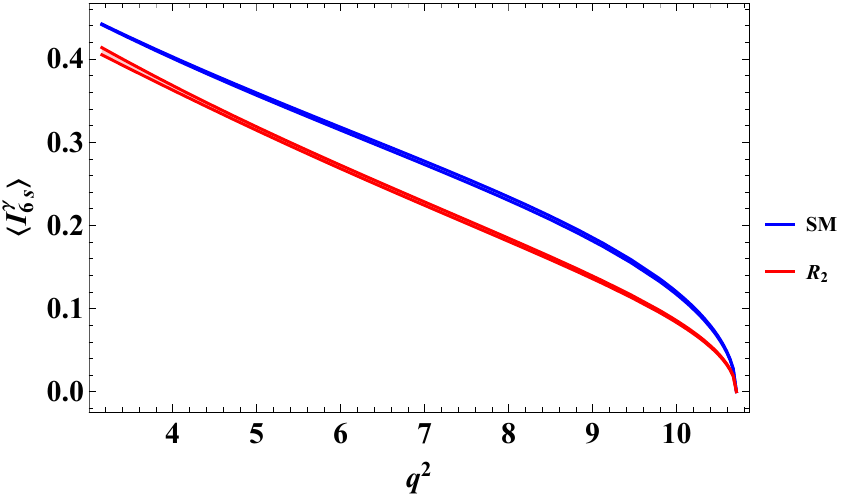}
\includegraphics[scale=0.55]{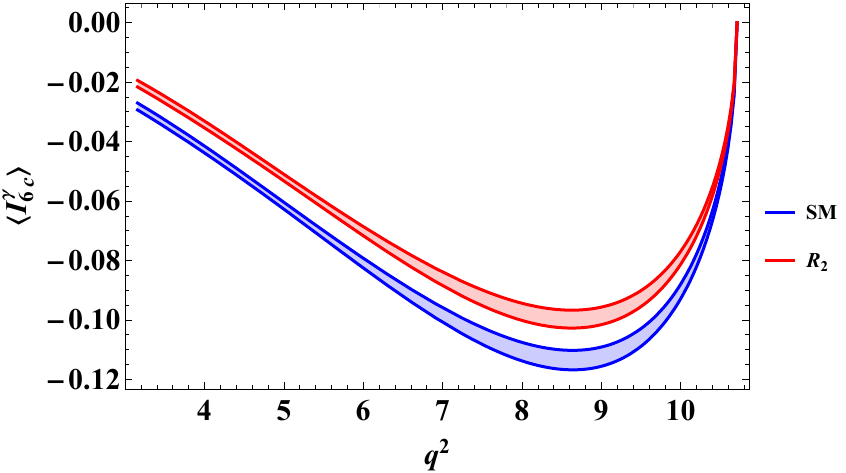}
\caption{Angular observables $\langle I^{\gamma}_{1s}\rangle$, $\langle I^{\gamma}_{1c}\rangle$, $\langle I^{\gamma}_{2s}\rangle$, $\langle I^{\gamma}_{2c}\rangle$, $\langle I^{\gamma}_{3}\rangle$, $\langle I^{\gamma}_{4}\rangle$, $\langle I^{\gamma}_{5}\rangle$, $\langle I^{\gamma}_{6s}\rangle$, and  $\langle I^{\gamma}_{6c}\rangle$ for the decay $B\to D^{\ast}(\to D\gamma)\tau^{-}\bar{\nu}$, in SM and in the $R_{2}$ Leptoquark model.}
\label{fig:angcoeffDgammaSMR2LQ}
\end{center}
\end{figure}

\begin{figure}[!htbp]
\begin{center}
\includegraphics[scale=0.55]{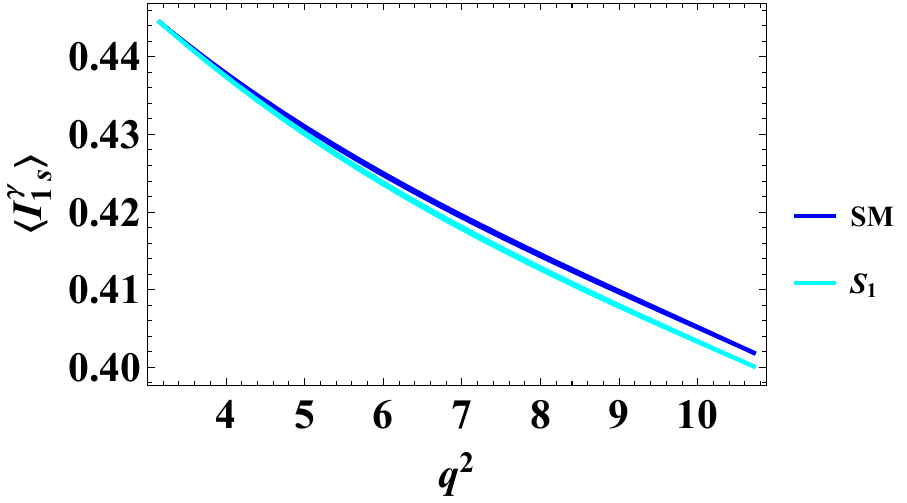}
\includegraphics[scale=0.55]{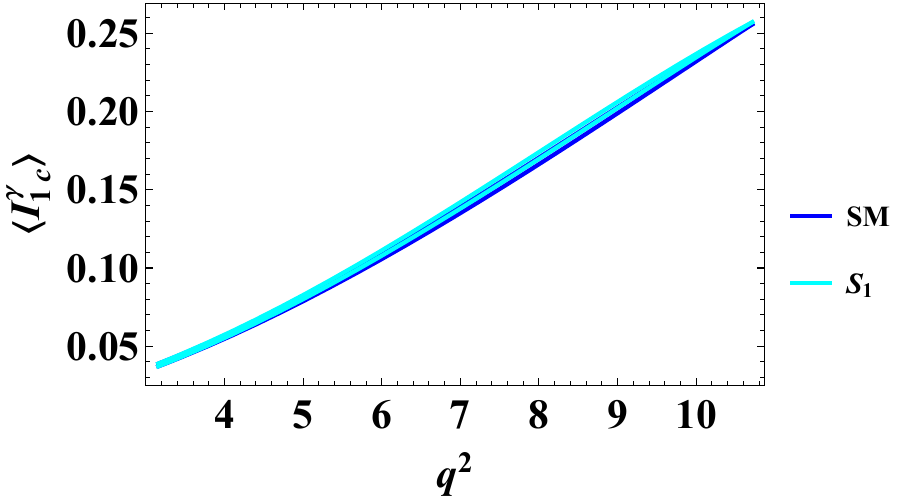}
\includegraphics[scale=0.55]{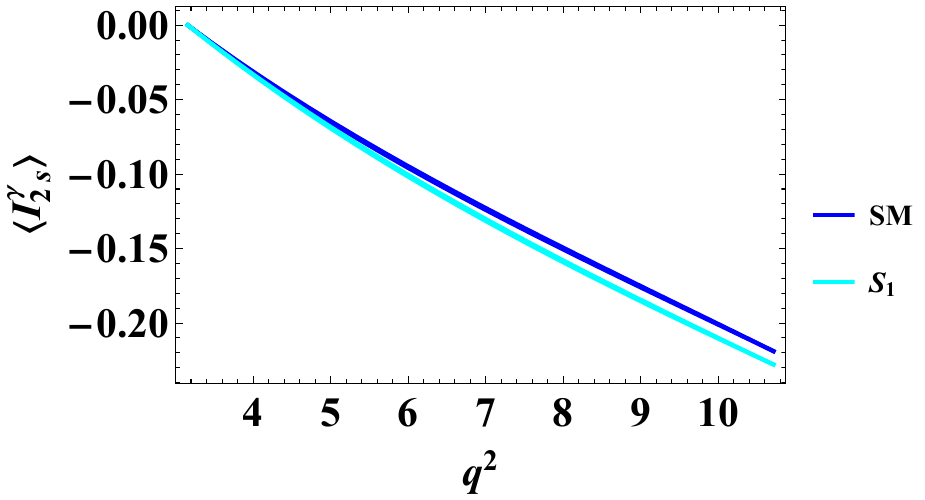}
\includegraphics[scale=0.55]{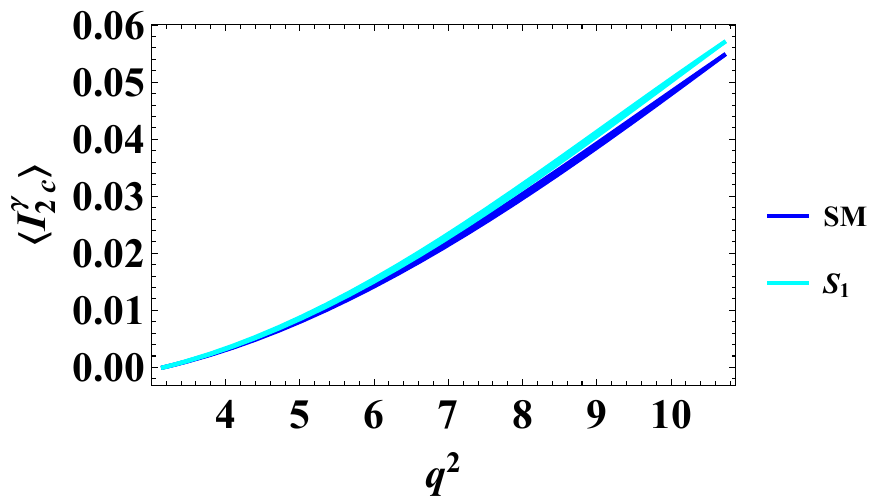}
\includegraphics[scale=0.55]{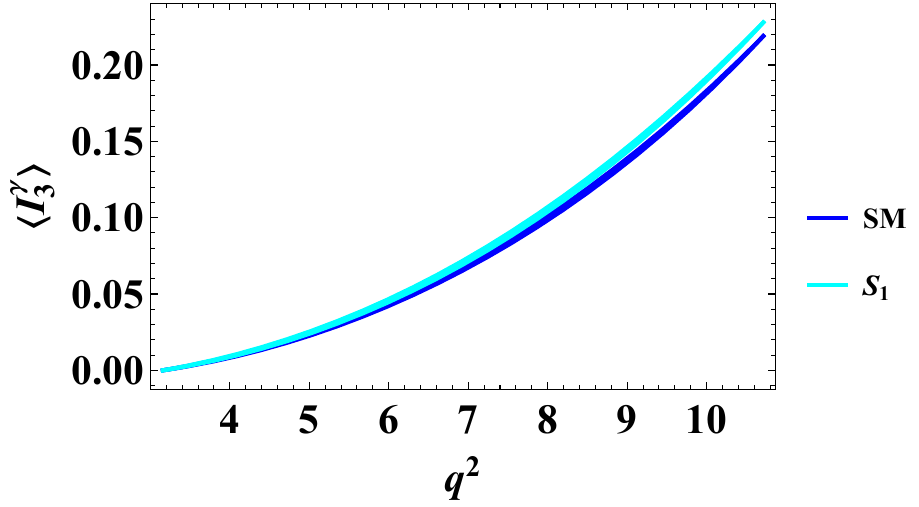}
\includegraphics[scale=0.55]{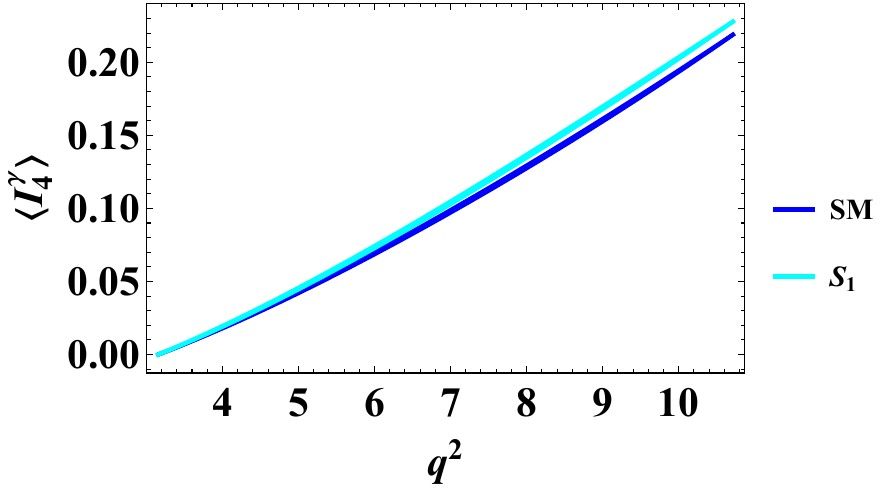}
\includegraphics[scale=0.55]{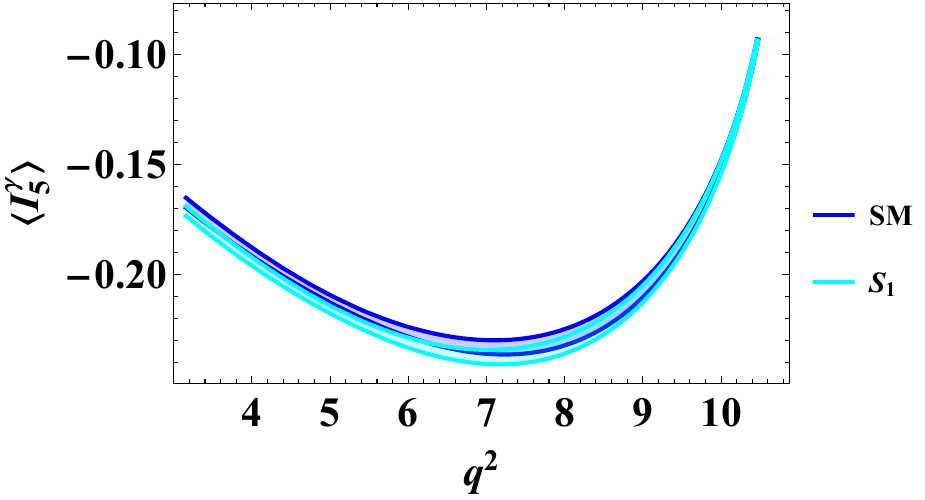}
\includegraphics[scale=0.55]{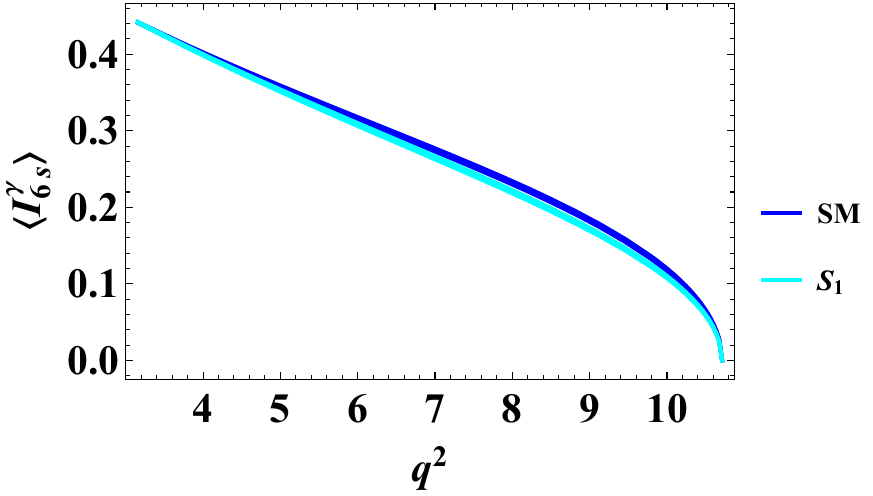}
\includegraphics[scale=0.55]{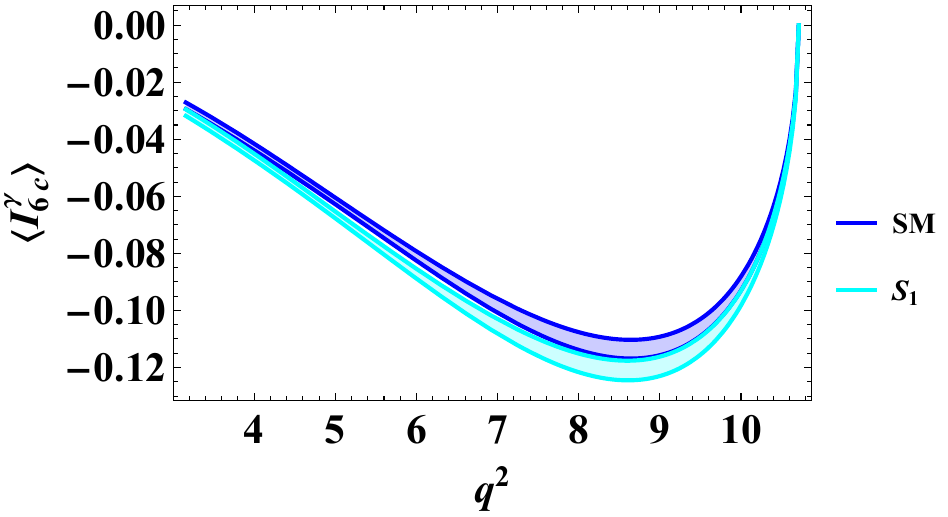}
\caption{Angular observables $\langle I^{\gamma}_{1s}\rangle$, $\langle I^{\gamma}_{1c}\rangle$, $\langle I^{\gamma}_{2s}\rangle$, $\langle I^{\gamma}_{2c}\rangle$, $\langle I^{\gamma}_{3}\rangle$, $\langle I^{\gamma}_{4}\rangle$, $\langle I^{\gamma}_{5}\rangle$, $\langle I^{\gamma}_{6s}\rangle$, and  $\langle I^{\gamma}_{6c}\rangle$  for the decay $B\to D^{\ast}(\to D\gamma)\tau^{-}\bar{\nu}$, in SM and in the $S_{1}$ Leptoquark model.}
\label{fig:angcoeffDgammaSMS1LQ}
\end{center}
\end{figure}

For the $R_{2}$ LQ model, the best-fit values accommodate $\mathcal{B}(B_{c}\to\tau\nu_{\tau})<30\%$ and by incorporating the constraints we obtain $\chi^{2}_{\text{min}}\approx 22.45$. The pull value presented in Table~\ref{tab:wcoef3} suggests that the $R_{2}$ LQ model is not excluded by the $2\sigma$ exclusion limit. The LQ models $S_{1}$ and $U_{1}$ also accommodate $\mathcal{B}(B_{c}\to\tau\nu_{\tau})<30\%$, and the $\chi^{2}_{\text{min}}$ values obtained from the constraint are $\chi^{2}_{\text{min}}\approx 23.01$ and $\chi^{2}_{\text{min}}\approx 23.22$, respectively. The best-fit values for the models $S_{1}$ and $U_{1}$ are not excluded at $2\sigma$. Using the best-fit values of the Wilson coefficients in Table~\ref{tab:wcoef3}, we obtain the predictions of the physical observables such as $R_{D^{(\ast)}}$, $P_{\tau}(D^{(\ast)})$, $F_{L}^{D^{\ast}}$, $\mathcal{A}^{\text{FB}}_{D^{(\ast)}}$, and the normalized angular coefficients $\langle I^{n}_{\lambda}\rangle$ in the three LQ models, which are presented in Tables~\ref{tab:obserLQ1}-\ref{table:10b}. Furthermore, the plots of the normalized angular coefficients $\langle I^{n}_{\lambda}\rangle$ as a function of $q^{2}$ in the SM and in the LQ models are also presented in Figures~\ref{fig:angcoeffDpiSMR2LQ}-\ref{fig:angcoeffDgammaSMS1LQ}.

\begin{table*}[!htbp]\small%
\centering
\caption{Predictions for $R_D$, $R_{D^*}$, $P_\tau(D)$ and $P_\tau(D^*)$ in the leptoquark models. The first and second uncertainties result from the form factors and the best-fit leptoquark couplings.}\label{tab:obserLQ1}
\begin{tabular}{ccccc}
\hline
\hline
LQ type & $R_D$& $R_{D^*}$   &$P_\tau(D)$ & $P_\tau(D^*)$ \\
\hline
$S_1$   &$0.314(5)(21)$ & $0.299(10)(13)$ &$0.131(7)(122)$ &$-0.464(7)(19)$ \\
$R_2$   &$0.235(4)(38)$ & $0.284(11)(19)$ & $0.222(3)(10)$
& $-0.410(7)(21)$      \\
$U_1$   &$0.318(5)(20)$ & $0.296(10)(13)$ &$0.255(4)(73)$ &$-0.509(4)(12)$ \\
\hline
\hline
\end{tabular}
\end{table*}
\begin{table*}[!htbp]\small%
\centering
\caption{Predictions for $F_L^{D^*}$, $\mathcal{A}^{\text{FB}}_{D}$ and $\mathcal{A}^{\text{FB}}_{D^{\ast}}$ in the leptoquark models. The first and second uncertainties result from the form factors and the best-fit leptoquark couplings.}\label{tab:obserLQ2}
\begin{tabular}{cccc}
\hline
\hline
LQ type &$F_L^{D^*}$ &$\mathcal{A}^{\text{FB}}_{D}$& $\mathcal{A}^{\text{FB}}_{D^{\ast}}$\\
\hline
$S_1$   &$0.485(4)(10)$ & $0.377(1)(9)$& $-0.063(5)(4)$\\
$R_2$   &$0.470(6)(2)$ & $0.325(3)(9)$ & $0.011(8)(19)$ \\
$U_1$    &$0.461(4)(4)$ & $0.368(1)(4)$& $-0.067(6)(7)$\\
\hline
\hline
\end{tabular}
\end{table*}

From Tables~\ref{tab:obserLQ1} and \ref{tab:obserLQ2}, it is found that the most of the observables shows a departure from SM predictions. In $S_{1}$ LQ type model, the values of $R_D$, $R_{D^{\ast}}$, $F^{D^{\ast}}_{L}$, and $\mathcal{A}^{\text{FB}}_{D}$ indicate slight departure from SM predictions. However, the values of $P_{\tau}(D),P_{\tau}(D^{\ast})$ and $\mathcal{A}^{\text{FB}}_{D^{\ast}}$ in the $S_{1}$ LQ type model show a significant deviation from the SM predictions. For $R_{2}$ LQ type model, the values of $R_D$, $P_{\tau}(D)$, $P_{\tau}(D^{\ast})$, and $\mathcal{A}^{\text{FB}}_{D^{\ast}}$ show a significant departure from SM predictions, but the values of observables such as $R_{D^{\ast}}$, $F^{D^{\ast}}_{L}$, and $\mathcal{A}^{\text{FB}}_{D^{\ast}}$ are close to SM predictions. In $U_{1}$ LQ type model, the values of observables $R_D$, $R_{D^{\ast}}$, $P_{\tau}(D^{\ast})$, $F^{D^{\ast}}_{L}$, and $\mathcal{A}^{\text{FB}}_{D}$ are similar to the SM values, whereas the observables $P_{\tau}(D)$ and $\mathcal{A}^{\text{FB}}_{D^{\ast}}$ deviate from SM predictions.

\begin{table}[ht!]
\begin{center}
\caption{Predictions of averaged values of angular observables for the $B\to D^{\ast}(\to D\pi)\tau\bar{\nu}_{\tau}$ decay in the three Leptoquark models. The first and second errors presented arise from the uncertainties of the form factors and Wilson coefficients, respectively.}\label{table:10}
\begin{tabular}{|c||c|c|c|c|c|}
 \hline
 \textbf{~}&\textbf{SM}& \textbf{${R_{2}}$ LQ model} &\textbf{${S_{1}}$ LQ model}&\textbf{${U}_{1}$ LQ model}\\
 \hline
 $\langle I^{\pi}_{1s}\rangle$   & $0.364(6)(0)$& $0.368(6)(4)$&   $0.366(6)(3)$&   $0.361(6)(3)$\\
 \hline
 $\langle I^{\pi}_{1c}\rangle$ & $0.594(12)(0)$   & $0.585(10)(3)$&   $0.588(11)(10)$&   $0.598(11)(0)$\\
 \hline
$\langle I^{\pi}_{2s}\rangle$ & $0.0583(0)(0)$   & $0.0482(0)(1)$&   $0.0617(0)(4)$&   $0.0583(0)(0)$\\
 \hline
$\langle I^{\pi}_{2c}\rangle$ & $-0.151(2)(0)$   & $-0.127(3)(4)$&   $-0.159(2)(11)$&   $-0.1503(2)(1)$\\
 \hline
  $\langle I^{\pi}_{3}\rangle$   & $-0.099(2)(0)$   & $-0.083(1)(2)$&   $-0.105(2)(7)$&   $-0.099(2)(0)$\\
 \hline
 $\langle I^{\pi}_{4}\rangle$ & $-0.126(0)(0)$   & $-0.106(1)(3)$&   $-0.133(1)(9)$&   $-0.126(1)(1)$\\
 \hline
$\langle I^{\pi}_{5}\rangle$ & $0.284(6)(0)$   & $0.229(7)(8)$&   $0.286(6)(5)$&   $0.284(7)(4)$\\
 \hline
$\langle I^{\pi}_{6s}\rangle$ & $-0.217(10)(0)$   & $-0.175(8)(8)$&   $-0.232(11)(20)$&   $-0.217(9)(2)$\\
 \hline
 $\langle I^{\pi}_{6c}\rangle$   & $0.404(11)(0)$   & $0.334(8)(10)$&   $0.389(9)(18)$&   $0.410(9)(14)$\\
 \hline
\end{tabular}
\end{center}
\end{table}
\begin{table}[!htb]
\begin{center}
\caption{Predictions of averaged values of angular observables for the $B\to D^{\ast}(\to D\gamma)\tau\bar{\nu}_{\tau}$ decay, in the three LQ models. The first and second errors presented arise from the uncertainties of the form factors and Wilson coefficients, respectively.}\label{table:10b}
\begin{tabular}{|c||c|c|c|c|}
 \hline
 \textbf{~}&\textbf{SM}& \textbf{${R_{2}}$ LQ model} &\textbf{${S_{1}}$ LQ model}&\textbf{${U}_{1}$ LQ model}\\
 \hline
 $\langle I^{\gamma}_{1s}\rangle$ & $0.542(2)(0)$ & $0.544(2)(0)$   & $0.540(3)(5)$&   $0.543(3)(2)$\\
 \hline
 $\langle I^{\gamma}_{1c}\rangle$& $0.165(5)(0)$& $0.170(5)(2)$   & $0.167(6)(5)$&   $0.163(5)(3)$\\
 \hline
$\langle I^{\gamma}_{2s}\rangle$ &$-0.137(1)(0)$& $-0.124(1)(2)$   & $-0.145(2)(10)$&   $-0.136(1)(2)$\\
 \hline
$\langle I^{\gamma}_{2c}\rangle$ &$0.026(0)(0)$& $0.023(0)(0)$   & $0.028(0)(0)$&   $0.026(0)(0)$\\
 \hline
  $\langle I^{\gamma}_{3}\rangle$  &$0.091(3)(0)$ & $0.081(2)(1)$   & $0.096(4)(7)$&   $0.090(3)(2)$\\
 \hline
 $\langle I^{\gamma}_{4}\rangle$ &$0.115(2)(0)$& $0.103(2)(2)$   & $0.121(3)(8)$&   $0.114(2)(2)$\\
 \hline
$\langle I^{\gamma}_{5}\rangle$ &$-0.257(7)(0)$ &$-0.217(7)(7)$   & $-0.260(7)(5)$&   $-0.257(7)(2)$\\
 \hline
$\langle I^{\gamma}_{6s}\rangle$ & $0.367(5)(0)$&$0.311(5)(10)$   & $0.00(0)(0)$&   $0.373(5)(9)$\\
 \hline
 $\langle I^{\gamma}_{6c}\rangle$   &$-0.098(5)(0)$& $-0.084(4)(3)$  & $-0.104(5)(9)$&   $-0.097(5)(2)$\\
 \hline
\end{tabular}
\end{center}
\end{table}


Now, we shall discuss the phenomenological analysis of the normalized angular coefficients $\langle I^{n}_{\lambda}\rangle$ for the $B\to D^{\ast}(\to D\pi)\tau\bar{\nu}_{\tau}$ decay in SM and LQ models. The angular observables $\langle I_{1s}^{\pi}\rangle$, $\langle I_{1c}^{\pi}\rangle$, $\langle I_{2s}^{\pi}\rangle$, $\langle I_{2c}^{\pi}\rangle$, $\langle I^{\pi}_{3}\rangle$, $\langle I^{\pi}_{4}\rangle$, $\langle I^{\pi}_{5}\rangle$, $\langle I_{6s}^{\pi}\rangle$ and $\langle I_{6c}^{\pi}\rangle$ in SM and in $R_{2}$ LQ model are presented in Figure~\ref{fig:angcoeffDpiSMR2LQ}. From the figure, it can be seen that there is a minimal discrimination between the values calculated in the SM and in $R_{2}$ LQ model for the observables $\langle I^{\pi}_{1s}\rangle$ and $\langle I^{\pi}_{1c}\rangle$ for all $q^{2}$ range. However, for the observables $\langle I^{\pi}_{2s}\rangle,\langle I^{\pi}_{2c}\rangle,\langle I^{\pi}_{3}\rangle,\langle I^{\pi}_{4}\rangle,\langle I^{\pi}_{5}\rangle,\langle I^{\pi}_{6s}\rangle$, and $\langle I^{\pi}_{6c}\rangle$, there is a significant deviation between the values calculated in the SM and the $R_{2}$ LQ model over the entire $q^{2}$ range.


Figure~\ref{fig:angcoeffDpiSMS1LQ}, presents results in SM and in the $S_{1}$ LQ model. For observables $\langle I_{1s}^{\pi}\rangle$ and $\langle I_{1c}^{\pi}\rangle$, it is observed that there is a very small distinction between the values calculated within the SM and the $S_{1}$ LQ model for the whole range $q^{2}$. For observables $\langle I_{2s}^{\pi}\rangle$, $\langle I_{2c}^{\pi}\rangle$, $\langle I^{\pi}_{3}\rangle$ and $\langle I^{\pi}_{4}\rangle$, overlapping is seen between the values calculated in the SM and in the $S_{1}$ LQ model for the lower range of $q^{2}$. On the other hand, for higher $q^{2}$ values both models show small discrimination among each other. For observables $\langle I^{\pi}_{5}\rangle$ and $\langle I_{6s}^{\pi}\rangle$, the values calculated within the SM and in the $S_{1}$ LQ model overlap for the upper $q^{2}$ range, while the values are distinguishable for the lower range of $q^{2}$. The observable $\langle I_{6c}^{\pi}\rangle$ shows less or no deviation between the values calculated in the SM and $S_{1}$ LQ model. For $U_{1}$ LQ model, all the angular observables show no or small deviations from the SM predictions, therefore, we have not presented their plots, however the numerical results of the averaged angular coefficients are presented in Tables~\ref{table:10}-\ref{table:10b}.

At last, we discuss the phenomenological analysis of the angular observables of the decay $B\to D^{\ast}(\to D\gamma)\tau\bar{\nu}$ in SM and LQ models. The results of $\langle I_{1s}^{\gamma}\rangle$, $\langle I_{1c}^{\gamma}\rangle$, $\langle I_{2s}^{\gamma}\rangle$, $\langle I_{2c}^{\gamma}\rangle$, $\langle I_{3}^{\gamma}\rangle$, $\langle I_{4}^{\gamma}\rangle$, $\langle I_{5}^{\gamma}\rangle$, $\langle I_{6s}^{\gamma}\rangle$ and $\langle I_{6c}^{\gamma}\rangle$ in SM and in $R_{2}$ LQ model, are given in Figure~\ref{fig:angcoeffDgammaSMR2LQ}. It is found that the predictions of angular observables $\langle I_{1s}^{\gamma}\rangle$, $\langle I_{2s}^{\gamma}\rangle$, $\langle I_{2c}^{\gamma}\rangle$, $\langle I^{\gamma}_{3}\rangle$, and $\langle I^{\gamma}_{4}\rangle$, in $R_{2}$ LQ model overlap with the SM values, in the low $q^2$ range, however, they start deviating from the SM predictions after $q^2\approx 5$ $\text{GeV}^{2}$ and become more and more distinct as the $q^2$ range increases towards higher values. On the other hand, observable $\langle I_{1c}^{\gamma}\rangle$, in $R_{2}$ LQ model, shows no or slight deviation from the SM in the entire range of $q^2$. Furthermore, $R_{2}$ LQ model predictions, for the observables $\langle I^{\gamma}_{5}\rangle$, $\langle I_{6s}\rangle$, and $\langle I_{6c}^{\gamma}\rangle$, in the whole $q^2$ range, show identical but distinct pattern as compared to SM values, where the predictions in both models get closer to each other for the higher $q^{2}$ range and tend to merge there. Similarly, the angular observables in SM and in $S_{1}$ LQ model are presented in Figure~\ref{fig:angcoeffDgammaSMS1LQ}. Observable $\langle I_{1s}^{\gamma}\rangle$, and $\langle I_{2s}^{\gamma}\rangle$ start to show discrimination between the values calculated within SM and $S_{1}$ LQ model from $q^{2}\sim 6$ $\text{GeV}^{2}$ and onward. For observables $\langle I_{1c}^{\gamma}\rangle$, $\langle I_{2c}^{\gamma}\rangle$, $\langle I^{\gamma}_{3}\rangle$, $\langle I^{\gamma}_{4}\rangle$, and $\langle I^{\gamma}_{6s}\rangle$, it can be seen that the values calculated within the SM and $S_{1}$ LQ model are very close to each other for the lower range of $q^{2}$. In contrast, they do indicate slight deviation from each other for the higher range of $q^{2}$.  For observables $\langle I^{\gamma}_{5}\rangle$ and $\langle I_{6c}^{\gamma}\rangle$, it is observed that for lower and higher range of $q^{2}$, the values calculated within the SM and $S_{1}$ LQ model overlap but for mid range they tend to deviate from each other. Finally, for the LQ models, we also summarize the numerical values of the most discriminated observables, w.r.t the SM, in Tables.\ref{tab:obserLQ1disc} and \ref{tableLQdis}.


\begin{table*}[!htbp]
\centering
\footnotesize  
\renewcommand{\arraystretch}{1.1}  
\setlength{\tabcolsep}{4pt}  
\caption{Predictions of observables $R_D$, $R_{D^*}$, $P_\tau(D)$, and $P_\tau(D^*)$, showing their most distinguishing values w.r.t the SM in the LQ models. The first and second uncertainties arise from the form factors and the fitted Wilson coefficients, respectively.} \label{tab:obserLQ1disc}
\begin{tabular}{lccccccc}
\hline\hline
\textbf{Scenario} & $R_D$ & $R_{D^*}$ & $P_\tau(D)$ & $P_\tau(D^*)$ & $F_L^{D^*}$ & $\mathcal{A}^{\text{FB}}_{D}$ & $\mathcal{A}^{\text{FB}}_{D^*}$ \\
\hline
SM        & $0.300(3)(0)$  & $0.251(1)(0)$  & $0.325(3)(0)$  & $-0.507(3)(0)$  & $0.453(3)(0)$  & $0.360(0)(0)$  & $-0.052(4)(0)$ \\
$S_1$     &$0.134(5)(21)$ & $0.299(10)(13)$ & $0.131(7)(122)$ & $-0.464(7)(19)$ & $0.485(4)(10)$ & $0.377(1)(9)$ & $--$\\
$R_2$     &$0.235(4)(38)$ & $0.284(11)(19)$ & $0.222(3)(10)$ & $-0.401(7)(21)$ & $0.470(6)(2)$ & $0.325(3)(9)$ & $0.011(8)(19)$\\
$U_1$      &$--$        & $0.296(10)(13)$ & $0.255(4)(73)$ & $--$ & $0.461(4)(4)$ & $--$ & $--$\\
\hline\hline
\end{tabular}
\end{table*}


\begin{table}[ht!]
\begin{center}
\caption{Predictions of the most discriminated average values of angular observables for the $B\to D^{\ast}(\to D\pi,D\gamma)\tau\bar{\nu}_{\tau}$ decay in the three Leptoquark models. The first and second errors presented arise from the uncertainties of the form factors and Wilson coefficients, respectively.}\label{tableLQdis}
\begin{tabular}{|c||c|c|c|c|c|}
 \hline
 \textbf{~}&\textbf{SM}& \textbf{${R_{2}}$ LQ model} &\textbf{${S_{1}}$ LQ model}&\textbf{${U}_{1}$ LQ model}\\
 \hline
$\langle I^{\pi}_{2s}\rangle$ & $0.0583(0)(0)$   & $0.0482(0)(1)$&   $0.0617(0)(4)$&   $--$\\
 \hline
$\langle I^{\pi}_{2c}\rangle$ & $-0.151(2)(0)$   & $-0.127(3)(4)$&   $--$&   $--$\\
 \hline
  $\langle I^{\pi}_{3}\rangle$   & $-0.099(2)(0)$   & $--$&   $-0.105(2)(7)$&   $--$\\
 \hline
 $\langle I^{\pi}_{4}\rangle$ & $-0.126(0)(0)$   & $-0.106(1)(3)$&   $-0.133(1)(9)$&   $--$\\
 \hline
$\langle I^{\pi}_{5}\rangle$ & $0.284(6)(0)$   & $0.229(7)(8)$&   $--$&   $0.284(7)(4)$\\
 \hline
$\langle I^{\pi}_{6s}\rangle$ & $-0.217(10)(0)$   & $-0.175(8)(8)$&   $--$&   $--$\\
 \hline
 $\langle I^{\pi}_{6c}\rangle$   & $0.404(11)(0)$   & $0.334(8)(10)$&   $--$&   $--$\\
 \hline
 $\langle I^{\gamma}_{3}\rangle$  &$0.091(3)(0)$ & $0.081(2)(1)$   & $--$&   $--$\\
 \hline
 $\langle I^{\gamma}_{4}\rangle$ &$0.115(2)(0)$& $0.103(2)(2)$   & $--$&   $--$\\
 \hline
$\langle I^{\gamma}_{5}\rangle$ &$-0.257(7)(0)$ &$-0.217(7)(7)$   & $--$&   $--$\\
 \hline
$\langle I^{\gamma}_{6s}\rangle$ & $0.367(5)(0)$&$0.311(5)(10)$   & $--$&   $--$\\
 \hline
 $\langle I^{\gamma}_{6c}\rangle$   &$-0.098(5)(0)$& $-0.084(4)(3)$  & $--$&   $--$\\
 \hline
\end{tabular}
\end{center}
\end{table}


\section{Conclusions}\label{Conc}

The study of $B$ meson decays provides an opportunity to probe new physics (NP) while testing the parameters of the Standard Model (SM). In particular, several semileptonic decays involving $b \to c$ transitions exhibit significant deviations from SM predictions. In this work, we analyze the  $B \to D^{(*)}\tau\bar{\nu}_{\tau} $ decay within model independent framework as well as in the context of three leptoquark models. We calculate the physical observables, including the lepton flavor universality ratio $R_{D^{(*)}}$, the lepton polarization asymmetry $P_{\tau}(D^{(\ast)})$, the longitudinal helicity fraction $F^{D^{\ast}}_{L}$ of the $D^*$ meson, the forward-backward asymmetry $\mathcal{A}^{\text{FB}}_{D^{(*)}}$, and the normalized angular coefficients $\langle I^{n}_{\lambda}\rangle$ associated with the decay $B \to D^*(\to D\pi, D\gamma)\tau\bar{\nu}_{\tau}$. These observables are evaluated using the effective Hamiltonian formalism. We present numerical predictions for these observables within SM, model independent scenarios, and the three leptoquark models. Our findings offer valuable insights into potential new physics effects in semileptonic $b \to c$, charged current $B$-meson decays.


Based on this study, the findings reveal significant deviations in new physics scenarios from the predictions of SM for the analyzed physical observables. In model independent scenarios, it is observed that scenarios $S_{2}$ and $T$ exhibit the largest deviations from SM predictions for the observables $P_\tau(D)$, $P_\tau(D^*)$, $F_L^{D^*}$, and $\mathcal{A}^{FB}_{D}$. For the observable $\mathcal{A}^{FB}_{D^\ast}$, also scenario $V_{2}$, along with $S_{2}$ and $T$, show notable deviation from SM. Regarding the normalized angular coefficients in the decay $B\to D^{\ast}(\to D\pi, D\gamma)\tau^{-}\bar{\nu}$, scenarios $P$ and $T$ show significant deviations from SM predictions. It is important to mention here that in the performed global fit $V_{1},V_{2},S_{2}$ and $T$ scenarios are still allowed while the scenario $S_{1}$ is excluded at $2\sigma$. Also, Leptoquark models, $R_{2}$ and $S_{1}$, demonstrate substantial deviations from SM for various angular observables in both decay channels, $B\to D^{\ast}(\to D\pi)\tau^{-}\bar{\nu}$ and $B\to D^{\ast}(\to D\gamma)\tau^{-}\bar{\nu}$. Among these models, the $R_{2}$ leptoquark model exhibit more profound difference, whereas the $S_{1}$ model shows mild deviation from the SM prediction, in comparison. Lastly, $U_{1}$ model mostly indicates an overlap or less deviations compared to SM predictions. Future measurements of these observables at the LHC and upcoming colliders will be crucial for exploring potential new physics effects associated with these decays.
\section*{Acknowledgements}
We would like to thank Marco Fedele, Andreas Juettner, Ying-Nan Mao, Soumitra Nandi and Ipsita Ray for
helpful communications. Y.L is supported in part by the National Science Foundation of
China under the Grants No. 11925506, 12375089, 12435004, and the Natural Science
Foundation of Shandong province under the Grant No. ZR2022ZD26. One of us M.A.P.
would also like to express gratitude for the hospitality provided by Yantai University, as well
as acknowledge the support provided by the Higher Education Commission of Pakistan
through Grant no. NRPU/20-15142 during the research visit. Z.R.H would like to thank
Yantai University for its hospitality and acknowledge the support from the National Natural
Science Foundation of China under Grant 12305104 and the Education Department of Hunan Province under Grant No. 24B0503.
\bibliographystyle{refstyle}
\bibliography{references1}
\end{document}